\documentstyle[12pt,epsf]{article}

\newcommand{\be}{\begin{equation}}
\newcommand{\ee}{\end{equation}}
\newcommand{\ba}{\begin{eqnarray}}
\newcommand{\ea}{\end{eqnarray}}
\newcommand{\ep}{\varepsilon}
\newcommand{\nn}{\nonumber}

\begin{document}

\pagestyle{empty}


\vglue 2cm

\begin{center} \begin{Large} \begin{bf}
Laurent series expansion of a class of massive scalar one-loop
integrals to ${\cal O}(\ep^2)$.
\end{bf} \end{Large} \end{center}
\vglue 0.35cm
{\begin{center}
J.G.\ K\"{o}rner$^{a,\dag}$,
Z.\ Merebashvili$^{b,\ddag}$ and
M.\ Rogal$^{a,*}$
\end{center}}
\parbox{6.4in}{\leftskip=1.0pc
{\it a.\ Institut f\"{u}r Physik, Johannes Gutenberg-Universit\"{a}t,
         D-55099 Mainz, Germany}\\
\vglue -0.25cm
{\it b.\ High Energy Physics Institute,
Tbilisi State University, 0186 Tbilisi, Georgia}
}
\begin{center}
\vglue 1.0cm
\begin{bf} ABSTRACT \end{bf}
\end{center}
\vglue 1.0cm
{\rightskip=1.5pc
\leftskip=1.5pc
\tenrm\baselineskip=12pt
 \noindent
We use dimensional regularization
to calculate the ${\cal O}(\ep^2)$ expansion of all
scalar one-loop one-, two-, three- and four-point integrals that are needed
in the calculation of hadronic heavy quark production. The Laurent series
up to ${\cal O}(\ep^2)$ is needed as input to that part of the NNLO
corrections to heavy flavor production at hadron colliders where the
one-loop integrals appear in the loop-by-loop contributions. The four-point
integrals are the most complicated. The ${\cal O}(\ep^2)$ expansion of the
three- and four-point integrals contains in general polylogarithms up to
${\rm Li}_4$ and functions related to multiple polylogarithms of maximal
weight and depth four.

\vglue 2cm
PACS number(s): 12.38.Bx, 13.85.-t, 13.85.Fb, 14.65.Ha
}

\renewcommand{\thefootnote}{\dag}
\footnotetext{e-mail address: koerner@thep.physik.uni-mainz.de}
\renewcommand{\thefootnote}{\ddag}
\footnotetext{e-mail address: zaza@thep.physik.uni-mainz.de}
\renewcommand{\thefootnote}{*}
\footnotetext{e-mail address: rogal@thep.physik.uni-mainz.de}

\newpage

\pagestyle{plain}
\setcounter{page}{1}
\renewcommand{\theequation}{2.\arabic{equation}}
\renewcommand{\thefootnote}{\arabic{footnote}}

\begin{center}\begin{large}\begin{bf}
I. INTRODUCTION
\end{bf}\end{large}\end{center}
\vglue .3cm

The full next-to-leading order (NLO) corrections to the hadroproduction of heavy
flavors have been completed in the late eighties
$[$\ref{Dawson:1988},\ref{Been}$]$. They
have raised the leading order (LO) estimates $[$\ref{LO:1978}$]$ but several
initial analysis' showed a serious disagreement with experimental results
$[$\ref{CDF}, \ref{D0}$]$.
Recently the situation has considerably improved in that a more refined NLO
analysis (due to considerably more precise experimental input for the
$b$-quark fragmentation function as well as other QCD parameters) now
shows signs of rapprochement between theory and the new
experimental data (see $[$\ref{Italians}$]$ and references therein for the
new CDF measurements).
However, the NLO predictions are still slightly
below the experimental numbers. Moreover, the theoretical NLO predictions
suffer from the usual large uncertainty resulting from the freedom in the
choice of renormalization and factorization scales of perturbative
QCD. In this light there are hopes that a next-to-next-to-leading
order (NNLO) calculation will bring theoretical predictions even closer to
the experimental data. Also,
the dependence on the factorization and renormalization scales of the
physical process is expected to be greatly reduced at NNLO. This would
reduce the theoretical uncertainty and therefore make the comparison
between theory and experiment much more significant.

In Fig.~1 we show one generic diagram each for the four classes
of gluon-induced contributions that need to be calculated for the NNLO
corrections to hadroproduction of heavy flavors.
They involve the two-loop
contribution (1a), the loop-by-loop contribution (1b), the one-loop
gluon emission contribution (1c) and, finally, the two gluon
emission contribution (1d). A similar classification holds for the
quark-induced contributions.
\\
\\
\[
\mbox{ \epsfysize = 4.0cm  \epsffile {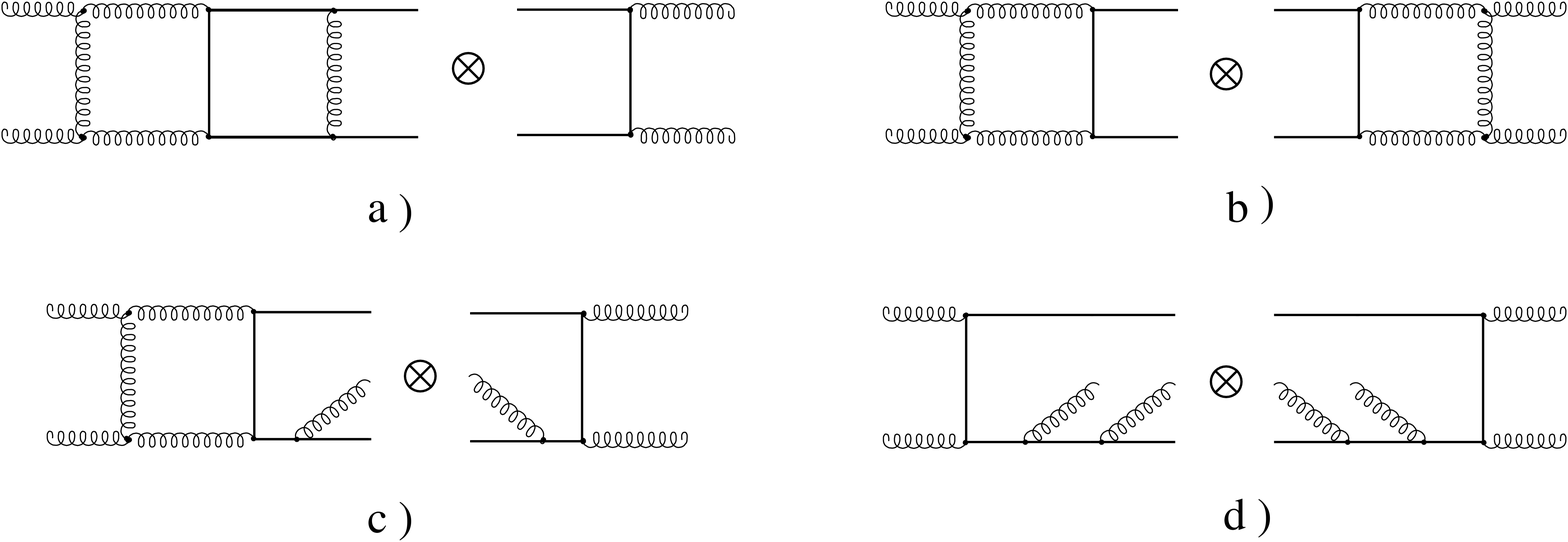}}
\]
{\begin{center}
\tenrm\vglue -0.2cm
FIG.~1. Exemplary gluon fusion diagrams for the NNLO calculation
of heavy hadron production.
\end{center}
}
\vglue .1cm
In this paper we concentrate on the
loop-by-loop contribution Fig.~1b. Specifically, working in the framework
of the dimensional regularization scheme $[$\ref{DREG}$]$, we shall
present ${\cal O}(\ep^2)$ results on
all scalar one-loop one-, two-, three- and four-point integrals that are
needed in the calculation of hadronic heavy flavour production.
We generate the coefficients of the $\ep$-expansion in a rather direct
way. Let us briefly describe our procedure. We introduce Feynman parameter
representations
for each of the two-, three- and four-point integrals. The
two-point case is straightforward. In case of the three- and four-point
integrals we generally integrate over one and two Feynman parameters,
respectively, keeping the full $\ep$-dependence of the result. Before
doing the last Feynman parameter integration we expand the respective
integrands up to ${\cal O}(\ep^2)$ and then integrate the expanded
integrand term by term.

Because the one-loop integrals exhibit ultraviolet (UV) and infrared
(IR)/collinear (or mass (M))
singularities up to ${\cal O}(\ep^{-2})$ one needs to know the one-loop
integrals up to ${\cal O}(\ep^2)$ because the one-loop contributions appear
in product form in the loop-by-loop
contributions\footnote{In a more
general setting the Laurent-series expansion of the scalar integrals is
needed if the integration-by-parts technique $[$\ref{ibyparts}$]$ is
employed. The reason is that the solution of the recursion relations induced by
the integration-by-parts technique can bring in negative powers of $\ep$.}.
It is clear that the spin
algebra and the Passarino-Veltman decomposition of tensor integrals in the
one-loop contributions also have to be done up to ${\cal O}(\ep^2)$.
This task will be left to a companion paper where we present complete
results on the one-loop amplitudes up to ${\cal O}(\ep^2)$ including
spin algebra and Passarino-Veltman decomposition effects.

The general case of massive one-loop integrals
was studied some time ago $[$\ref{Andrei}$]$, where a general
one-loop $N$-point integral was expressed in terms of hypergeometric
functions of several variables.
Recently, there have been a number of papers where the authors took a more
general attitude to calculate the $\ep$-expansion of massive one-loop
integrals. They write down general representations of the $\ep$-expansion
of one-loop integrals for general kinematic configurations. We have
attempted to compare our results with the results of the more general
approaches whenever possible.
In papers $[$\ref{AndreiExp}, \ref{Binomial}$]$ the all-order
$\ep$-expansion of one-loop two-point and of certain
three-point functions was done explicitly by expanding the relevant hypergeometric
functions.
One-fold integral representations for general three- and four-point
functions, as well as ways to get expansion terms of order $\ep$ for
3-point
functions, were worked out in a recent paper $[$\ref{Tarasov}$]$.
Publications $[$\ref{Andrei}, \ref{Binomial}, \ref{Tarasov}$]$
also contain a comprehensive list of references on the subject.

However, in general, the required $\ep$-expansion (including $\ep^2$-terms)
is not readily available for all the integrals needed in the hadronic heavy
flavour production process. Also, the analytic continuation of the above
mentioned hypergeometric functions in $[$\ref{Andrei}, \ref{Tarasov}$]$
to the appropriate kinematical regions of
validity is not always possible. This mainly concerns the four-point
functions. In addition, it is more convenient to
present results for the $\ep$-expansion in terms of simpler special
functions, in the form convenient for numerical evaluation. And finally,
collecting together all the necessary scalar integrals needed for the
derivation of tensor integrals entering the loop-by-loop contribution
constitutes
a first step in the difficult task of obtaining the NNLO
corrections to heavy flavor hadroproduction cross section.

In our notation we shall remain very close to the notation introduced and
used in $[$\ref{Been},\ref{KM}$]$. In particular, we use dimensional
regularization working in $n=4-2\ep$ dimensions, as e.g. in $[$\ref{KM}$]$.
For the calculation of the NNLO virtual
corrections to hadroproduction of heavy flavors one needs the same set of
scalar master integrals as given in the Appendix~A of $[$\ref{Been}$]$
(the relevant set of master integrals is listed in Table~\ref{t:tab1}).
However, as explained above, knowledge of their singular and finite terms
is not sufficient for the calculation of NNLO loop-by-loop corrections.
For that purpose one needs to know the one-loop integrals up to
${\cal O}(\ep^2)$ including also their imaginary parts which equally well
contribute to the modulus squared of the one-loop amplitudes.
The imaginary parts of the one-loop integrals are really needed only up to
${\cal O}(\ep)$ since the highest singularity of the imaginary parts is
only ${\cal O}(\ep^{-1})$ compared to ${\cal O}(\ep^{-2})$ for the real
parts. We have nevertheless decided to include ${\cal O}(\ep^2)$ results
also for the imaginary parts which may be of interest in other
applications. Consequently, in this paper we
present the relevant expressions for all scalar integrals needed in the
calculation of the NNLO loop-by-loop corrections to hadroproduction of
heavy flavors. For reasons of comprehensiveness we have decided to include also
the singular and finite (i.e. ${\cal O}(\ep^{0})$) parts of the scalar
integrals in our presentation. They agree with the results of the real
contributions presented in $[$\ref{Been}$]$.

A comment on the length of the formula expressions in our paper is appropriate.
The untreated computer output of the integrations is generally quite lengthy.
The hard work is to simplify these expressions. We have written semi--automatic
computer codes that achieve the simplifications using known identities among
polylogarithms and using a number of identities for the $L$--functions introduced
in this paper which are derived in an Appendix.

The paper is organized as follows. In Sec.~II we deal with one- and
two-point functions. Sec.~III contains our results on three-point functions,
while in Sec.~IV we present our results for the $\ep$-expansion of the
four-point functions. In Sec.~V we give our summary and conclusions.
We collect some technical material in three Appendixes. In Appendix A we
discuss the Taylor series expansion around $p^2=m^2$ for the self--energy
insertion two--point function which is needed for the calculation of the
heavy quark wave function renormalization constant. In Appendix B we define
multiple polylogarithms and demonstrate that our analytical results can all
be expressed in terms of multiple polylogarithms. In Appendix C we derive
a number of identities for the so--called $L$--functions introduced in the
main text. A judicious use of these identities has allowed us to considerably
reduce the length of our final analytical results for the three-- and four--point
functions.

\begin{table}
\caption{List of one-, two-, three- and four-point massive one-loop functions
calculated in this paper up to ${\cal O}(\ep^2)$.}

\begin{tabular}{lclcccl}    \hline\hline
& &Nomenclature of $[$\ref{Been}$]$ & Our nomenclature & Novelty && Comments
\\  \hline
1-point & &       $A(m)$     &  $A$  & -- && Re  \\
\hline
2-point & & $B(p_4-p_2,0,m)$ & $B_1$ & -- && Re  \\
        & & $B(p_3+p_4,m,m)$ & $B_2$ & -- && Re, Im   \\
        & & $B(p_4,0,m)$     & $B_3$ & -- && Re   \\
        & & $B(p_2,m,m)$     & $B_4$ & -- && Re   \\
        & &  $B(p_3+p_4,0,0)$ & $B_5$ & -- && Re, Im   \\
\hline
3-point & & $C(p_4,p_3,0,m,0)$ & $C_1$ & new && Re, Im   \\
        & & $C(p_4,-p_2,0,m,m)$ & $C_2$ & new && Re  \\
        & & $C(-p_2,p_4,0,0,m)$ & $C_3$ & -- && Re  \\
        & & $C(-p_2,-p_1,0,0,0)$ & $C_4$ & -- && Re, Im  \\
        & & $C(-p_2,-p_1,m,m,m)$ & $C_5$ & -- && Re, Im  \\
        & & $C(p_3,p_4,m,0,m)$ & $C_6$ & -- && Re, Im  \\
\hline
4-point & & $D(p_4,-p_2,-p_1,0,m,m,m)$ & $D_1$ & new && Re, Im   \\
        & & $D(-p_2,p_4,p_3,0,0,m,0)$ & $D_2$ & new && Re, Im   \\
        & & $D(-p_2,p_4,-p_1,0,0,m,m)$ & $D_3$ & new && Re  \\ \hline\hline
\end{tabular}
\label{t:tab1}
\end{table}

\vglue 1.0cm
\begin{center}\begin{large}\begin{bf}
II. ONE- AND TWO-POINT FUNCTIONS
\end{bf}\end{large}\end{center}
\vglue .3cm

We start with the one-loop one-point function which is defined by
\be
A(m)=\mu^{2\ep} \int \frac{d^nq}{(2\pi)^n} \frac{1}{q^2-m^2}.
\ee
Working in $n=4-2\ep$ dimensions,
the expansion for this one-point scalar integral can be written in the
general form:
\be
A(m) = iC_{\ep}(m^2)\frac{m^2}{\ep (1-\ep)} =
        iC_{\ep}(m^2) m^2 \left\{ \frac{1}{\ep} + 1 +
                          \ep + \ep^2 + {\cal O}(\ep^3) \right\}.
\ee
where $m$ is the internal loop mass and where we have defined
\be
\label{ceps}
C_{\ep}(m^2)\equiv\frac{\Gamma(1+\ep)}{(4\pi)^2}
\left(\frac{4\pi\mu^2}{m^2}\right)^\ep .
\ee

The one-loop two-point functions are defined by $[$\ref{Been}$]$
\be
\label{two-point}
B(q_1,m_1,m_2)=\mu^{2\ep} \int \frac{d^nq}{(2\pi)^n}
\frac{1}{(q^2-m_1^2)[(q+q_1)^2-m_2^2]},
\ee
where the $m_i\,\, (i=1,2)$ can be either $m$ or $0$. In the
denominators of the relevant functions we always
imply the ``causal'' $+i\delta$ prescription to deal with singularities in
pseudo-Euclidean space.

In what follows, we will always present our
results for the scalar functions separately for the real and imaginary
contributions. We introduce the Mandelstam-type variables
\be
\label{s-t-u}
s\equiv (p_1+p_2)^2, {\rm \hspace{.3in}}  t\equiv T-m^2
\equiv (p_1-p_3)^2-m^2,
{\rm \hspace{.3in}}  u\equiv U-m^2\equiv (p_2-p_3)^2-m^2,
\ee
with the kinematical condition on external momenta being $p_1+p_2=p_3+p_4$
(i.e. $s+t+u=0$) and the on-shell conditions are $p_1^2=p_2^2=0, \,\,
p_3^2=p_4^2=m^2$. Note that the variables $t$ and $u$ defined in
$({\ref{s-t-u}})$ are {\it not} the usual Mandelstam variables.

There are altogether five different two-point scalar functions $B_i \,\,
(i=1,2,...,5)$ needed for hadronic heavy flavor production $[$\ref{Been}$]$.
Again we choose to extract a common factor $i C_{\ep}(m^2)$, where
$C_{\ep}(m^2)$ is defined in (\ref{ceps}).
The coefficients of the $\ep$-expansion are denoted by $B_i^{(j)}$, i.e. we
write
\be
B_i = i\,C_{\ep}(m^2) \left\{\frac{1}{\ep} B_i^{(-1)} +
B_i^{(0)} + \ep B_i^{(1)} + \ep^2
B_i^{(2)} + {\cal O}(\ep^3) \right\}.
\ee
The $\ep$-expansion of the two-point functions starts at $\ep^{-1}$. It
turns out that $B_i^{(-1)}=1$ for all $i$.
The first two-point function
\be
B_1 \equiv B(p_4-p_2,0,m)
\ee
is real for our kinematics which can be seen by drawing the appropriate Feynman
diagram for $B_1$ and applying the Landau-Cutkosky rules. The same statement holds
true for the two-point functions $B_3$ and $B_4$ to be discussed later on. One has
\ba
\label{B1}  &&
{\rm Re}\,B_1^{(-1)}= 1,         \\
\nn   &&
{\rm Re}\,B_1^{(0)}= 2 - \frac{t}{T} \ln\frac{-t}{m^2},    \\
\nn   &&
{\rm Re}\,B_1^{(1)} = 2 {\rm Re}\,B_1^{(0)} +
\frac{t}{T} \ln^2\frac{-t}{m^2} + \frac{t}{T} {\rm
Li}_2(\frac{T}{m^2}),                 \\
\nn  &&
{\rm Re}\,B_1^{(2)}= 2 {\rm Re}\,B_1^{(1)} - \frac{t}{T} \left[
\frac{1}{3} \ln^3\frac{-t}{m^2}
+ 2 {\rm Li}_3(\frac{T}{t}) + {\rm Li}_3(\frac{T}{m^2}) \right];  \\
&&
{\rm Im}\,B_1^{(j)}= 0.
\ea

\noindent
The second scalar two-point function
\be
B_2 \equiv B(p_3+p_4,m,m)
\ee
has both real and imaginary parts. Defining
$\beta=(1-4m^2/s)^{1/2}$ and $x=(1-\beta)/(1+\beta)$ we obtain
\ba
\label{B2re} &&
{\rm Re}\,B_2^{(-1)} = 1,          \\
\nn  &&
{\rm Re}\,B_2^{(0)} = 2 + \beta\ln x,             \\
\nn  &&
{\rm Re}\,B_2^{(1)} = \beta \left[ \frac{4}{\beta} - 2
\ln(\frac{s\beta^2}{m^2})
+ \frac{1}{2}\ln^2 (\frac{s\beta^2}{m^2})
+ 4 \ln (1-x) - 2 \ln^2 (1-x) - 4\zeta(2) - 2{\rm Li}_2(x) \right],   \\
\nn  &&
{\rm Re}\,B_2^{(2)} = 2 {\rm Re}\,B_2^{(1)} + \beta \left[
              4\zeta(2)\ln (\frac{s\beta^2}{m^2})
       - \frac{1}{6}\ln^3(\frac{s \beta^2}{m^2})
       + 4 \zeta(2) \ln(1-x) - 2\ln x \ln^2(1-x)
   \right.   \\
\nn
&& {\rm \hspace{2in}}      \left.
               + \frac{4}{3} \ln^3(1-x)
              + 2\zeta(3)
           - 4{\rm Li}_3(1-x) - 2{\rm Li}_3(x)
\right];                                 \\
\label{B2im}  &&
{\rm Im}\,B_2^{(0)} = \pi\beta, {\rm \hspace{.4in}}
{\rm Im}\,B_2^{(1)}= \pi\beta\left[
                   2 - \ln(\frac{s\beta^2}{m^2})\right],        \\
\nn  &&
{\rm Im}\,B_2^{(2)} =   2 {\rm Im}\,B_2^{(1)} + \pi\beta\left[
     \frac{1}{2}\ln^2(\frac{s\beta^2}{m^2}) - 2\zeta(2) \right].
\ea

The remaining three two-point functions have a simple structure:
\ba
\label{B3}
\nonumber
B_3 \equiv B(p_4,0,m) &=&  iC_{\ep}(m^2) \frac{1}{\ep (1-2\ep)}     \\
        &=&  iC_{\ep}(m^2) \left\{ \frac{1}{\ep} + 2 + 4\ep + 8\ep^2
+ {\cal O}(\ep^3)   \right\};     \\
\label{B4}
B_4 \equiv B(p_2,m,m) &=&  iC_{\ep}(m^2)  \frac{1}{\ep};            \\
\label{B5}
B_5 \equiv B(p_3+p_4,0,0) &=&  iC_{\ep}(m^2)
\frac{\Gamma^2(1-\ep)}{\Gamma(2-2\ep)}
\left(-\frac{s+i\delta}{m^2}\right)^{-\ep}
\frac{1}{\ep}.
\ea
The results for $B_3$ and $B_4$ in $(\ref{B3}),(\ref{B4})$ are not separately
listed in the standard
format $B_i^{(j)}$ which can of course be read off from the relevant
expressions $(\ref{B3}),(\ref{B4})$.
The two-point function $B_5$ $(\ref{B5})$ has both real
and imaginary parts:
\ba   &&
{\rm Re}\,B_5^{(-1)} = 1,          \\
\nn  &&
{\rm Re}\,B_5^{(0)}= 2 - \ln\frac{s}{m^2},  \\
\nn   &&
{\rm Re}\,B_5^{(1)}= 2 {\rm Re}\,B_5^{(0)} - 4\zeta(2) + \frac{1}{2}
                                     \ln^2\frac{s}{m^2},    \\
\nn   &&
{\rm Re}\,B_5^{(2)} = 2 {\rm Re}\,B_5^{(1)} + 4 \zeta(2) \ln\frac{s}{m^2}
- \frac{1}{6}\ln^3\frac{s}{m^2} - 2\zeta(3);         \\
\nn   \\    &&
{\rm Im}\,B_5^{(0)} = \pi, {\rm \hspace{.4in}}
{\rm Im}\,B_5^{(1)} = \pi \left[ 2 - \ln\frac{s}{m^2} \right],  \\
\nn   &&
{\rm Im}\,B_5^{(2)} = 2 {\rm Im}\,B_5^{(1)} + \pi \left[
                \frac{1}{2}\ln^2\frac{s}{m^2} - 2\zeta(2) \right].
\ea

We have done various checks on the
above results. First of all, they were double-checked, i.e. the results
were obtained by two independent calculations. Secondly, they were
checked numerically by verifying that the original integrals (after
Feynman parametrization and integrating out the loop momentum, and for $B_1$
and $B_2$ also expanding in $\ep$) are equal numerically to the final
integrals.
We have also verified our results by extracting the relevant expressions from
general formulae given in $[$\ref{Andrei},\ref{AndreiExp}$]$.
In particular, our coefficients
$(\ref{B1})$ may be obtained from Eq.~(10) of the first reference of
$[$\ref{Andrei}$]$ and
then using Eq.~(2.14) of $[$\ref{AndreiExp}$]$. Our result $(\ref{B2re}),
(\ref{B2im})$ can be obtained from Eqs.~(2.10) and (2.14) of
$[$\ref{AndreiExp}$]$. Finally, our expressions $(\ref{B3})-(\ref{B5})$
can also be obtained from Eqs.~(10), (17) and (8), respectively, of the first
reference of $[$\ref{Andrei}$]$.

There is one more special case of the two-point integral which is needed for the
calculation of a self-energy insertion into external massive
fermion lines. This integral is used for the definition of the fermion mass
and wave function renormalization constants in the on-shell renormalization
scheme. This specific two-point function is given in the Appendix~A of this paper.

\vglue 1cm
\renewcommand{\theequation}{3.\arabic{equation}}
\begin{center}\begin{large}\begin{bf}
III. THREE-POINT FUNCTIONS
\end{bf}\end{large}\end{center}
\vglue .3cm

The one-loop three-point functions are defined by $[$\ref{Been}$]$
\be
C(q_1,q_2,m_1,m_2,m_3)=\mu^{2\ep} \int \frac{d^nq}{(2\pi)^n}
\frac{1}{(q^2-m_1^2)[(q+q_1)^2-m_2^2][(q+q_1+q_2)^2-m_3^2]}.
\ee
The three masses $m_1,m_2$ and $m_3$ come in various combinations of zero
and nonzero masses where all nonzero masses are equal to $m$ as
before.
There are six different types of three-point functions $C_i\,\, (i=1,2,...,6)$
needed for our purposes $[$\ref{Been}$]$.
They have both real and imaginary parts except for $C_2$ and $C_3$ which are real.
Again, this can be seen from the Feynman diagrams representing $C_2$ and $C_3$
and applying the Landau-Cutkosky rules. Their $\ep$-expansion is again written in
the following universal format:
\be
C_i = i\,C_{\ep}(m^2) \left\{\frac{1}{\ep^2} C_i^{(-2)} + \frac{1}{\ep}
C_i^{(-1)} + C_i^{(0)} + \ep C_i^{(1)} + \ep^2
C_i^{(2)} + {\cal O}(\ep^3) \right\},
\ee
where the $\ep$-expansion now starts at $\ep^{(-2)}$. Note that the $
C_i^{(-2)}$ are purely real.

It turns out that the ${\cal O}(\ep^2)$ results for the three-point functions can
no longer be presented in terms of classical polylogarithms but require a new
class of functions given by the one-fold integral representations defined below.
To write down our results in a short and convenient form, we introduce the
following functions:
\be
\label{Lfunction}
L_{\sigma_1\sigma_2\sigma_3}(\alpha_1,\alpha_2,\alpha_3,\alpha_4)=
\int_0^1 dy \frac{\ln (\alpha_1+\sigma_1 y) \ln (\alpha_2+\sigma_2 y)
\ln (\alpha_3+\sigma_3 y)}{\alpha_4+y},
\ee
and
\be
\label{Lpfunction}
L_{\sigma_1}(\alpha_1,\alpha_2,\alpha_3,\alpha_4)=
\int_0^1 dy \frac{\ln (\alpha_1+\sigma_1 y) {\rm Li}_2(\alpha_2+\alpha_3 y)
}{\alpha_4+y}.
\ee
Here the $\sigma_i\,\, (i=1,2,3)$ take values $\pm 1$ and the $\alpha_j$'s are
either integers $\{1,0,-1\}$ or else
kinematical variables. The above $L$-functions arise naturally in our
calculational framework
\footnote{As A.~Davydychev informs us the functions analogous to our
triple-index functions $L_{\sigma_1\sigma_2\sigma_3}$
also arise in the approach of $[$\ref{AndreiExp}$]$ when one analytically
continues their Eq.~(3.2) for the order $\ep^2$ terms.}.
They can all be
expressed in terms of so-called multiple polylogarithms of maximum weight four
$[$\ref{Multilogs}$]$ (see Appendix~B for details).
However, we choose to write our results in terms of
the above single- and triple-index $L$-functions for several reasons.
The results look simpler, e.g.
they can be expressed as one-fold integrals of products of logarithms and
dilogarithms, and are shorter. We have
also found that the $L$-functions are much easier to evaluate
numerically than the corresponding multiple polylogarithms (see $[$\ref{Weinzierl}$]$
for relevant details).

There exist simple algebraic relations between these $L$-functions based on
either symmetry relations regarding permutations of indices and change of
integration variables
or on relations based on integration-by-parts techniques. We describe them
in Appendix~C. In particular this means that our results on the three- and
four-point functions can all be written in terms of the
$L_{-++}$ and $L_{+++}$
variants of the triple--index $L_{\sigma_1\sigma_2\sigma_3}$ functions in
Eq.~(\ref{Lfunction}),
and of the $L_+$ variant of the single--index $L_{\sigma_1}$ function of
Eq.~(\ref{Lpfunction}).

We start with the three-point function $C_1$ defined by
\[
          C_1 \equiv C(p_4,p_3,0,m,0).
\]
One obtains
\ba
\label{c1real}
{\rm Re}\,C_1^{(-2)} &=& {\rm Re}\,C_1^{(-1)} = 0,         \\
\nn
{\rm Re}\,C_1^{(0)} &=& \frac{1}{2s\beta} [\ln^2x + 4{\rm Li}_2(-x) +
2\zeta(2)],                                     \\
\nn
{\rm Re}\,C_1^{(1)} &=&
    - \frac{1}{s\beta} \left[ \frac{1}{6} \ln^3\frac{s}{m^2}
    + 2\ln\frac{s}{m^2}\ln(1-x)\ln x + \ln(1-x)\ln^2x
    - 4\zeta(2)\ln\frac{s}{m^2}        \right.    \\
\nn
    &+& 2\ln\frac{s}{m^2}{\rm Li}_2(x) + 5\zeta(3)
     - 4{\rm Li}_3(1-x) + 2{\rm Li}_3(x)
    +  2{\rm Li}_3(1-x^2)                          \\
\nn
         &-&  \left.  8{\rm Li}_3(\frac{1}{1+x})  \right],       \\
\nn
{\rm Re}\,C_1^{(2)} &=&   \frac{1}{s\beta} \left[
     \frac{7}{48} \ln^4 \frac{s}{m^2} -
    \frac{11}{24} \ln^3 \frac{s}{m^2}\,\ln x
-  \frac{1}{4} \ln^2 \frac{s}{m^2} \left(
        \ln^2 (1 - x) + 6 \ln (1 - x) \ln x
                                                    \right.  \right.   \\
\nn
&+& \left.   5 \ln^2 x \right)      + \ln \frac{s}{m^2}
     \left( \frac{1}{3} \ln^3 (1 - x) - \frac{7}{2} \ln^2 (1 - x) \ln x
      + \frac{11}{24} \ln^3 x \right)                         \\
\nn
&-& \frac{1}{3} \ln^3 (1 - x) \ln x -
    \frac{1}{4} \ln^2 (1 - x) \ln^2 x +
    \frac{5}{2} \ln (1 - x) \ln^3 x + \frac{55}{48} \ln^4 x           \\
\nn
&-& \frac{7}{8}  \left( 3 \ln^2 \frac{s}{m^2} + 10 \ln \frac{s}{m^2} \ln x +
       16 \ln (1 - x) \ln x + 3 \ln^2 x \right)  \zeta(2)          \\
\nn
&+& \frac{1}{2} \left( \ln \frac{s}{m^2} - 11 \ln x \right) \zeta(3) +
    \frac{5}{2}\, \zeta(4) - 2 {\rm Li}_2^2(-x)
+ {\rm Li}_2(x) \left( -\ln^2 \frac{s}{m^2}        \right.                \\
\nn
&+&  \left. 2 \ln (1 - x) \ln x + 7 \ln^2 x - 2\ln \frac{s}{m^2}
\left( \ln (1 - x) - \ln x \right)  - 10\, \zeta(2) \right)         \\
\nn
&+&    {\rm Li}_2(-x)
     \left( -\frac{13}{4} \ln^2 \frac{s}{m^2}
    - 2 \ln \frac{s}{m^2}\left( \ln (1 - x) - \ln x \right)
                                      + 2\ln (1 - x)\ln x      \right.  \\
\nn
&+&  \left. \frac{31}{4} \ln^2 x
 - 6\,\zeta(2) \right) + 4{\rm Li}_3(1 - x) \left( \ln \frac{s}{m^2}
                  + 2\ln (1 - x)  - \ln x \right)            \\
\nn
&-& {\rm Li}_3(-x) \left( \frac{21}{2} \ln \frac{s}{m^2} +
       20\ln (1 - x) + \frac{29}{2}\ln x \right)
- 4 {\rm Li}_3(x) \left( \ln \frac{s}{m^2} + 2\ln (1 - x)   \right.     \\
\nn
&+&   \left.    \frac{5}{2} \ln (x) \right)
-  {\rm Li}_3 \left( \frac{1}{1 + x} \right) \left( 3\ln \frac{s}{m^2} + 4\ln (1 -
x) - 2\ln x \right)
- 2 {\rm Li}_3 \left( 1 - x^2 \right)  \times    \\
\nn        &&
 \left( \ln \frac{s}{m^2} + 2 \ln (1 - x) - \ln x \right)
- 4\, {\rm Li}_4(1 - x) - 4\,{\rm Li}_4 \left( \frac{-x}{1 - x} \right)
   + 12\,{\rm Li}_4(-x)                                        \\
\nn
&+& 28 {\rm Li}_4(x) +  17 {\rm Li}_4 \left( \frac{1}{1 + x} \right) -
    17 {\rm Li}_4 \left( \frac{x}{1 + x} \right) +
    4 {\rm Li}_4 \left( 1 - x^2 \right) +
    4 {\rm Li}_4 \left( \frac{-x^2}{1 - x^2} \right)           \\
\nn
&-& L_{-++} \left( 1,x^{-1},x^{-1},0 \right) +
    L_{-++} \left( 1,x^{-1},x^{-1},x \right) +
    2\,L_{-++} \left( 1,x,x^{-1},0 \right)           \\
\nn
&-& 2\,L_{-++} \left( 1,x,x^{-1},x^{-1} \right) -
    2\,L_{-++} \left( 1,x,x^{-1},x \right) +
    L_{-++}(1,x,x,0)                  \\
\nn
&-& L_{-++} \left( 1,x,x,x^{-1} \right) +
    2\,L_{-++} \left( 1 + x^{-1},0,0,-1  \right) -
    2\,L_{-++}(1 + x,0,0,-1)         \\
\nn
&-& 2\,L_{-++} \left( 1 + x^{-1},0,0,-1 - x \right) +
    2\,L_{-++} \left( 1 + x,0,0,-1 - x^{-1} \right)    \\
\nn
&+&
    2\,L_{+} \left( 0,0,-x^{-1},x \right) -
    2\,L_{+} \left( 0,0,-x,x^{-1} \right) -
    4\,L_{+}(0,0,-x,x)                       \\
\nn
&-& 4\,L_{+} \left( 0,-x^{-1},x^{-1},-1 \right) +
    4\,L_{+} \left( 0,-x^{-1},x^{-1}, -1 - x \right) +
    4\,L_{+}(0,-x,x,-1)                 \\
\nn
&+& 4\,L_{+} \left( 0,-x^{-1},x^{-1},-1 - x^{-1} \right) -
    4\,L_{+}(0,-x,x,-1 - x)                        \\
\nn
&-& 4\,L_{+} \left( 0,-x,x,-1 - x^{-1} \right) -
    4\,L_{+} \left( 0,\frac{-x}{1 - x},\frac{x}{1 - x^2},-1 \right)  \\
\nn
&+& 4\,L_{+} \left( 0,\frac{-x}{1 - x},\frac{x}{1 - x^2},-1 - x \right) +
    4\,L_{+} \left( 0,\frac{-x}{1 - x},\frac{x}{1 - x^2},-1 - x^{-1} \right) \\
\nn
&+& 2\,L_{+} \left( x^{-1},0,-x^{-1},x \right) +
    2\,L_{+} \left( x^{-1},0,-x,x \right) -
    3\,L_{+++} \left( 0,0,x^{-1},x \right)       \\
\nn
&-& L_{+++} \left( 0,0,x,x^{-1} \right) +
    \frac{3}{2}\,L_{+++} \left( 0,x^{-1},x^{-1},x \right) +
    3\,L_{+++}(0,x^{-1},x,x)          \\
\nn
&+&   \left.   L_{+++} \left( 0,x,x^{-1},x^{-1} \right) +
    \frac{1}{2}\,L_{+++} \left( 0,x,x,x^{-1} \right) -
    2\,L_{+++} \left( x^{-1},x,x^{-1},x \right)       \right];
\nn          \\
\label{c1imag}
{\rm Im}\,C_1^{(-1)}&=&0, {\rm \hspace{.6in}}
{\rm Im}\,C_1^{(0)}=\frac{\pi}{s\beta} \ln x,        \\
\nn
{\rm Im}\,C_1^{(1)}&=&-\frac{\pi}{2s\beta}   \left[
       2\ln\frac{s}{m^2}\ln x - 4\ln(1-x)\ln x + \ln^2x + 4\zeta(2)
     - 4{\rm Li}_2(x)   \right],                    \\
\nn
{\rm Im}\,C_1^{(2)}&=&\frac{\pi}{6s\beta}     \left[
   3\ln^2\frac{s}{m^2}\ln x - 12\ln\frac{s}{m^2}\ln(1-x)\ln x +
3\ln\frac{s}{m^2}\ln^2x - 6\ln(1-x)\ln^2x    \right.  \\
\nn
&+&     \left.    \ln^3x
- 12\ln\frac{s}{m^2}{\rm Li}_2(x) - 24{\rm Li}_3(1-x) - 12{\rm Li}_3(x)
+ 12\zeta(2)\ln\frac{s}{m^2} + 12\zeta(3)     \right].
\ea
We have not been able to derive the corresponding result from the known
general
hypergeometric function that represents the above integral in
$[$\ref{Andrei}$]$. On the other
hand, for a general three-point function, an expression for
the order $\ep$-terms was obtained in
$[$\ref{Nierste}$]$ in terms of simple polylogarithms up to
${\rm Li}_3$. However, we believe that
the result Eq.~(5.21) in $[$\ref{Nierste}$]$ is not applicable to our case
as one faces
singularities resulting from vanishing denominators in the arguments of the
relevant logarithms and
polylogarithms.
We have checked our final result numerically against the original two-fold
and one-fold Feynman parameter integrals (after $\ep$-expanding
the corresponding integrand). This was done term by term for coefficients at the
corresponding orders in $\ep$.
Although the result for the $\ep^2$-coefficient looks lengthy, our final
analytic results (\ref{c1real}), (\ref{c1imag}) for the three-point function $C_1$
integrate out numerically very fast (in fraction of a
second on a desktop computer for a chosen numerical point) and without any
problems. In comparison, the numerical integration of the one-fold integral by
Mathematica took eight times longer, and that of the two-fold integral even
200 times longer to evaluate. In addition, because of various branch cuts, the
one- and two-fold integrals
would only allow integrations in the complex plane of kinematical variables,
while for the physical region they have severe problems.

The integral $C_2$ is real and finite:
\[
               C_2 \equiv C(p_4,-p_2,0,m,m);
\]
\ba
{\rm Re}\,C_2^{(-2)} &=& {\rm Re}\,C_2^{(-1)} = 0,     \\
\nn
{\rm Re}\,C_2^{(0)} &=& \frac{1}{t} \left[ \zeta(2) - {\rm
Li}_2(\frac{T}{m^2})     \right],                 \\
\nn
{\rm Re}\,C_2^{(1)} &=&
     \frac{1}{3t} \left[ \ln^3\frac{-t}{m^2} + 6 \ln\frac{-t}{m^2} {\rm
Li}_2(\frac{T}{m^2}) + 9 \zeta(3) - 6 {\rm Li}_3(\frac{T}{t}) - 9 {\rm
Li}_3(\frac{T}{m^2}) \right],           \\
\nn
{\rm Re}\,C_2^{(2)} &=&
\frac{1}{24t} \left[ - 5 \ln^4\frac{-t}{m^2} + 12 \zeta(2)
\ln^2\frac{-t}{m^2} + 12 \zeta(2) \ln^2\frac{-T}{m^2}
- 12 \ln^2\frac{-t}{m^2} {\rm Li}_2(\frac{T}{m^2})  \right.    \\
\nn
&&
- 24 \ln\frac{-t}{m^2} \ln\frac{-T}{m^2} {\rm Li}_2(\frac{T}{m^2}) - 24
\zeta(3) \ln\frac{-t}{m^2} + 24 \zeta(3) \ln\frac{-T}{m^2}     \\
\nn
&&
+ 24 \ln\frac{-t}{m^2} {\rm Li}_3(\frac{T}{t})
+ 24 \ln\frac{-T}{m^2} {\rm  Li}_3(\frac{T}{t}) + 24 \ln\frac{-t}{m^2}
{\rm Li}_3(\frac{T}{m^2})               \\
\nn
&&
+ 48 \ln\frac{-T}{m^2} {\rm  Li}_3(\frac{T}{m^2}) +
192 \zeta(4) - 24 {\rm  Li}_4(\frac{m^2}{-t})               \\
\nn
&&     \left.
+ 12 L_{-++}(1,-\frac{m^2}{T},-\frac{m^2}{T},0)
- 12 L_{-++}(\frac{t}{T},0,0,-1)      \right];               \\
{\rm Im}\, C_2^{(j)} &=& 0.
\ea
This result was checked numerically against the original double parametric
representation (obtained after doing Feynman parametrization) of this
integral expanded in powers of $\ep$. We could not obtain similar
expressions from known general results for this integral, as the
$\ep$-expansion of the relevant hypergeometric function is problematic.
In addition, it turns out that the general result for the order $\ep$-terms
for the massive three-point function of $[$\ref{Nierste}$]$ does not allow
for a straightforward extraction of the corresponding expression for this
particular case. More exactly, the equation $Q_3(y)=0$ originating from the
table in $[$\ref{Nierste}$]$ (on page~608), does not have solutions for the
relevant kinematics.
In this sense, our expressions for the coefficients of the $\ep$- and
$\ep^2$-terms for $C_2$ represent a new result.

The integration of the function $C_3$ defined by
\[
          C_3 \equiv C(-p_2,p_4,0,0,m)
\]
requires the construction of a subtraction term since an $\ep$-expansion of the
relevant integrand does not straightforwardly lead to the desired $\ep$-expansion
of the integral. This is best illustrated in a simple example which nevertheless
captures the essential idea of the subtraction method. Consider the integral
\be
\label{test1}
\int_0^1 dx \, x^{-1+\ep} f(x,\ep)
\ee
where $f(x,\ep)$ is an integrable function in the interval $[0,1]$ and has
derivatives in $\ep$. For the sake
of the argument take $f(x,\ep)$ to have a Laurent series expansion starting at the
zeroth order in $\ep$, i.e. $f(x,\ep)=f^{(0)}(x) + \ep f^{(1)}(x) + \ldots$. It is
clear that expanding the integrand
\[
x^{-1+\ep} f(x,\ep) = \frac{1}{x} f^{(0)}(x) + \ep\left( \frac{\ln x}{x}
f^{(0)}(x) + \frac{1}{x} f^{(1)}(x) \right) + \ldots
\]
does not in general render the integral integrable. However, if we write
\be
\label{test2}
\int_0^1 dx x^{-1+\ep} f(x,\ep) = \int_0^1 dx \left(x^{-1+\ep}\right)_{+} f(x,\ep)
+ f(0,\ep)
\int_0^1 dx x^{-1+\ep}
\ee
the terms on the r.h.s. of (\ref{test2}) are now integrable. In (\ref{test2}) we
have introduced a ``plus'' prescription
\[
\int_0^1 dx \left(x^{-1+\ep}\right)_{+} f(x,\ep) = \int_0^1 dx x^{-1+\ep} (
f(x,\ep) - f(0,\ep)
)
\]
not unlike the ``plus'' prescription usually introduced when discussing parton
splitting
functions. The $\ep$-expansion of the integral (\ref{test1}) can now be obtained
since the first integrand on the r.h.s. of (\ref{test2}) can be expanded in $\ep$
and then be integrated term by term whereas the second integral can be computed in
closed form.
The task is then to find the appropriate subtraction terms for the integrals
encountered in our calculation.
This is required for the three-point function $C_3$ and the
three four-point functions to be discussed in the next section.

As exemplified above we derive the
subtraction terms by substituting the value of
the integration variable (usually the lower
or upper limit of integration) at which the given integrand diverges into
the nonsingular part of the singular integrand.
Adding and
subtracting the subtraction term does all the job: e.g. the subtraction term
contains all the poles
in a given Feynman parameter but can be
easily integrated due to its simpler analytic structure, while the rest of
the integrand is now finite with respect to
the same parameter and can therefore be integrated as well.
When dealing with such a finite but complicated integration we often make
use of the integration-by-parts method to evaluate and simplify our expressions.

Applying the subtraction method to the evaluation of the three-point function
$C_3$ one obtains:
\ba
\nn
{\rm Re}\,C_3^{(-2)}&=&\frac{1}{2t}, {\rm \hspace{.2in}}
{\rm Re}\,C_3^{(-1)}=-\frac{1}{t} \ln\frac{-t}{m^2}, {\rm \hspace{.2in}}
{\rm Re}\,C_3^{(0)}=\frac{1}{t} \left[ \ln^2\frac{-t}{m^2} + {\rm
Li}_2(\frac{T}{m^2}) \right],                   \\
{\rm Re}\,C_3^{(1)}&=& -\frac{1}{t} \left[
\frac{1}{3} \ln^3\frac{-t}{m^2} + 2 {\rm Li}_3(\frac{T}{t}) +
{\rm Li}_3(\frac{T}{m^2}) \right],         \\
\nn
{\rm Re}\,C_3^{(2)}&=&
\frac{1}{3t} \left[ \ln^4\frac{-t}{m^2} - \ln^3\frac{-t}{m^2}
\ln\frac{-T}{m^2} - 3 \zeta(2) \ln^2\frac{-t}{m^2}
+ 6 \zeta(3) \ln\frac{-t}{m^2} - 6 \zeta(4)    \right.       \\
\nn
&&    \left.
- 3 {\rm Li}_4(\frac{T}{m^2}) + 6 {\rm Li}_4(\frac{m^2}{-t}) - 6 {\rm
Li}_4(\frac{T}{t}) ]    \right];                     \\
{\rm Im}\, C_3^{(j)} &=& 0.
\ea
Note that one can obtain corresponding expressions
in terms of generalized Nielsen polylogarithms
from Eq.~(27) of $[$\ref{Andrei}$]$.
The corresponding hypergeometric function of three variables $\Phi_1$ can
be reduced to a hypergeometric
function $_2F_1$ of one variable and one can then use Eq.~(2.14) of
$[$\ref{AndreiExp}$]$ to get the relevant $\ep$-expansion. We have verified
agreement with $[$\ref{Andrei}$]$ analytically up to
${\cal O}(\ep)$. The agreement for the $\ep^2$-terms was verified
numerically.

The three-point function $C_4$ has a closed form solution:
\be
\label{C4}
C_4 \equiv C(-p_2,-p_1,0,0,0) =\frac{iC_{\ep}(m^2)}{s}
\frac{\Gamma^2(-\ep)}{\Gamma(1-2\ep)}
\left(-\frac{s+i\delta}{m^2}\right)^{-\ep}
\ee
which is straightforward to obtain.
For the $\ep$-expansion of the real and imaginary parts of (\ref{C4})
we get:
\ba
\nn
{\rm Re}\,C_4^{(-2)}&=&\frac{1}{s}, {\rm \hspace{.2in}}
{\rm Re}\,C_4^{(-1)}=-\frac{1}{s} \ln\frac{s}{m^2}, {\rm \hspace{.2in}}
{\rm Re}\,C_4^{(0)}=\frac{1}{2s} \left[ \ln^2\frac{s}{m^2} - 8 \zeta(2)
                                        \right],                   \\
{\rm Re}\,C_4^{(1)}&=&\frac{1}{s} \left[
                4 \zeta(2) \ln\frac{s}{m^2} - \frac{1}{6}
                \ln^3\frac{s}{m^2} - 2 \zeta(3) \right],             \\
\nn
{\rm Re}\,C_4^{(2)}&=&\frac{1}{s}\left[
        \frac{1}{24} \ln^4\frac{s}{m^2} - 2 \zeta(2) \ln^2\frac{s}{m^2}
         + 2 \zeta(3) \ln\frac{s}{m^2} + 9 \zeta(4) \right];     \\
\nn          \\
{\rm Im}\,C_4^{(-1)}&=&\frac{\pi}{s}, {\rm \hspace{.2in}}
{\rm Im}\,C_4^{(0)}=-\frac{\pi}{s} \ln\frac{s}{m^2},        \\
\nn
{\rm Im}\,C_4^{(1)}&=& \frac{\pi}{2s} \left[ \ln^2\frac{s}{m^2}
                                              - 4 \zeta(2) \right],    \\
\nn
{\rm Im}\,C_4^{(2)}&=& \frac{\pi}{s} \left [2 \zeta(2) \ln\frac{s}{m^2} -
                    \frac{1}{6} \ln^3\frac{s}{m^2} - 2 \zeta(3) \right].
\ea

For the fifth three-point integral $C_5$ defined by
\[
          C_5 \equiv C(-p_2,-p_1,m,m,m)
\]
we first obtain a one-fold integral representation similar to
Eq.~(3.13) of $[$\ref{AndreiExp}$]$. As before, the main difficulty is the
derivation of
the coefficient for the $\ep^2$-term. The corresponding coefficient has a
complicated singularity structure as well as two branch points on its
integration path. Therefore, in order to analytically separate the real and
imaginary
parts for our final result, we have divided the integration regions for the
relevant terms into three parts. After analytical integration these terms
are free of numerical instabilities and converge very fast.
One obtains
\ba
{\rm Re}\,C_5^{(-2)}&=&
{\rm Re}\,C_5^{(-1)}=0,                 \\
\nn
{\rm Re}\,C_5^{(0)}&=& \frac{1}{2s} \left[ \ln^2x - 6\zeta(2) \right],  \\
\nn
{\rm Re}\,C_5^{(1)}&=& \frac{1}{2s} \left[
               \frac{1}{3}\ln^3x - 8\zeta(2)\ln x + 4\ln x {\rm Li}_2(x)
                - 6\zeta(3) - 8{\rm Li}_3(x) \right],                 \\
\nn
{\rm Re}\,C_5^{(2)}&=& \frac{1}{2s}     \left[
       4 \ln x \ln^3 (1-x) - \frac{9}{2} \ln^2 x \ln^2 (1-x)  -
       \frac{1}{3} \ln^3 x \ln (1 - x) + \frac{1}{12} \ln^4 x     \right.  \\
\nn   &&
  -    \zeta(2) \left( 3 \ln^2 x  - 10 \ln x \ln (1 - x) +
          6 \ln^2 (1-x) \right)  - \ln^2 x \,{\rm Li}_2(-x)           \\
\nn   &&
  +     2
        \left( \ln^2 x - 6 \ln x \ln (1 - x) + 3 \ln^2 (1-x)
                  - 6 \zeta(2) \right) {\rm Li}_2(x) + 10 \zeta(3) \ln x   \\
\nn   &&
  -   6 \zeta(3) \ln (1 - x)
  +   2 \left( \ln x + 3 \ln (1 - x) \right) {\rm Li}_3(x)
  +   2 \ln x \,{\rm Li}_3(-x)                               \\
\nn   &&
  - 12 \left( \ln x - \ln (1 - x) \right) {\rm Li}_3(1-x)
  + \frac{45}{2}\,\zeta(4) - L_{-++}\left(1,0,0,\frac{1}{-1 + x}\right)      \\
\nn   &&
  + L_{-++}\left(1,0,0,\frac{x}{1 - x}\right)
  - L_{+++}\left(0,0,\frac{1-x}{x},-1\right)
  + L_{+++}\left(0,0,\frac{1-x}{x},\frac{1}{x}\right)                        \\
\nn   &&        \left.
  - L_{+++}\left(0,\frac{1-x}{x},\frac{1-x}{x},-1\right)
  + L_{+++}\left(0,\frac{1-x}{x},\frac{1-x}{x},\frac{1}{x}\right)         \right];
\nn       \\
{\rm Im}\,C_5^{(-1)}&=&0, {\rm \hspace{.2in}}
{\rm Im}\,C_5^{(0)}=\frac{\pi}{s}\ln x,                \\
\nn
{\rm Im}\,C_5^{(1)}&=& \frac{\pi}{2s} \left[
              \ln^2x - 4\zeta(2) + 4 {\rm Li}_2(x) \right],   \\
\nn
{\rm Im}\,C_5^{(2)}&=& \frac{\pi}{6s} \left[
       - 12\ln x\ln^2(1-x) + \ln^3x
          - 12 (\ln x - 2\ln(1-x)) (\zeta(2) - {\rm Li}_2(x))  \right.  \\
\nn
&&  \left.
    + 12\zeta(3)
    - 12 {\rm Li}_3(x)
    - 24 {\rm Li}_3(1-x)    \right].
\ea
Explicit result for this integral was given very recently in Eq.~(4.4)
of $[$\ref{Binomial}$]$. We have checked
agreement with $[$\ref{Binomial}$]$ analytically up to
${\cal O}(\ep)$. The agreement for the $\ep^2$-terms was verified
numerically.

Finally, we write down real and imaginary parts for the last required
three-point function $C_6$ defined by
\[
           C_6 \equiv C(p_3,p_4,m,0,m).
\]
One has
\ba
{\rm Re}\,C_6^{(-2)}&=&0, {\rm \hspace{.2in}}
{\rm Re}\,C_6^{(-1)}=\frac{1}{s\beta}\ln x,                 \\
\nn
{\rm Re}\,C_6^{(0)}&=&\frac{1}{2s\beta} \left[
      -4\ln x\ln(1-x) + \ln^2x - 8\zeta(2) - 4 {\rm Li}_2(x) \right],  \\
\nn
{\rm Re}\,C_6^{(1)}&=& \frac{1}{6s\beta} \left[
      - 6\ln^2x\ln(1-x) + \ln^3x - 24\zeta(2)\ln x + 72\zeta(2)\ln(1-x)
      + 12\zeta(3)
                        \right.      \\
\nn   && \left.
      - 12 {\rm Li}_3(x) - 24 {\rm Li}_3(1-x)      \right],
\\   \nn
{\rm Re}\,C_6^{(2)}&=& \frac{1}{24s\beta} \left[
                - 16\ln x\ln^3(1-x) + 24\ln^2x\ln^2(1-x)
                - 8\ln^3x\ln(1-x) + \ln^4x      \right.      \\
\nn  &&
      + 4\ln^4(1-x)
      + 192\zeta(2)\ln x\ln(1-x) - 48\zeta(2)\ln^2x - 240\zeta(2)\ln^2(1-x)
\\  \nn   && \left.
      - 48\zeta(3)\ln x + 120\zeta(4) + 48 {\rm Li}_4(x)
      + 96 {\rm Li}_4(\frac{-x}{1-x})
      + 96 {\rm Li}_4(1-x)   \right];
\nn       \\
{\rm Im}\,C_6^{(-1)}&=&\frac{\pi}{s\beta}, {\rm \hspace{.2in}}
{\rm Im}\,C_6^{(0)}=\frac{\pi}{s\beta}\left[\ln x - 2\ln(1-x)\right],    \\
\nn
{\rm Im}\,C_6^{(1)}&=&\frac{\pi}{6s\beta} \left[ - 12\ln x\ln(1-x) +
3\ln^2x + 12\ln^2(1-x) - 12\zeta(2) \right],                \\
\nn
{\rm Im}\,C_6^{(2)}&=&\frac{\pi}{6s\beta} \left[
                   12\ln x\ln^2(1-x) - 6\ln^2x\ln(1-x) + \ln^3x
                 - 8\ln^3(1-x) - 12\zeta(2)\ln x         \right.      \\
\nn  &&  \left.
                 + 24\zeta(2)\ln(1-x) - 12\zeta(3) \right].
\ea
Corresponding results for $C_6$ may be obtained from
Eqs.~(3.5), (3.7), (2.10) and (2.14) of $[$\ref{AndreiExp}$]$. We have
done an order by order numerical comparisons for the coefficients of the $\ep$-
and $\ep^2$-terms, while other terms can be easily compared analytically.
We obtain exact agreement.

We mention that we have checked all our analytical results for the
three--point functions against numerical results provided to us by
M.M.~Weber $[$\ref{Private}$]$ (see also $[$\ref{Weber}$]$). We have found
agreement.

\vglue 1cm
\renewcommand{\theequation}{4.\arabic{equation}}
\begin{center}\begin{large}\begin{bf}
IV. FOUR-POINT FUNCTIONS
\end{bf}\end{large}\end{center}
\vglue .3cm

The scalar four-point one-loop integrals with one, two or three heavy
quarks running in the loop are the most difficult to evaluate.
The one-loop four-point functions are defined by $[$\ref{Been}$]$
\ba
\lefteqn{D(q_1,q_2,q_3,m_1,m_2,m_3,m_4)=}    \\
\nn        &&
             \mu^{2\ep} \int \frac{d^nq}{(2\pi)^n}
           \frac{1}{(q^2-m_1^2)[(q+q_1)^2-m_2^2][(q+q_1+q_2)^2-m_3^2]
                    [(q+q_1+q_2+q_3)^2-m_4^2]}.
\ea
As before, the $+i\delta$ terms in the denominators have not been written out.
Again, there is only one internal mass scale for our purposes.

For heavy flavor production one needs three different types of four-point
functions $D_i\,\, (i=1,2,3)$ which are expanded as
\be
D_i = i\,C_{\ep}(m^2) \left\{\frac{1}{\ep^2} D_i^{(-2)} + \frac{1}{\ep}
D_i^{(-1)} + D_i^{(0)} + \ep D_i^{(1)} + \ep^2
D_i^{(2)} + {\cal O}(\ep^3) \right\}.
\ee
Again the coefficient of the most singular part of the four-point functions
is purely real, i.e. ${\rm Im}\, D_i^{(-2)}=0$.

Before we give our results for the four-point functions it is necessary to
discuss
some general technical features. After applying Feynman
parametrization, we are left with a three-fold parametric integral for
the four-point functions:
\be
\label{Dparam}
D = iC_{\ep}(m^2) (1+\ep) (m^2)^{-2} \int_0^1 dx_1dx_2dx_3 \,\,
x_1^2x_2 K^{-2-\ep},
\ee
with the kernel $K$ given by
\ba
\label{kernel}
\nn
m^2 K &=& -abq_1^2 - ac(q_1+q_2)^2 - ad(q_1+q_2+q_3)^2 - bcq_2^2 \\
&&    - bd(q_2+q_3)^2 - cdq_3^2 + am_1^2 + bm_2^2 + cm_3^2 + dm_4^2 -
i\delta,
\ea
where the $m_i\,\, (i=1,\ldots,4)$ can be either $m$ or $0$.
The set of parameters $\{a,b,c,d\}$ above corresponds to an arbitrary
choice
from the set of original parameters $\{x_1x_2x_3, \,x_1x_2(1-x_3),
\,x_1(1-x_2), \,1-x_1\}$. For each particular four-point
function we make a judicious choice of Feynman parameters which enables us to get
the most convenient
kernel for the subsequent integrations.
\\
\\
\[
\mbox{ \epsfysize = 3.0cm  \epsffile {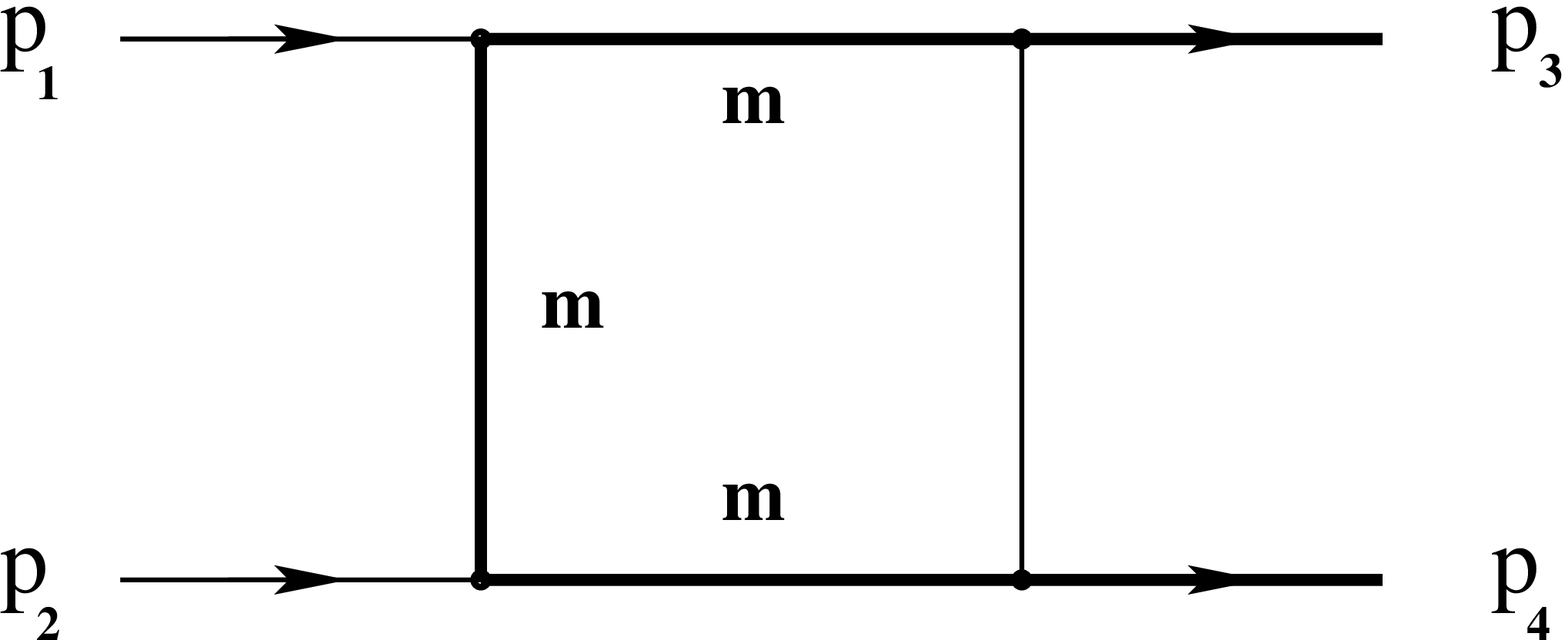}}
\]
{\begin{center}
\tenrm\vglue -0.1cm
FIG.~2. Massive box $D_1$ with three massive propagators.
\end{center}
}
\vglue .1cm

First we consider the four-point function $D_1$ with three massive propagators shown
in  Fig.~2 which is defined by
\[
D_1 \equiv D(p_4,-p_2,-p_1,0,m,m,m).
\]
Substitution of the corresponding values of momenta and masses for $D_1$ into
the expression for our kernel (\ref{kernel}) gives
\be
K = ac\tilde{t} - bd\tilde{s} + (1-a)^2 - i\delta,
\ee
where we have introduced positive valued dimensionless variables
\be
\tilde{s} \equiv \frac{s}{m^2},
{\rm \hspace{.4in}}
\tilde{t} \equiv -\frac{t}{m^2}.
\ee
The kinematical conditions $\tilde{s}\ge 4, \, \tilde{t}\ge 1, \, \tilde{s} \ge
\tilde{t}$ constrain the allowable region of phase space for the present physical
$2\rightarrow 2$ process.
Our choice for the parameters $\{a,b,c,d\}$ is $\{1-x_1, \, x_1x_2x_3,
\, x_1x_2(1-x_3), \, x_1(1-x_2) \}$.
For $D_1$, the integration of the corresponding integrand over $x_3$ results in
two terms:
\ba
\label{ID1x1x2}
I^{D_1}_{x_1x_2}=-\frac{x_1^{-\ep}\left[ x_1 + \tilde t(1-x_1) x_2)
\right]^{-1-\ep}}{(1+\ep)
[\tilde s x_1(1-x_2)+\tilde t (1-x_1)]},         \\
\label{IID1x1x2}
II^{D_1}_{x_1x_2}=\frac{x_1^{-1-2\ep}\left[ 1 - \tilde s x_2 (1-x_2) - i\delta
\right]^{-1-\ep}} {(1+\ep)[\tilde s x_1 (1-x_2)+\tilde t (1-x_1)]},
\ea
which then have to be integrated over the remaining parameters $x_1$ and $x_2$.
Eqs.~(\ref{ID1x1x2}) and (\ref{IID1x1x2}) correspond to the indefinite integral
(or primitive) evaluated at the upper and lower limit of $x_3$, respectively.
The term $I^{D_1}_{x_1x_2}$ in
(\ref{ID1x1x2}) does not change sign on the integration path,
e.g. does not have a branch cut in the interval $[0,1]$ for both variables $x_1$
and $x_2$.
Consequently, it does not give an imaginary contribution and it is thus safe
to drop the $i\delta$ shift in $I^{D_1}_{x_1x_2}$. Furthermore, since
$I^{D_1}_{x_1x_2}$ does not have poles in $\ep$, we expand it up to $\ep^2$ and
straightforwardly integrate over the second variable $x_2$ to obtain
$I^{D_1}_{x_1}$.
Concerning the second term $II^{D_1}_{x_1x_2}$ in (\ref{IID1x1x2}), one can see
that there is a branch cut for the variable $x_2$ in its numerator as well as a
divergence due to the factor $x_1^{-1-2\ep}$ at the lower limit of the
integration $x_1=0$ (we have dropped  the $i\delta$ shift in the denominator
as it does not affect our further calculation).
At this point we introduce a subtraction term for
$II^{D_1}_{x_1x_2}$ in the simplest possible way: we set $x_1=0$ in
$II^{D_1}_{x_1x_2}$ everywhere except for the divergent term $x_1^{-1-2\ep}$.
This results in the following subtraction term:
\be
II^{D_1,s}_{x_1x_2}=\frac{x_1^{-1-2\ep}\left[ 1 - \tilde s x_2 (1-x_2) - i\delta
\right]^{-1-\ep}} {(1+\ep) \tilde t},
\ee
which, in the framework of the dimensional regularization scheme, integrates over
$x_1$ to
\be
II^{D_1,s}_{x_2}=-\frac{\left[ 1 - \tilde s x_2 (1-x_2) - i\delta
\right]^{-1-\ep}} {2\ep (1+\ep) \tilde t}.
\ee
Then we expand the above expression up to $\ep^2$ and reexpress the argument of
subsequent logarithms as
\be
\label{roots1}
1 - \tilde s x_2 (1-x_2) - i\delta =
\frac{(x_2-x_2^{(0)})(x_2-1+x_2^{(0)})}{x_2^{(0)}(1-x_2^{(0)})},
\ee
with
\be
x_2^{(0)}=\frac{1+\sqrt{1-4/\tilde s}}{2} + i\delta = \frac{1+\beta}{2} +
i\delta
\ee
A final integration of the subsequent series can be done analytically in the
complex plane and its result is expressed in terms of logarithms and
classical polylogarithms up to ${\rm Li}_4$. Analytic continuation of the
result for $\delta \rightarrow 0$ is then straightforward.

Lastly, we calculate the finite difference
\[
\Delta II^{D_1}_{x_1x_2} = II^{D_1}_{x_1x_2} - II^{D_1,s}_{x_1x_2}
= \frac{x_1^{-1-2\ep} \left( \tilde t - \tilde s (1-x_2) \right)
\left[ 1 - \tilde s x_2 (1-x_2) - i\delta \right]^{-1-\ep}}
{(1+\ep) \, \tilde t \, [\tilde s x_1 (1-x_2)+\tilde t (1-x_1)]}
\]
by again expanding the difference up to $\ep^2$
and using (\ref{roots1}) for the arguments of the logarithms.
Then we
first integrate over the variable $x_2$, leading to a reduction of the integrand to
simple
fractions w.r.t. $x_2$. In this way we avoid spurious poles in the remaining
integral which would otherwise arise in case of integration over $x_1$ first.

To complete the derivation of the first four-point (box) integral, we combine
the two terms
\[
I^{D_1}_{x_1} + \Delta II^{D_1}_{x_1}
\]
and perform the last integration over the variable $x_1$.

At this point we
would like to comment on some technical details of our calculation which are used
throughout this work. For instance, the integrand for the last integration
contains expressions such as
\be
\label{partfrac}
f(x_1) \cdot {\rm Li}_{2,3} \left( \frac{a_1 x_1^2 + a_2 x_1 + a_3}{a_4 x_1^2
+ a_5 x_1 + a_6} \right),
\ee
where $f(x_1)$ is a rational function or a product of a rational function and
a logarithm.
Using recursively the method of integration-by-parts as much as necessary
we render their arguments to be linear functions of $x_1$.
In addition, in the case of ${\rm Li}_3$, we can reduce the weight of
${\rm Li}_3$ by one.
At the same time,
the sources of imaginary contributions are transferred into logarithms (or
remain in ${\rm Li}_2$'s and ${\rm Li}_3$'s with arguments that are
independent of the integration variable). Finally, performing the last
integration and adding up all the relevant contributions we arrive at the
result for the box integral with three massive lines, containing polylogs up to
${\rm Li}_4$ and the single- and triple-index $L$-functions introduced in
Eqs.~(\ref{Lfunction}) and (\ref{Lpfunction}).

In order to keep our results at reasonable length we introduce the
abbreviations
\ba
\label{notations}
\nn  &&
z_3 \equiv (s + 2 t + s\beta)/2,  {\rm \hspace{.4in}}
z_4 \equiv (s + 2 t - s\beta)/2,             \\
&&    z_5 \equiv (2 m^2 + t + t \beta)/2,  {\rm \hspace{.4in}}
z_6 \equiv (2 m^2 + t - t \beta)/2,     \\
\nn &&
l_s \equiv \ln \frac{s}{m^2},  {\rm \hspace{.2in}}
l_t \equiv \ln \frac{-t}{m^2},  {\rm \hspace{.2in}}
l_T \equiv \ln \frac{-T}{m^2},  {\rm \hspace{.2in}}
l_x \equiv \ln x,        {\rm \hspace{.2in}}                         \\
\nn &&
l_{\beta}  \equiv \ln \beta,    {\rm \hspace{.2in}}
l_{z3}  \equiv  \ln \frac{z_3}{m^2},     {\rm \hspace{.2in}}
l_{z4}  \equiv  \ln \frac{-z_4}{m^2}
\ea
and obtain:
\be
\label{d1real}
{\rm Re}\,D_1^{(-2)}=0, {\rm \hspace{.6in}}
            {\rm Re}\,D_1^{(-1)}=\ln x/(s t \beta),   {\rm \hspace{2.63in}}
\ee
\vspace{-.4in}
\ba
\nn
{\rm Re}\,D_1^{(0)}&=& - \left[ 2 \ln x \ln(-t\beta/m^2) + 2 {\rm Li}_2(x)
                         - 2 {\rm Li}_2(-x) + 3\zeta(2) \right]/(st\beta),
\\
\nn
{\rm Re}\,D_1^{(1)}&=& \frac{1}{st\beta} \left[ \frac{l_s^3}{12}
- \frac{l_s^2\, l_x}{2} - \frac{5\,l_x^3}{6} - l_x^2\,l_{z3} +
l_x^2\,l_{z4} - {l_x}\,l_{z4}^2
- \frac{l_t^3}{3} + l_t^2 \,( 3\,{l_x} + {l_{z4}} ) -   \right.  \\
\nn  &&
{l_t}\,\left( l_T^2 + l_x^2 - l_{z3}^2 + 2{l_T}\, ( {l_x} + {l_{z3}} - {l_{z4}} )
         + l_{z4}^2 + {l_x}\,( 2\,{l_{z3}} - 4\,{l_{\beta}} )  \right)  +
{l_x}\,l_{\beta}^2 - l_{\beta}^3 +                   \\
\nn   &&
      {l_s}\,\left( {l_t}\,{l_x} + l_x^2/4 + {l_x}\,
\left( {l_{z3}} + {l_{z4}} \right) - l_{\beta}^2 \right)  +
       \left( -{l_s} + 10 {l_t} - {l_x} + 6 {l_{\beta}} \right) \,\zeta(2) - \\
\nn   &&
5 \zeta(3) + 4 {\rm Li}_2(x)\,{l_t} -
       2 {\rm Li}_2\left( \frac{m^2\,x}{-T} \right) \,{l_t} +
2 {\rm Li}_2\left(\frac{T}{z_3}\right)\,{l_t} -
       2 {\rm Li}_2\left(\frac{m^2}{z_5}\right)\,{l_x} -  \\
\nn   &&
2 {\rm Li}_2\left(\frac{-t\,(1 - \beta)}{2\,m^2}\right)\,{l_x} -
       2 {\rm Li}_3(-x) - 2 {\rm Li}_3\left(\frac{-x^2}{1 - x^2}\right) -
2 {\rm Li}_3\left(\frac{z_3}{t}\right) +              \\
\nn   &&
       2 {\rm Li}_3\left(\frac{z_4}{t}\right)
     + 4 {\rm Li}_3\left(\frac{1 - \beta}{2}\right)
     + 2 {\rm Li}_3\left( \frac{m^2\,(1 - \beta)}{2\,z_5} \right)
     - 2 {\rm Li}_3\left(\frac{-t\,(1 - \beta)}{2\,z_5}\right) +   \\
\nn   &&        \left.
8 {\rm Li}_3\left(\frac{-1 + \beta }{2\,\beta}\right) -
       2 {\rm Li}_3\left(\frac{2\,z_6}{m^2 (1 + \beta}\right) +
       2 {\rm Li}_3\left( \frac{2\,z_6}{-t (1 + \beta)}\right) \right],  \\
\nn
{\rm Re}\,D_1^{(2)}&=& \frac{1}{st\beta} \left[ -\frac{5\,l_s^4}{64}  -
\frac{43}{24}\,l_s^3\,{l_t} - \frac{7}{4}\,l_s^2\,l_t^2 +
       \frac{2}{3}\,{l_s}\,l_t^3 + \frac{5\,l_t^4}{12} -
\frac{11}{24}\,l_s^3\,{l_T} - \frac{3}{2}\,l_s^2\,{l_t}\,{l_T} -
                                                    \right.                \\
\nn   &&
       \frac{5}{2}\,{l_s}\,l_t^2\,{l_T} - \frac{4}{3}\,l_t^3\,{l_T}
- \frac{5}{16}\,l_s^2\,l_T^2 + \frac{1}{4}\,{l_s}\,{l_t}\,l_T^2 -
       \frac{1}{2}\,{l_s}\,l_T^3 + {l_t}\,l_T^3
+ \frac{5}{8}\,l_s^3\,{l_x} - \frac{3}{8}\,l_s^2\,{l_t}\,{l_x} -     \\
\nn   &&
4\,{l_s}\,l_t^2\,{l_x} -
       l_t^3\,{l_x} - \frac{7}{8}\,l_s^2\,{l_T}\,{l_x}
+ \frac{3}{2}\,{l_s}\,{l_t}\,{l_T}\,{l_x}
- \frac{3}{2}\,l_t^2\,{l_T}\,{l_x} +
       \frac{3}{8}\,{l_s}\,l_T^2\,{l_x}
+ \frac{13}{4}\,{l_t}\,l_T^2\,{l_x} -            \\
\nn   &&
\frac{1}{2} l_T^3\,{l_x}
- \frac{13}{32}\,l_s^2\,l_x^2 +
       \frac{1}{8}\,{l_s}\,{l_t}\,l_x^2
- \frac{1}{4}\,l_t^2\,l_x^2 + \frac{17}{8}\,{l_s}\,{l_T}\,l_x^2
+ \frac{1}{2}\,{l_t}\,{l_T}\,l_x^2 +
       \frac{7}{16}\,l_T^2\,l_x^2 + \frac{1}{2} {l_s}\,l_x^3 +    \\
\nn   &&
\frac{3}{8}\,{l_t}\,l_x^3 + \frac{17}{24}\,{l_T}\,l_x^3 +
       \frac{7\,l_x^4}{64} + \frac{29}{48}\,l_s^3\,{l_{z3}}
+ 3 l_s^2\,{l_t}\,{l_{z3}} +
       \frac{5}{2}\,{l_s}\,l_t^2\,{l_{z3}}
- \frac{3}{2}\,l_t^3\,{l_{z3}} - \frac{3}{4}\,l_s^2\,{l_T}\,{l_{z3}} +   \\
\nn   &&
       \frac{17}{2}\,{l_s}\,{l_t}\,{l_T}\,{l_{z3}}
+ \frac{5}{2}\,l_t^2\,{l_T}\,{l_{z3}} + 2\,{l_t}\,l_T^2\,{l_{z3}} -
       \frac{19}{16}\,l_s^2\,{l_x}\,{l_{z3}}
+ {l_s}\,{l_t}\,{l_x}\,{l_{z3}} + \frac{5}{2}\,l_t^2\,{l_x}\,{l_{z3}} +   \\
\nn   &&
       \frac{9}{2}\,{l_s}\,{l_T}\,{l_x}\,{l_{z3}}
- \frac{1}{2}\,{l_t}\,{l_T}\,{l_x}\,{l_{z3}} -
       \frac{1}{2}\,l_T^2\,{l_x}\,{l_{z3}}
+ \frac{13}{16}\,{l_s}\,l_x^2\,{l_{z3}} + {l_t}\,l_x^2\,{l_{z3}} -
       \frac{3}{4}\,{l_T}\,l_x^2\,{l_{z3}} -      \\
\nn   &&
\frac{11}{48}\,l_x^3\,{l_{z3}}
- \frac{1}{8}\,l_s^2\,l_{z3}^2 -
       4\,{l_s}\,{l_t}\,l_{z3}^2
+ \frac{1}{2}\,l_t^2\,l_{z3}^2
+ {l_s}\,{l_T}\,{{l_{z3}}}^2 -
       5\,{l_t}\,{l_T}\,l_{z3}^2
- \frac{3}{4}\,{l_s}\,{l_x}\,l_{z3}^2 +              \\
\nn   &&
{l_t}\,{l_x}\,l_{z3}^2 -
       2\,{l_T}\,{l_x}\,l_{z3}^2
- \frac{1}{8}\,l_x^2\,l_{z3}^2 + {l_t}\,l_{z3}^3 +
       \frac{61}{48}\,l_s^3\,{l_{z4}}
+ \frac{27}{4}\,l_s^2\,{l_t}\,{l_{z4}}
+ \frac{1}{2}\,{l_s}\,l_t^2\,{l_{z4}} -     \\
\nn   &&
       \frac{13}{6}\,l_t^3\,{l_{z4}}
+ \frac{9}{4}\,l_s^2\,{l_T}\,{l_{z4}}
- \frac{1}{2}\,{l_s}\,{l_t}\,{l_T}\,{l_{z4}} +
       \frac{5}{2}\,l_t^2\,{l_T}\,{l_{z4}}
+ {l_s}\,l_T^2\,{l_{z4}} - \frac{5}{2}\,{l_t}\,l_T^2\,{l_{z4}} +   \\
\nn   &&
       \frac{9}{16}\,l_s^2\,{l_x}\,{l_{z4}}
+ \frac{11}{2}\,{l_s}\,{l_t}\,{l_x}\,{l_{z4}} +
       \frac{3}{2}\,l_t^2\,{l_x}\,{l_{z4}}
- \frac{3}{2}\,{l_s}\,{l_T}\,{l_x}\,{l_{z4}} -
       \frac{5}{2}\,{l_t}\,{l_T}\,{l_x}\,{l_{z4}}
+ \frac{1}{2} l_T^2\,{l_x}\,{l_{z4}} -                   \\
\nn   &&
       \frac{11}{16}\,{l_s}\,l_x^2\,{l_{z4}}
- \frac{1}{4}\,{l_t}\,l_x^2\,{l_{z4}} -
       \frac{9}{4}\,{l_T}\,l_x^2\,{l_{z4}}
- \frac{5}{16}\,l_x^3\,{l_{z4}} -
       2\,l_s^2\,{l_{z3}}\,{l_{z4}}
- 6\,{l_s}\,{l_t}\,{l_{z3}}\,{l_{z4}} -      \\
\nn   &&
       l_t^2\,{l_{z3}}\,{l_{z4}} - 3 {l_s}\,{l_T}\,{l_{z3}}\,{l_{z4}} -
       6 {l_t}\,{l_T}\,{l_{z3}}\,{l_{z4}}
- 4 {l_t}\,{l_x}\,{l_{z3}}\,{l_{z4}} -
       3 {l_T}\,{l_x}\,{l_{z3}}\,{l_{z4}}
- \frac{1}{2} l_x^2\,{l_{z3}}\,{l_{z4}} +               \\
\nn   &&
       {l_s}\,l_{z3}^2\,{l_{z4}}
+ \frac{5}{2}\,{l_t}\,l_{z3}^2\,{l_{z4}} +
       {l_x}\,l_{z3}^2\,{l_{z4}}
- \frac{77}{16}\,l_s^2\,l_{z4}^2 -
       \frac{13}{2}\,{l_s}\,{l_t}\,l_{z4}^2
+ 3\,l_t^2\,l_{z4}^2 - \frac{5}{4}\,{l_s}\,{l_T}\,l_{z4}^2 +   \\
\nn   &&
       2\,{l_t}\,{l_T}\,l_{z4}^2
+ \frac{1}{4}\,l_T^2\,l_{z4}^2
- \frac{33}{8}\,{l_s}\,{l_x}\,l_{z4}^2 -
       \frac{5}{2}\,{l_t}\,{l_x}\,l_{z4}^2
+ \frac{7}{4}\,{l_T}\,{l_x}\,l_{z4}^2 +
       \frac{15}{16}\,l_x^2\,l_{z4}^2
+ 3\,{l_s}\,{l_{z3}}\,l_{z4}^2 +                 \\
\nn   &&
       \frac{7}{2}\,{l_t}\,{l_{z3}}\,l_{z4}^2
+ 3\,{l_T}\,{l_{z3}}\,l_{z4}^2 +
       2\,{l_x}\,{l_{z3}}\,l_{z4}^2 - l_{z3}^2\,l_{z4}^2 +
       \frac{71}{12}\,{l_s}\,l_{z4}^3
+ \frac{2}{3}\,{l_t}\,l_{z4}^3 - \frac{5}{6}\,{l_T}\,l_{z4}^3 +     \\
\nn   &&
       \frac{31}{12}\,{l_x}\,l_{z4}^3
- 2 {l_{z3}}\,l_{z4}^3 - \frac{49}{24}l_{z4}^4 -
       \frac{13}{24}\,l_s^3\,{l_{\beta}}
- \frac{7}{2}\,l_s^2\,{l_t}\,{l_{\beta}}
- 2\,{l_s}\,l_t^2\,{l_{\beta }} -
       \frac{5}{3}\,l_t^3\,{l_{\beta}}
- \frac{9}{4}\,l_s^2\,{l_T}\,{l_{\beta}} -   \\
\nn   &&
3\,{l_s}\,{l_t}\,{l_T}\,{l_{\beta }} - 4\,l_t^2\,{l_T}\,{l_{\beta}} +
       \frac{13}{8}\,l_s^2\,{l_x}\,{l_{\beta}}
- 2\,{l_s}\,{l_t}\,{l_x}\,{l_{\beta }} - 5\,l_t^2\,{l_x}\,{l_{\beta}} -
       \frac{1}{2}\,{l_s}\,{l_T}\,{l_x}\,{l_{\beta}} -          \\
\nn   &&
 3\,{l_t}\,{l_T}\,{l_x}\,{l_{\beta }} + l_T^2\,{l_x}\,{l_{\beta}} -
       \frac{9}{8}\,{l_s}\,l_x^2\,{l_{\beta}}
+ \frac{1}{2}\,{l_t}\,l_x^2\,{l_{\beta}}
+ \frac{7}{4}\,{l_T}\,l_x^2\,{l_{\beta}} +
       \frac{17}{24}\,l_x^3\,{l_{\beta}}
+ \frac{9}{4}\,l_s^2\,{l_{z3}}\,{l_{\beta}} +          \\
\nn   &&
4\,{l_s}\,{l_t}\,{l_{z3}}\,{l_{\beta }} +
       4\,l_t^2\,{l_{z3}}\,{l_{\beta }}
+ 2\,{l_s}\,{l_T}\,{l_{z3}}\,{l_{\beta }}
+ 6\,{l_t}\,{l_T}\,{l_{z3}}\,{l_{\beta }} -
       \frac{5}{2}\,{l_s}\,{l_x}\,{l_{z3}}\,{l_{\beta }}
+ 4\,{l_t}\,{l_x}\,{l_{z3}}\,{l_{\beta }} +                    \\
\nn   &&
       4 {l_T}\,{l_x}\,{l_{z3}}\,{l_{\beta }}
+ \frac{5}{4} l_x^2\,{l_{z3}}\,{l_{\beta }} -
       {l_s}\,l_{z3}^2\,{l_{\beta }}
- 3 {l_t}\,l_{z3}^2\,{l_{\beta }} - {l_x}\,l_{z3}^2\,{l_{\beta }} +
       \frac{11}{4} l_s^2\,{l_{z4}}\,{l_{\beta }}
+ 7 {l_s}\,{l_t}\,{l_{z4}}\,{l_{\beta }} +                 \\
\nn   &&
       l_t^2\,{l_{z4}}\,{l_{\beta }}
+ 3\,{l_s}\,{l_T}\,{l_{z4}}\,{l_{\beta }}
+ 2\,{l_t}\,{l_T}\,{l_{z4}}\,{l_{\beta }} +
       \frac{1}{2}\,{l_s}\,{l_x}\,{l_{z4}}\,{l_{\beta }}
+ 3\,{l_t}\,{l_x}\,{l_{z4}}\,{l_{\beta }} -
       {l_T}\,{l_x}\,{l_{z4}}\,{l_{\beta }} -    \\
\nn   &&
\frac{1}{4}\,l_x^2\,{l_{z4}}\,{l_{\beta }} -
       3\,{l_s}\,{l_{z3}}\,{l_{z4}}\,{l_{\beta }}
- 6\,{l_t}\,{l_{z3}}\,{l_{z4}}\,{l_{\beta }} -
       4\,{l_T}\,{l_{z3}}\,{l_{z4}}\,{l_{\beta }}
+ {l_x}\,{l_{z3}}\,{l_{z4}}\,{l_{\beta }} +
       2\,l_{z3}^2\,{l_{z4}}\,{l_{\beta }} -          \\
\nn   &&
5\,{l_s}\,l_{z4}^2\,{l_{\beta }} -
       {l_t}\,l_{z4}^2\,{l_{\beta }}
- 2\,{l_x}\,l_{z4}^2\,{l_{\beta }} +
       {l_{z3}}\,l_{z4}^2\,{l_{\beta }}
+ \frac{7}{3} l_{z4}^3\,{l_{\beta }}
- \frac{7}{8} l_s^2\,l_{\beta}^2 -
       \frac{5}{2} {l_s}\,{l_t}\,l_{\beta}^2
- 2\,l_t^2\,l_{\beta}^2 -                         \\
\nn   &&
\frac{5}{2}\,{l_s}\,{l_T}\,l_{\beta}^2 -
       3\,{l_t}\,{l_T}\,l_{\beta}^2
+ \frac{3}{4}\,{l_s}\,{l_x}\,l_{\beta}^2
- \frac{5}{2}\,{l_t}\,{l_x}\,l_{\beta}^2 -
       \frac{1}{2}\,{l_T}\,{l_x}\,l_{\beta}^2
- \frac{3}{8}\,l_x^2\,l_{\beta}^2
+ \frac{5}{2}\,{l_s}\,{l_{z3}}\,l_{\beta}^2 +           \\
\nn   &&
       3\,{l_t}\,{l_{z3}}\,l_{\beta}^2
+ 2\,{l_T}\,{l_{z3}}\,l_{\beta}^2 -
       \frac{1}{2}\,{l_x}\,{l_{z3}}\,l_{\beta}^2
- l_{z3}^2\,l_{\beta}^2 +
       \frac{5}{2}\,{l_s}\,{l_{z4}}\,l_{\beta}^2
+ 3\,{l_t}\,{l_{z4}}\,l_{\beta}^2 +
       2\,{l_T}\,{l_{z4}}\,l_{\beta}^2 +              \\
\nn   &&
\frac{1}{2}\,{l_x}\,{l_{z4}}\,l_{\beta}^2 -
       2\,{l_{z3}}\,{l_{z4}}\,l_{\beta}^2
- \frac{3}{2}\,l_{z4}^2\,l_{\beta}^2 -
       \frac{1}{3}\,{l_s}\,l_{\beta}^3
- \frac{2}{3}\,{l_t}\,l_{\beta}^3 - {l_T}\,l_{\beta}^3
- \frac{1}{3}\,{l_x}\,l_{\beta}^3 +
       {l_{z3}}\,l_{\beta}^3 +              \\
\nn   &&
{l_{z4}} l_{\beta}^3
+ \frac{l_{\beta}^4}{6} +
       \left(  \frac{19}{8} l_x^2 -\frac{l_s^2}{8}
- 9 l_t^2 + 5 {l_T} {l_x} + {l_T} {l_{z3}} +
          \frac{11}{2} {l_x} {l_{z3}} - l_{z3}^2
- 6 {l_T} {l_{z4}} + 2 {l_x} {l_{z4}} -            \right.   \\
\nn   &&
          3\,{l_{z3}}\,{l_{z4}} - 7\,l_{z4}^2 +
          2\,{l_t}\,\left( 3\,{l_T} + 7\,{l_{z3}} + 5\,{l_{z4}}
- 12\,{l_{\beta }} \right)  + 6\,{l_T}\,{l_{\beta }} -
          10\,{l_x}\,{l_{\beta }} - 2\,{l_{z3}}\,{l_{\beta}} +       \\
\nn   &&  \left.
14\,{l_{z4}}\,{l_{\beta }} - 13\,l_{\beta}^2 -
          \frac{1}{4}\,{l_s}\,\left( 64\,{l_t} - 8\,{l_T} + 35\,{l_x}
+ 18\,{l_{z3}} - 44\,{l_{z4}} + 32\,{l_{\beta}} \right)  \right) \,
        \zeta(2) +                           \\
\nn   &&
\left( -2\,{l_s} - 4\,{l_t} - {l_T} + {l_x}
+ 7\,{l_{z3}} + {l_{z4}} \right) \,\zeta(3) -
       \frac{35}{4}\zeta(4) - 2\,{{\rm Li}_2^2\left(\frac{m^2}{z_5}\right)} + \\
\nn   &&
2\,{{\rm Li}_2^2\left(\frac{-t (1 - \beta) }{2\,m^2}\right)} +
       {\rm Li}_2(x)\, \left( 2\,{\rm Li}_2\left(\frac{m^2}{z_5}\right)
+ 2\,{\rm Li}_2\left(\frac{-t (1 - \beta) }{2\,m^2}\right) +
          \frac{l_s^2}{4} -   \right.   \\
\nn   &&
2\,l_t^2 + l_T^2 + {l_T}\,{l_x}
+ \frac{l_x^2}{4} + 2\,{l_T}\,{l_{z3}} + 3\,{l_x}\,{l_{z3}} +
          {l_t}\,\left( -4\,{l_T} - 3\,{l_x} + 2\,{l_{z3}} \right) +   \\
\nn   &&   \left.
          {l_s}\,\left( {l_t} - {l_T} - \frac{9}{2}{l_x} - {l_{z3}}
- {l_{z4}} \right)  + 5\,{l_x}\,{l_{z4}} +
          l_{z4}^2 - 4\,{l_x}\,{l_{\beta }} - 6\,\zeta(2) \right)  +
       {\rm Li}_2\left(\frac{z_3}{z_4}\right) \times      \\
\nn   &&
        \left( -2 {\rm Li}_2\left(\frac{m^2}{z_5}\right)
           - \frac{l_s^2}{4} + 4 l_t^2
+ {l_T}\,{l_x} + \frac{3}{4} l_x^2 -
             \frac{l_s}{2} \left( 6 {l_t} + 2 {l_T} + 3 {l_x}
- 4 {l_{z3}} - 4 {l_{z4}} \right) +   \right.  \\
\nn   &&   \left.
2\,{l_x}\,{l_{z4}} -
             l_{z4}^2 - {l_t}\,\left( {l_x} - 2\,{l_T}
+ 2\,{l_{z3}} + 2\,{l_{z4}}  \right)  -
             2\,{l_x}\,{l_{\beta}}  - 6\,\zeta(2) \right)  +  \\
\nn   &&
       {\rm Li}_2\left(\frac{-t (1 - \beta)}{2\,m^2}\right)\,
        \left( -2 {\rm Li}_2\left(\frac{z_3}{z_4}\right) +
          \frac{1}{4} \left( l_s^2 - 12 l_t^2
- 2 l_T^2 - 6 {l_T}\,{l_x} + l_x^2 - 12 {l_T}\,{l_{z3}} +  \right.\right.  \\
\nn   &&
             8\,{l_x}\,{l_{z3}} - 4\,{l_x}\,{l_{z4}} - 4\,l_{z4}^2 +
             {l_s}\,\left( -4\,{l_t} + 6\,{l_T} + 2\,{l_x}
- 8\,{l_{z3}} + 4\,{l_{z4}} - 4\,{l_{\beta}} \right)  +            \\
\nn   &&   \left.\left.
             4\,{l_t}\,\left( {l_x} + 4\,{l_{z3}} + 2\,{l_{z4}}
- 2\,{l_{\beta }} \right)  + 12\,{l_x}\,{l_{\beta }} +
             8\,{l_{z4}}\,{l_{\beta }} - 4\,l_{\beta}^2 \right)
+ 12\,\zeta(2) \right)  +                          \\
\nn   &&
       \frac{1}{8}\,{\rm Li}_2\left( \frac{m^2 x}{-T} \right)
        \left( -l_s^2 + 16 {l_t} {l_T} - {l_x} \left( 4 {l_T}
+ {l_x} - 4 {l_{z4}} \right)  -
          2 {l_s} \left( 2 {l_T} + {l_x}
- 2 {l_{z4}} \right)  \right)  +
       \frac{1}{4}   \times              \\
\nn   &&
{\rm Li}_2(\frac{m^2}{z_5})
        \left( 3 l_s^2 + 4 l_t^2 + 2 l_T^2
+ 2 {l_T} {l_x} + 7 l_x^2 + 8 {l_T} {l_{z3}} -
          2 {l_s}  ( 8 {l_t} + 3 {l_T} - {l_x} + 4 {l_{z3}}
- 6 {l_{z4}} )          \right.  \\
\nn   &&    \left.
+ 4 {l_T} {l_{z4}} +
          4 {l_x} {l_{z4}} - 12 l_{z4}^2
+ 8 {l_t} ( {l_{z3}} + {l_{z4}} - {l_{\beta }} )  +
          8 {l_x} {l_{\beta }} - 8 {l_{z3}} {l_{\beta }}
+ 8 {l_{z4}} {l_{\beta }} - 4 l_{\beta}^2 \right)  +     \\
\nn   &&
       \frac{1}{8}\,{\rm Li}_2\left(\frac{T}{z_3}\right)\,
        \left( 9\,l_s^2 - 16\,l_t^2
+ {l_x}\,\left( -4\,{l_T} + {l_x} + 12\,{l_{z3}} + 8\,{l_{z4}}
- 8\,{l_{\beta }} \right)  -                 \right.         \\
\nn   &&    \left.
          2 {l_s} ( 4 {l_t} - 2 {l_T} + 5 {l_x}
+ 6 {l_{z3}} + 4 {l_{z4}} - 4 {l_{\beta }} )  -
          8 {l_t} ( 2 {l_T} + {l_x} - 2 {l_{z3}}
- 2 {l_{z4}} + 2 {l_{\beta}} )  \right)  +              \\
\nn   &&
       \frac{1}{2} {\rm Li}_2(\frac{T}{m^2}) \left( 2 l_s^2
- 2 {l_T} {l_x} - 2 {l_T} {l_{z3}} + 3 {l_x} {l_{z3}} +
          5 l_{z3}^2 - 2 {l_t} ( {l_x}
+ 2 {l_{z3}} )  + 2 {l_T} {l_{z4}} + 3 {l_x} {l_{z4}} +   \right. \\
\nn   &&    \left.
          4 {l_{z3}} {l_{z4}} - l_{z4}^2
- 2 {l_x} {l_{\beta }} - 4 {l_{z3}} {l_{\beta }} +
          {l_s} ( 2 {l_t} - 3 {l_x} - 7 {l_{z3}}
- {l_{z4}} + 2 {l_{\beta }} )  \right)  +
       \frac{1}{4} {\rm Li}_2(-x)  \times        \\
\nn   &&
\left( 5 l_s^2
- 16 l_t^2 + l_x^2 - 4 {l_T} {l_{z3}} + 6 {l_x} {l_{z3}} -
          6 l_{z3}^2 - 4 {l_T} {l_{z4}} + 2 {l_x} {l_{z4}}
- 8 {l_{z3}} {l_{z4}} +
          2 l_{z4}^2 - 4 {l_x}\,{l_{\beta}} +     \right. \\
\nn   &&     \left.
 8 {l_{z3}} {l_{\beta}} +
          2 {l_s} ( -6 {l_t} + 2 {l_T} - 2 {l_x} + {l_{z3}}
- 3 {l_{z4}} + 2 {l_{\beta}} )  -
          4 {l_t} ( {l_x} - 6 {l_{z3}} - 4 {l_{z4}}
+ 4 {l_{\beta }} )  \right)  +                \\
\nn   &&
       {\rm Li}_3\left(\frac{z_4}{T}\right)\,( -{l_T} + {l_{z4}} )  +
  2 \left(  {\rm Li}_3\left(\frac{z_3}{s \beta}\right) +
            {\rm Li}_3\left(\frac{z_5}{t \beta}\right) \right)
\,( 2\,{l_t} + 2\,{l_T} - {l_{z3}} - {l_{z4}}  )  +      \\
\nn   &&
      2 {\rm Li}_3\left(\frac{-1 + \beta}{2\beta}\right)\,( 2\,{l_s}
- 2\,{l_t} - 2\,{l_T} - {l_{z3}} - {l_{z4}} )  +    \\
\nn   &&
{\rm Li}_3\left(\frac{m^2}{z_5}\right)\,
        \left( -\frac{17}{2} {l_s}
- 8\,{l_t} + 4\,{l_T} - \frac{{l_x}}{2} + 8\,{l_{z3}} + 9\,{l_{z4}}
- 8\,{l_{\beta }} \right)  +                      \\
\nn   &&
\frac{1}{2}\,{\rm Li}_3\left(\frac{T}{z_6}\right)\,
        \left( -5\,{l_s} - 4\,{l_t} - 2\,{l_T} + {l_x}
+ 6\,{l_{z3}} + 4\,{l_{z4}} - 4\,{l_{\beta }} \right)  +         \\
\nn   &&
       2\,{\rm Li}_3\left(\frac{z_3}{z_4}\right)\,( -{l_s} + {l_t}
+ {l_T} + {l_x} - {l_{\beta}} )  +
       {\rm Li}_3\left(\frac{z_3}{t}\right)\,( -{l_s} - 2\,{l_T}
- 3\,{l_x} + 2\,{l_{z3}} )  +                 \\
\nn   &&
       \frac{1}{2} {\rm Li}_3\left(\frac{z_6}{m^2}\right)
        ( {l_s} + 8 {l_t} - 8 {l_T} - {l_x} - 6 {l_{z3}}
- 4 {l_{z4}} )  +
       {\rm Li}_3\left(\frac{z4}{t}\right) ( {l_s} + 4 {l_t}
+ 2 {l_T} - 3 {l_x} -                        \\
\nn   &&
2 {l_{z3}} - 4 {l_{z4}} )  +
       \frac{1}{2} {\rm Li}_3\left(\frac{z_5}{T}\right) ( {l_s}
+ 2 {l_T} + {l_x} - 2 {l_{z4}} )  +
2 \left(  {\rm Li}_3\left(\frac{-x^2}{1 - x^2}\right) -     \right.       \\
\nn   &&   \left.
     2 {\rm Li}_3\left(\frac{1 - \beta}{2}\right) -
       {\rm Li}_3\left( \frac{m^2 (1 - \beta)}{2\,z_5}\right) +
       {\rm Li}_3\left(\frac{-t(1 - \beta)}{2\,z_5}\right)
   - 2 {\rm Li}_3\left(\frac{T}{m^2}\right) \right)\,{l_T}   +   \\
\nn   &&
     2 \left( 2 {\rm Li}_3\left(\frac{T}{m^2}\right) +
                {\rm Li}_3\left(\frac{2\,z_6}{m^2(1 + \beta)}\right) -
                {\rm Li}_3\left( \frac{-2\,z_6}{t( 1 + \beta)}\right) +
                {\rm Li}_3\left(\frac{z_6}{z_5}\right)  \right)\,
        ( {l_s} + {l_t} +                                          \\
\nn   &&
                           {l_T} - {l_{z3}} - {l_{z4}} + {l_{\beta}})  -
              2 {\rm Li}_3\left(\frac{z_6}{z_5}\right) {l_x}
          - {\rm Li}_3\left(\frac{T}{z_3}\right) {l_{z3}} +
    \left(  {\rm Li}_3\left(\frac{T}{z_3}\right)
       + 2\,{\rm Li}_3(-x) \right) \times                 \\
\nn   &&
( 2 {l_s} + 2 {l_t} + {l_T} - 2 {l_{z3}} - 2 {l_{z4}} + 2 {l_{\beta}} )  +
       2 {\rm Li}_3(x)\, ( 4 {l_s} + 3 {l_t} + {l_T} + {l_x}
- 2 {l_{z3}} - 6 {l_{z4}} +                             \\
\nn   &&
 3 {l_{\beta}} )  +
       2 {\rm Li}_4(x) - {\rm Li}_4\left(\frac{T}{z_3}\right) -
       4 {\rm Li}_4\left(\frac{z_3}{t}\right) +
       4 {\rm Li}_4\left(\frac{z_4}{t}\right) -
         {\rm Li}_4\left(\frac{z_4}{T}\right) +
         {\rm Li}_4\left(\frac{z_5}{T}\right) +                      \\
\nn   &&
       2 {\rm Li}_4\left(\frac{s(1 - \beta)}{-2 t}\right) +
       3 {\rm Li}_4\left(\frac{s(1 - \beta) }{2\, z_4}\right) +
       4 {\rm Li}_4\left(\frac{-1 + \beta}{2\,\beta}\right) +
       2 {\rm Li}_4\left(\frac{-2 t}{s(1 + \beta)}\right) +            \\
\nn   &&
       4 {\rm Li}_4\left(\frac{2\,\beta }{1 + \beta}\right) +
         {\rm Li}_4\left(\frac{s T( 1 + \beta )}{2\,m^2\,z_3}\right) +
       3 {\rm Li}_4\left(\frac{2\,z_3}{s(1 + \beta)}\right) +    \\
%
%
\nn   &&
       \frac{1}{2} L_{-++}\left(1,\frac{m^2}{-T},
                     \frac{m^2}{-T},\frac{s(1 - \beta)}{-2 z_4}\right) -
       \frac{1}{2} L_{-++}\left(1,\frac{m^2}{-T},
                     \frac{m^2}{-T},\frac{s(1 + \beta)}{-2 z_3} \right) +
\\
\nn   &&
       2 L_{-++}\left(1,\frac{m^2}{-T},\frac{s(1 - \beta)}{-2 z_4},0\right) +
       L_{-++}\left(1,-\frac{t}{z_3},-\frac{t}{z_3},-\frac{t}{T}\right) -  \\
\nn   &&
       L_{-++}\left(1,\frac{s(1 - \beta)}{-2 z_4},\frac{m^2}{-T},
        \frac{s(1 - \beta)}{-2 z_4}\right) -
L_{-++}\left(1,\frac{s(1 - \beta)}{-2 z_4},\frac{m^2}{-T},
        \frac{s(1 + \beta)}{-2 z_3} \right) +                    \\
\nn   &&
 L_{-++}\left(1,\frac{s(1 - \beta)}{- 2 z_4},\frac{s(1 - \beta)}{- 2 z_4},0\right) -
       \frac{1}{2} L_{-++}\left(1,\frac{s(1 - \beta)}
{- 2 z_4},\frac{s(1 - \beta)}{- 2 z_4},
         \frac{s(1 + \beta)}{- 2 z_3} \right) +        \\
\nn   &&
\frac{1}{2} L_{-++}\left(\frac{t}{T},0,0,-\frac{t}{z_3}\right) -
\frac{1}{2} L_{-++}\left(\frac{t}{T},0,0,-\frac{t}{z_4} \right) +
       2 L_{-++}\left(\frac{t}{T},0,-\frac{t}{z_3},-1\right) -    \\
\nn   &&
       L_{-++}\left(\frac{t}{T},0,-\frac{t}{z_3},-\frac{t}{z_3} \right) -
       L_{-++}\left(\frac{t}{T},0,-\frac{t}{z_3},-\frac{t}{z_4} \right) -
       L_{-++}\left(\frac{t}{T},-\frac{t}{z_3},-\frac{t}{z_3},
                                   -\frac{t}{z_4} \right) +            \\
\nn   &&
3 L_{-++}(\frac{t}{z_4},0,0,-1) -
       \frac{3}{2} L_{-++}\left(\frac{t}{z_4},0,0,-\frac{t}{z_3} \right) -
       2 L_{-++}\left(\frac{t}{z_4},0,-\frac{t}{z_3},-1\right) +         \\
\nn   &&
       L_{-++}\left(\frac{t}{z_4},0,-\frac{t}{z_3},-\frac{t}{z_3} \right) +
     3 L_{-++}\left(\frac{t}{z_4},0,-\frac{t}{z_3},-\frac{t}{z_4} \right) +  \\
\nn   &&
\frac{1}{2} L_{-++}\left(\frac{t}{z_4},-\frac{t}{z_3},-\frac{t}{z_3},
         -\frac{t}{T} \right) +
2 L_{-++}\left(\frac{s(1 + \beta)}{2\,z_3},\frac{m^2}{-T},0,
          \frac{m^2}{-T} \right) -                               \\
\nn   &&
2 L_{-++}\left(\frac{s(1 + \beta)}{2\,z_3},\frac{m^2}{-T},0,
         \frac{s(1 - \beta)}{- 2 z_4}\right) -
       2 L_{-++}\left(\frac{s(1 + \beta)}{2\,z_3},\frac{m^2}{-T},0,
         \frac{s(1 + \beta)}{- 2 z_3} \right) +             \\
\nn   &&
       \frac{1}{2} L_{-++}\left(\frac{s(1 + \beta)}{2\,z_3},
    \frac{m^2}{-T},\frac{m^2}{-T},\frac{s(1 - \beta)}{- 2 z_4}\right) +
       2 L_{-++}\left(\frac{s(1 + \beta)}{2\,z_3},
\frac{s(1 - \beta)}{- 2 z_4},0,\frac{m^2}{-T} \right) -         \\
\nn   &&
\frac{1}{2} L_{-++}\left(\frac{s(1 + \beta)}{2\,z_3},
    \frac{s(1 - \beta)}{-2z_4},\frac{s(1 - \beta)}{-2z_4},\frac{m^2}{-T}\right) -
4 L_{+}\left(0,\frac{z_3}{s\beta},-\frac{z_3 z_4}{st\beta},-1\right) +    \\
\nn   &&
    4 L_{+}\left(0,\frac{z_3(1 - \beta)}{-2t\beta},
\frac{z_3 z_4}{st\beta},\frac{m^2}{-T}\right) +
2 L_{+}\left(\frac{m^2}{-T},\frac{z_3(1 - \beta)}{-2t\beta},
         \frac{z_3 z_4}{st\beta},\frac{s(1 - \beta)}{-2z_4}\right) +    \\
\nn   &&
       2 L_{+}\left(-\frac{t}{z_3},0,\frac{z_4}{t},-\frac{t}{z_4} \right) +
       2 L_{+}\left(-\frac{t}{z_3},0,\frac{z_3}{t},-\frac{t}{z_4} \right)  +
       2 L_{+}\left(-\frac{t}{z_3},\frac{z_3}{s\beta},
         -\frac{z_3 z_4}{st\beta},-\frac{t}{T} \right) +       \\
\nn   &&
       \frac{3}{2} L_{+++}\left(0,0,-\frac{t}{z_3},-\frac{t}{z_4} \right) -
       2 L_{+++}\left(0,\frac{m^2}{-T},
     \frac{s(1 - \beta)}{-2z_4},\frac{m^2}{-T} \right) +             \\
\nn   &&
       2 L_{+++}\left(0,\frac{m^2}{-T},\frac{s(1 - \beta)}{-2z_4},
         \frac{s(1 - \beta)}{-2z_4}\right) +
       2 L_{+++}\left(0,\frac{m^2}{-T},\frac{s(1 - \beta)}{-2z_4},
         \frac{s( 1 + \beta)}{-2z_3} \right) +           \\
\nn   &&
       L_{+++}\left(0,-\frac{t}{z_3},-\frac{t}{z_3},-1\right) -
L_{+++}\left(\frac{m^2}{-T},\frac{m^2}{-T},0,\frac{s(1 - \beta)}{-2z_4}\right) +
\\
\nn   &&
L_{+++}\left(\frac{m^2}{-T},\frac{m^2}{-T},0,\frac{s(1 + \beta)}{-2z_3} \right) -
     \frac{1}{2} L_{+++}\left(\frac{m^2}{-T},
\frac{m^2}{-T},\frac{s(1 - \beta)}{-2z_4},\frac{s(1 + \beta)}{-2z_3} \right) -
\\
\nn   &&
3 L_{+++}\left(-\frac{t}{z_3},0,0,-1\right) +
\frac{3}{2} L_{+++}\left(-\frac{t}{z_3},-\frac{t}{z_3},0,-\frac{t}{z_4} \right) +
\\
\nn   &&  \left.
L_{+++}\left(\frac{s(1 - \beta)}{-2z_4},
               \frac{s(1 - \beta)}{-2z_4},0,\frac{m^2}{-T} \right)
 \right];
\ea
\ba
\label{D1imag}
{\rm Im}\,D_1^{(-1)}&=&\pi/(st\beta), {\rm \hspace{.6in}}
{\rm Im}\,D_1^{(0)}= - 2 \pi \ln (-t\beta/m^2) /(st\beta),         \\
\nn
{\rm Im}\,D_1^{(1)}&=& \frac{\pi}{st\beta} \left[ -\frac{3}{4}\,
         l_s^2 + 2\, l_t^2 - \frac{3}{4}\, l_x^2 -
           {l_x} {l_{z3}} + {l_x} {l_{z4}} - l_{z4}^2 +
           {l_s}\left( {l_t} + {l_x}/2 +
              {l_{z3}} + {l_{z4}} \right)  -      \right.     \\
\nn   &&       \left.
{l_t}\,( {l_x} + 2\,{l_{z3}} - 4\,{l_{\beta}} )
+ 2\, l_{\beta}^2   - 2\zeta(2) -
        2 {\rm Li}_2\left(\frac{m^2}{z_5}\right) -
        2 {\rm Li}_2\left(\frac{t( -1 + \beta)}
           {2 m^2} \right)   \right],                     \\
\nn
{\rm Im}\,D_1^{(2)}&=&\frac{\pi}{st\beta} \left[
\frac{1}{12}  \left(
        7\,l_s^3 - l_x^3 -
           3\,l_s^2 \left( 3\,{l_x} + 4\, {l_{z3}} + 4\,{l_{z4}}
 - 6\,{l_{\beta}} \right)  +
           6\,l_x^2 \left( -2\,{l_{z3}} + 3\,{l_{\beta }} \right)  +
\right. \right.        \\
\nn   &&
           12\,{l_x} \left( l_{z4}^2 -
              2\,\left( -{l_{z3}} + {l_{z4}} \right) \,{l_{\beta }}
+ 2\,{l_t} \left( -{l_{z3}} + {l_{\beta }} \right)  \right) -    \\
\nn   &&
 8\,{l_{\beta}} \left( 6\,l_t^2 -
              3\,l_{z4}^2 + 2\,l_{\beta}^2 +
              6\,{l_t}\,( -{l_{z3}} + {l_{\beta }} )  \right)
- 3\,{l_s} \left( 8\,l_t^2 - 5\,l_x^2 -
              4\,l_{z4}^2 -       \right.  \\
\nn   &&  \left.\left.
              8\,{l_t} \left( {l_x} + {l_{z3}} - {l_{\beta }} \right)  +
              8\,{l_{z3}}\,{l_{\beta }} +
              8\,{l_{z4}}\,{l_{\beta }} +
              4\,{l_x} \left( -2\,{l_{z3}} +
                 2\,{l_{z4}} + {l_{\beta}} \right)  \right) \right) +   \\
\nn   &&
4\,\left( {l_t} + {l_x} + {l_{\beta }} \right) \,\zeta(2) +
        2\,{\rm Li}_2(x)\,{l_x} +
        2\,{\rm Li}_2\left(\frac{z3}{z4}\right)\,{l_x} +  \\
\nn   &&
        2\,{\rm Li}_2\left(\frac{t(-1 + \beta)}{2\,m^2}\right)
              \left( {l_s} - {l_x} + 2\,{l_{\beta}} \right)  +
        2\,{\rm Li}_2\left(\frac{m^2}{z_5}\right)
              \left( {l_s} + {l_x} + 2\,{l_{\beta }} \right)  -    \\
\nn   &&     \left.
        2\,{\rm Li}_3(x) -
        4\,{\rm Li}_3\left(\frac{z_3}{t}\right) +
        2\,{\rm Li}_3\left(\frac{z_3}{z_4}\right) -
        4\,{\rm Li}_3\left(\frac{z_4}{t}\right) -
        2\,{\rm Li}_3\left(\frac{z_6}{z_5}\right)
\right].
\ea
\\
\\
\[
\mbox{ \epsfysize = 3.0cm  \epsffile {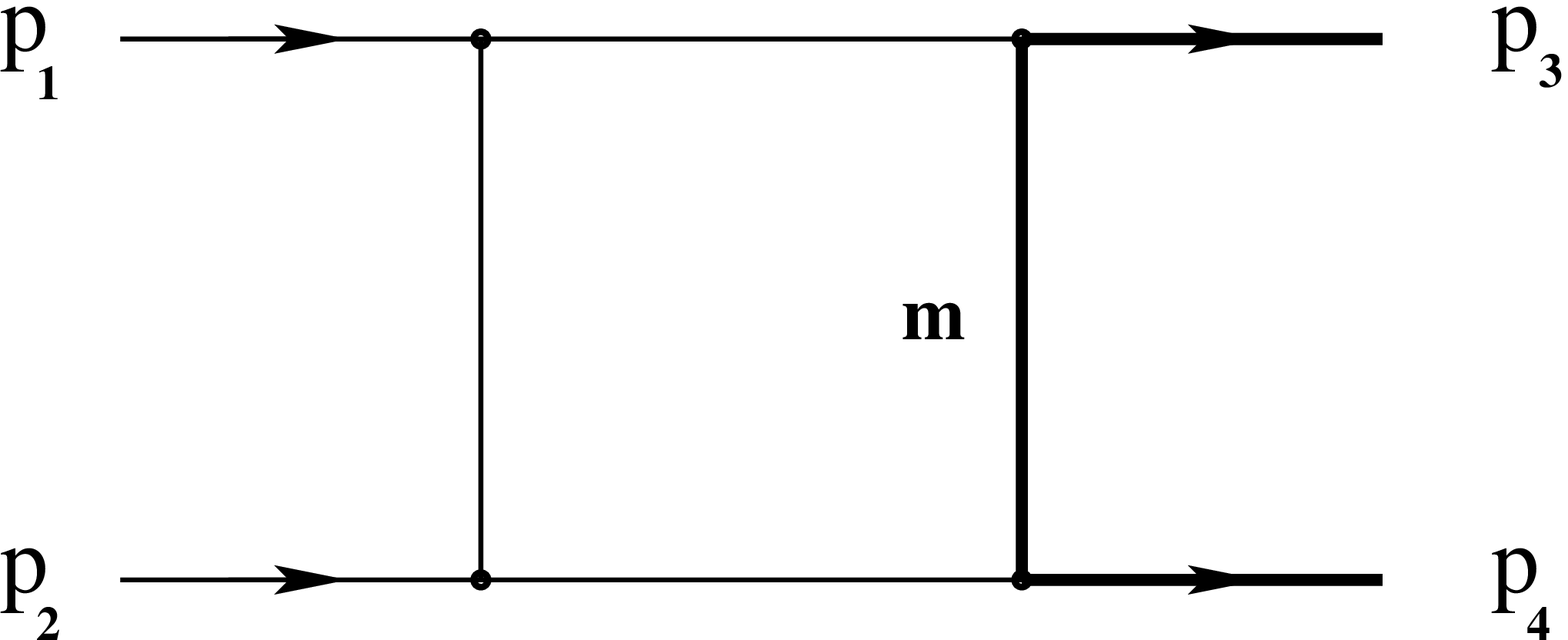}}
\]
{\begin{center}
\tenrm\vglue -0.1cm
FIG.~3. Massive box $D_2$ with one massive propagator.
\end{center}
}
\vglue .1cm

\noindent
Next we turn to the second four-point function $D_2$ with one massive
propagator shown in Fig.~3 which is defined by
\[
D_2 \equiv D(-p_2,p_4,p_3,0,0,m,0).
\]
We substitute the appropriate values of momenta and masses for the $D_2$ integral
into the general kernel expression (\ref{kernel}) and obtain
\be
K = ac\tilde{t} - bd\tilde{s} + c^2 - i\delta.
\ee
In order to simplify the first integration over the Feynman parameter $x_3$ we
choose $\{a,b,c,d\}$ as $\{x_1x_2(1-x_3), \, x_1x_2x_3, \,
x_1(1-x_2), \, 1-x_1\}$.
After $x_3$-integration we write the result for the integrand as a sum of two
parts $I^{D_2}_{x_1x_2}+II^{D_2}_{x_1x_2}$:
\ba
I^{D_2}_{x_1x_2}=-\frac{x_1\left[ x_1^2(1-x_2) (1+(\tilde t -1) x_2)- i
\delta \right]^{-1-\ep}}{(1+\ep)
[\tilde s (1-x_1)+\tilde t x_1 (1-x_2)]},         \\
\label{IID2x1x2o}
II^{D_2}_{x_1x_2}=\frac{x_1\left[ -\tilde s x_1 x_2 + x_1^{2}(1+(\tilde s -
2) x_2 + x_2^{2}) - i\delta \right]^{-1-\ep}} {(1+\ep)[\tilde s
(1-x_1)+\tilde t x_1 (1-x_2)]},
\ea
where again the two terms derive from the indefinite integral (or primitive)
evaluated at the upper and lower boundary of $x_3$.
First consider the integration of $I^{D_2}_{x_1x_2}$. One notes
that its numerator is not negative on the integration path since $
\tilde t > 1$, which implies there is no imaginary contribution coming from
$I^{D_2}_{x_1x_2}$. Therefore, we omit the $i\delta$ term for the
remaining integration. After integration over $x_1$ we arrive at
\ba
\label{id2x2}
I^{D_2}_{x_2}=\frac{\left[ (1 - x_2) \,\left( 1 + (\tilde t - 1)
\,x_2 \right)  \right]^
     {-1 -\ep}\, _{2} F_{1}(1, -2\ep, 1 - 2\ep, 1-A)}{2\tilde s\,(1 + \ep)
\,\ep },
\ea
where we have defined $A\equiv\tilde t\,(1- x_2)/\tilde s$
and $_{2}F_{1}$ is a hypergeometric function. The above expression
is singular at the upper integration limit $x_2 = 1$.
In order to regularize this singularity we
have to find a suitable subtraction term.

First note that the $\ep$-expansion of the
hypergeometric function reads
\ba
\nn
\label{expd21}
F_{1}(1,-2\ep, 1 - 2\ep, 1-A) =  1 +2 \ep \, \ln A - 4{\ep}^2 \, {\rm
Li}_{2}(1-A) - 8{\ep}^3 \, {\rm Li}_{3}(1-A)         \\
- 16{\ep}^4 \, {\rm Li}_{4}(1-A) + {\cal O}(\ep^5).
\ea
To obtain a suitable subtraction term we substitute (\ref{expd21}) into
(\ref{id2x2}) and replace $x_2$ by 1 everywhere in $I^{D_2}_{x_2}$ except
for terms that diverge. Therefore, our subtraction term can be defined as
\ba
I_{x_2}^{D_2,s}=\frac{( 1 - x_2 )^{-1 - \ep} \,  \tilde t^{-1 - \ep}
 \,[ 1 + 2\ep\, \ln A
- 4\ep^2 \zeta(2)  - 8\ep^3\, \zeta(3)
- 16\ep^4 \zeta(4) ]}{2\tilde s\,( 1 + \ep ) \,\ep }.
\ea
Our subtraction term is simple enough to be integrated analytically over
$x_2$ giving the result
\ba
I^{D_2,s}=\frac{\tilde t^{-1-\ep}\,\left( -3 -
2\ep\,\ln \frac{\tilde t}{\tilde s} +
4\ep^2\,\zeta(2) + 8\ep^3\,\zeta(3) +
16\ep^4\,\zeta(4)  \right) }
  {2\, (1 + \ep) \,\ep^2\,\tilde s}
\ea
which can readily be expanded in $\ep$.

Next we turn to the remaining finite integral
$I^{D_2}_{x_2}-I_{x_2}^{D_2,s}$.
We expand  $I^{D_2}_{x_2}-I_{x_2}^{D_2,s}$
up to second order in $\ep$ and integrate over $x_2$ after the expansion.

Next consider the second integrand $II^{D_2}_{x_1x_2}$ (\ref{IID2x1x2o}).
The term in the numerator in square brackets
raised to the power $(-1-\ep)$ changes sign on the integration path. It
means that the corresponding integral has an imaginary contribution.
We can rewrite $II^{D_2}_{x_1x_2}$ as follows:
\ba
\label{iid2x1x2}
II^{D_2}_{x_1x_2}=\frac{x_1^{-\ep} (\tilde s x_2)^{-1-\ep}
\left( -1-i\delta + x_1(\frac{(1-x_2)^{2}}{\tilde s x_2} + 1)
\right)^{-1-\ep} }{(1+\ep) \tilde s
\left( 1-x_1(1-\frac{\tilde t}{\tilde s} (1-x_2))\right)},
\ea
The integration of $II^{D_2}_{x_1x_2}$ over $x_1$ is more difficult because
of the additional term $x_1^{-\ep}$.
We proceed by expanding $x_1^{-\ep}$  as $
(1-\ep \ln x_1 + \frac{ \ep^2 }{2} \ln^{2} x_1 + ...)$. One can see that
only the first term of this expansion gives rise to a divergence
in the subsequent integration. As we will need to find a subtraction term for
this term we will treat it separately. For the remaining terms we do
an overall $\ep$-expansion of (\ref{iid2x1x2}) and then perform the
remaining integrations.
We therefore substitute $x_1^{-\ep}$ by $1$. The integral
looks simpler (we denote it by $II_{x_1x_2}^{0}$). Integration over $x_1$
yields
\ba
II_{x_2}^{0}= \frac{ (\tilde s x_2)^{-1-\ep} x_2
(-1-i\delta)^{-\ep} \,
_{2}F_{1}(1,-\ep,1-\ep,-\frac{x_2(\tilde
s+\tilde t (-1+x_2))}{(1-x_2)(1+(\tilde t -1)x_2)} )}
{\ep(1+\ep)(1-x_2)(1+(\tilde t -1)x_2)}   \nn  \\
- \frac{ (\tilde s x_2)^{-1-\ep} x_2 \,
(1-x_2)^{-2\ep} \,
_{2}F_{1}(1,-\ep,1-\ep, \frac{(1-x_2)(\tilde
s+ \tilde t(-1+x_2) )}{\tilde s(1+(-1+\tilde t)x_2)} )  }{
\ep (1+\ep)(1-x_2)(1+(\tilde t -1)x_2)    (\tilde s
x_2)^{-\ep} }.
\ea
Note that we omit the imaginary shifts $i\delta$ in the arguments of the
hypergeometric
functions $_{2}F_{1}$, as the branch cuts of $_{2}F_{1}$ are never crossed
in the physical region.
If we would directly integrate the above expression we would have a
divergence at $x_2 \rightarrow 1$. We must therefore define a subtraction term.
If one uses $II_{x_2}^{0}$ in the present form the definition is rather
difficult: as $x_2 \rightarrow 1$ the argument of the first function
$_{2}F_{1}$ goes to infinity.
To circumvent this problem we can use one of the relations between
hypergeometric functions
to transform the argument of the function. As a result we pull out
the divergent term as an overall factor multiplying the hypergeometric
function with a transformed argument. The whole expression can be rewritten
as
    \ba \label{yy1}
II_{x_2}^{0}=
-\frac{ ( 1 - x_2 )^{-1 - 2\ep}\,
     _{2}F_{1}(1, -\ep, 1-\ep,
       \frac{\left( \tilde s + \tilde t\,\left( -1 + x_2 \right)  \right)
\,( 1 - x_2 ) }
         {\tilde s\,\left( 1 + \left( -1 + \tilde t \right) \,x_2 \right) }
)}{\tilde s\,\left( 1 + \left( -1 + \tilde t \right) \,x_2 \right) \,
     ( 1 + \ep ) \,\ep }  +      {\rm \hspace{1.9in}}           \\
\nn
  \frac{ \left( \tilde s  ( 1 - x_2 )
\right)^{-1 - \ep} x_2^{-\ep}
     \left( 1 + \left( -1 + \tilde t \right) x_2 \right)^{-1 - \ep}
     ( -1 - i \,\delta )^{-\ep} \,
     _{2}F_{1}(-\ep,-\ep,1-\ep,
      \frac{\left( \tilde s - \tilde t\,\left( 1 - x_2 \right)  \right)
\,x_2}{ ( 1 - x_2 )^2 + \tilde s\,x_2}) }{ {\left( \left( 1 - x_2
             \right)^2 + \tilde s\,x_2 \right) }^{-\ep}
\, ( 1 + \ep ) \,\ep } .
    \ea
Now all the poles arise from the factors $(1 - x_2)^{-1 - 2\ep}$ and
$(1 - x_2)^{-1 - \ep}$, and we can derive the necessary subtraction term
following the above procedure. We briefly mention that when
$x_2 = 1$ the first hypergeometric function (in the first line of
Eq.~(\ref{yy1})) takes the value $_{2}F_{1}=1$. The second hypergeometric
function takes the value
$-\ep\pi/\sin(-\ep\pi)$ which, in turn, can be expanded to order $\ep^4$ as
$1 + \ep^2\zeta(2) + 7\ep^4\zeta(4)/4$. Thus, the subtraction term reads
\ba
\label{yy2}
II_{x_2}^{0,s}=-\frac{ (1 - x_2)^{-1 - 2\ep} }{(1 + \ep) \ep\tilde s\tilde
t} + \frac{ \left( 1 +  \ep^2\zeta(2) +
\frac{7\ep^4\zeta(4)}{4} \right) \,\tilde t^{-1 - \ep}\,
     (1 - x_2)^{-1 - \ep}\,(-1 - i\delta)^{-\ep}}
     {( 1 + \ep ) \,\ep\,\tilde s}.
\ea
Integrating this subtraction term we arrive at
\ba
II^{0,s}=\frac{1}{2(1 + \ep)\ep^2\tilde s\tilde t} -
  \frac{\left( 1 + \ep^2\zeta(2) +
\frac{7\ep^4\zeta(4)}{4} \right) \,\tilde t^{-1 - \ep}\,
     ( -1 - i \delta )^{-\ep}}
     {( 1 + \ep ) \ep^2\tilde s},
\ea
which can finally be expanded up to $\ep^2$.

Now that we have found a suitable subtraction term we can proceed
with the remaining terms. We subtract from $II_{x_2}^{0}$
(Eq.~(\ref{yy1})) the subtraction term $II_{x_2}^{0,s}$ (Eq.~(\ref{yy2})).
Since the result is convergent with regard to
the integration over $x_2$, we can expand the result in terms of
$\ep$ before integration which greatly simplifies the problem. We then
do the last integration. The difference
$II_{x_2}^{0} - II_{x_2}^{0,s}$ must be expanded
up to third order in $\ep$. The reason for this is that we have
already one pole $\sim 1/\ep$ after the $x_1$-integration.
Therefore, in order to get results up to second order the
hypergeometric functions have to be expanded to third order. The
expansion for one of the hypergeometric functions is done using (\ref{expd21}).
For the
$\ep$-expansion of the second hypergeometric function one gets
\ba   &&
_{2}F_{1}(-\ep,-\ep,1-\ep,z) = 1 + \ep^2\, {\rm Li}_{2}(z)       \\
\nn   &&
- \ep^3\, \left( \frac{1}{2}\ln^2(1 - z) \ln z + \ln (1 - z)
{\rm Li}_{2}(1 - z) -
            {\rm Li}_{3}( 1 - z) - Li_{3}(z) + \zeta(3) \right) .
\ea
Using these results for the $\ep$-expansions we expand
$II_{x_2}^{0}-II_{x_2}^{0,s}$ up to $\ep^2$
and integrate the resulting expression. Finally, carefully collecting all
the relevant pieces, we arrive at the final result for our second
four-point function. In order to reduce the length of the final result for $D_2$
we introduce four more abbreviations. We write
\ba
\label{nots2}
&&
D \equiv  m^2 s - t u,             {\rm \hspace{.4in}}
l_D \equiv \ln \frac{-D}{m^4},  {\rm \hspace{.4in}}
l_u \equiv \ln \frac{-u}{m^2},  {\rm \hspace{.4in}}
l_U \equiv \ln \frac{-U}{m^2}
\ea
Our result for the four-point box diagram $D_2$ reads:
\be
\label{d2}
{\rm Re}\,D_2^{(-2)} = 2/(s t),          {\rm \hspace{.6in}}
   {\rm Re}\,D_2^{(-1)}=-[l_s + 2 l_t]/(s t),   {\rm \hspace{2.0in}}
\ee
\vspace{-.3in}
\ba
\nn
{\rm Re}\,D_2^{(0)} &=& [ 2\,{l_s}\,{l_t} - 5\,\zeta(2) ]/(s t),    \\
\nn
{\rm Re}\,D_2^{(1)} &=& \frac{1}{s\,t} \left[
          \frac{1}{2}\,l_s^3 + 3\,l_t^3 + \frac{l_x^3}{12} -
          l_t^2 \left( 4\,{l_T} + 3\,{l_x} +
             4\,l_{z3} - l_{z4} \right)  +
          \frac{1}{2}\,l_x^2\,l_{z4} -
          {l_x}\,l_{z4}^2 +
          \frac{2}{3}\,l_{z4}^3 -
\right. \\
\nn  &&
          \frac{1}{4} l_s^2 \left( 6 {l_t} + 3 {l_x} +
             4 l_{z3} + 2 l_{z4} \right)  -
          {l_s} \left( 2 l_t^2 -
             \frac{1}{2} l_x^2 - l_x l_{z3} -
             2 {l_t} \left( {l_x} + 2 {l_{z3}} +
                {l_{z4}} \right)  \right)  -     \\
\nn  &&
          {l_t} \left( l_T^2 + \frac{1}{2} l_x^2 +
             2 {l_x} l_{z3} - 2 l_x l_{z4}  -
             2 {l_T}\left( {l_x} + {l_{z3}} +
                {l_{z4}} \right)  +
              l_{z3}^2 + 3 l_{z4}^2 \right)  -
              \left( {l_s} - 2 {l_t} +     \right.    \\
\nn   &&  \left.
       l_x - 4 {l_{z4}} \right) \zeta(2) + \zeta(3) +
       2 \left( {\rm Li}_2 \left( \frac{m^2}{z_5} \right) +
       {\rm Li}_2 \left( \frac{-t (1 - \beta)}{2\,m^2} \right) \right)
                               \left( {l_s} - 2\,{l_t} \right)  -   \\
\nn   &&
       4 {\rm Li}_2 \left( \frac{T}{m^2} \right) {l_t} -
       2 {\rm Li}_2 \left( \frac{m^2 x}{-T} \right) {l_t} -
       2 {\rm Li}_2 \left( \frac{T}{z_3} \right) {l_t} -
       4 {\rm Li}_3 \left( \frac{m^2}{-t} \right) +
       2 {\rm Li}_3(-x) -         \\
\nn   &&
       2 {\rm Li}_3 \left( \frac{z_3}{t} \right) -
       2 {\rm Li}_3 \left( \frac{z_4}{t} \right) -
       4 {\rm Li}_3 \left( \frac{m^2}{z_5} \right) -
       4 {\rm Li}_3 \left( \frac{z_6}{m^2} \right) +
       2 {\rm Li}_3 \left( \frac{m^2 (1 - \beta) }{2\,z_5} \right) -   \\
\nn  &&   \left.
       2 {\rm Li}_3 \left( \frac{-t(1 - \beta)}{2\,z_5} \right) +
       2 {\rm Li}_3 \left( \frac{2\,z_6}{m^2(1 + \beta)} \right) -
       2 {\rm Li}_3 \left( \frac{2\,z_6}{-t(1 + \beta)} \right)
\right],     \\
\nn
{\rm Re}\,D_2^{(2)} &=& \frac{1}{s\,t} \left[
       -\frac{l_s^4}{48} -
       \frac{17}{24} l_s^3 {l_t} - \frac{9}{2} l_s^2 l_t^2 -
       3 {l_s} l_t^3
    +  \frac{13}{6} l_s^3 l_T +
       \frac{5}{2} l_s^2 l_t l_T +
       \frac{3}{2} l_t^3 l_T - \frac{1}{2} l_s^2 l_T^2 -
       \frac{7}{2} l_t^2 l_T^2     \right.  \\
\nn   &&
        - 3 l_t l_T^3
- \frac{25}{24} l_t^4 -
       \frac{3}{4} l_s^3 {l_u} -
       \frac{13}{4} l_s^2 {l_t} {l_u} + 7 {l_s} l_t^2 {l_u} -
       3 l_t^3 {l_u} - \frac{5}{4} l_s^2 {l_T} {l_u} +
       2 {l_s} {l_t} {l_T} {l_u} -      \\
\nn   &&
       2 l_t^2 {l_T} {l_u} +
       \frac{1}{2} l_s^2 l_u^2 -
       \frac{3}{2} {l_s} {l_t} l_u^2 + l_t^2 l_u^2 -
       \frac{1}{2} {l_s} {l_T} l_u^2 + {l_t} {l_T} l_u^2 +
       \frac{1}{3} {l_s} l_u^3 - \frac{1}{3} {l_t} l_u^3 -
       \frac{1}{4} l_s^3 {l_x} -                  \\
\nn  &&
       \frac{23}{8} l_s^2 {l_t} {l_x} + 3 {l_s} l_t^2 {l_x} -
       \frac{7}{3} l_t^3 {l_x} -
       \frac{7}{4} l_s^2 {l_T} {l_x} -
       3 {l_s} {l_t} {l_T} {l_x} -
       3 l_t^2 {l_T} {l_x} +
       3 {l_t} l_T^2 {l_x} - \frac{1}{2} l_s^2 {l_u} {l_x} +     \\
\nn   &&
       \frac{1}{2} {l_s} {l_t} {l_u} {l_x} +
       \frac{5}{2} {l_s} {l_T} {l_u} {l_x} +
       \frac{1}{2} {l_s} l_u^2 {l_x} -
       \frac{1}{2} {l_t} l_u^2 {l_x} -
       \frac{1}{2} {l_T} l_u^2 {l_x} +
       \frac{1}{4} l_s^2 l_x^2 +
       \frac{19}{8} {l_s} {l_t} l_x^2 +
       \frac{3}{2} l_t^2 l_x^2 -        \\
\nn   &&
       4 {l_s} {l_T} l_x^2 -
       \frac{9}{2} {l_t} {l_T} l_x^2 -
       \frac{1}{2} l_T^2 l_x^2 -
       \frac{3}{4} {l_s} {l_u} l_x^2 +
       \frac{3}{4} {l_t} {l_u} l_x^2 -
       \frac{1}{4} {l_T} {l_u} l_x^2 +
       \frac{5}{6} {l_s} l_x^3 + \frac{53}{24} {l_t} l_x^3 -   \\
\nn   &&
       \frac{37}{12} {l_T} l_x^3 - \frac{13}{16} l_x^4 -
       \frac{29}{24} l_s^3 l_{z3} +
       \frac{19}{4} l_s^2 {l_t} l_{z3} + 11 {l_s} l_t^2 l_{z3} +
       \frac{10}{3} l_t^3 l_{z3} -
       \frac{13}{4} l_s^2 {l_T} l_{z3} -       \\
\nn  &&
       10 {l_s} {l_t} {l_T} l_{z3} -
       9 l_t^2 {l_T} l_{z3} +
       7 {l_t} l_T^2 l_{z3} +
       2 l_s^2 {l_u} l_{z3} -
       2 {l_s} {l_t} {l_u} l_{z3} +
       \frac{5}{8} l_s^2 {l_x} l_{z3} +
       \frac{11}{2} {l_s} {l_t} {l_x} l_{z3} +       \\
\nn  &&
       l_t^2 {l_x} l_{z3} -
       \frac{5}{2} {l_s} {l_T} {l_x} l_{z3} -
       10 {l_t} {l_T} {l_x} l_{z3} +
       \frac{19}{8} {l_s} l_x^2 l_{z3} +
       \frac{7}{4} {l_t} l_x^2 l_{z3} -
       \frac{5}{4} {l_T} l_x^2 l_{z3} -
       \frac{19}{24} l_x^3 l_{z3} -       \\
\nn  &&
       \frac{5}{2} {l_s} {l_t} l_{z3}^2 -
       \frac{5}{2} l_t^2 l_{z3}^2 +
       \frac{7}{2} {l_s} {l_T} l_{z3}^2 +
       {l_t} {l_T} l_{z3}^2 -
       {l_s} {l_u} l_{z3}^2 +
       {l_t} {l_u} l_{z3}^2 -
       \frac{1}{2} {l_s} {l_x} l_{z3}^2 +
       \frac{7}{2} {l_t} {l_x} l_{z3}^2 +    \\
\nn  &&
       \frac{3}{2} {l_T} {l_x} l_{z3}^2 -
       \frac{1}{2} l_x^2 l_{z3}^2 +
       \frac{1}{3} {l_s} l_{z3}^3 +
       {l_t} l_{z3}^3 - {l_T} l_{z3}^3 -
       \frac{2}{3} {l_x} l_{z3}^3 -
       l_s^3 l_{z4} +
       5 l_s^2 {l_t} l_{z4} +
       7 {l_s} l_t^2 l_{z4} +    \\
\nn  &&
       \frac{16}{3} l_t^3 l_{z4} -
       \frac{5}{2} l_s^2 {l_T} l_{z4} -
       4 {l_s} {l_t} {l_T} l_{z4} -
       2 l_t^2 {l_T} l_{z4} +
       8 {l_t} l_T^2 l_{z4} +
       3 l_s^2 {l_u} l_{z4} -
       3 {l_s} {l_t} {l_u} l_{z4} -     \\
\nn  &&
       {l_s} {l_T} {l_u} l_{z4} +
       {l_s} l_u^2 l_{z4} -
       {l_t} l_u^2 l_{z4} -
       {l_T} l_u^2 l_{z4} +
       4 l_s^2 {l_x} l_{z4} -
       3 {l_s} {l_t} {l_x} l_{z4} +
       5 l_t^2 {l_x} l_{z4} +
       4 {l_t} {l_T} {l_x} l_{z4}      \\
\nn  &&
       -
       {l_s} {l_u} {l_x} l_{z4} +
       {l_t} {l_u} {l_x} l_{z4} -
       3 {l_T} {l_u} {l_x} l_{z4} +
       \frac{3}{2} {l_s} l_x^2 l_{z4} +
       {l_t} l_x^2 l_{z4} +
       \frac{5}{2} {l_T} l_x^2 l_{z4} +
       \frac{1}{6} l_x^3 l_{z4} +     \\
\nn  &&
       2 l_s^2 l_{z3} l_{z4} -
       14 {l_s} {l_t} l_{z3} l_{z4} -
       4 l_t^2 l_{z3} l_{z4} +
       6 {l_s} {l_T} l_{z3} l_{z4} +
       6 {l_t} {l_T} l_{z3} l_{z4} -
       2 {l_s} {l_u} l_{z3} l_{z4} +      \\
\nn   &&
       2 {l_t} {l_u} l_{z3} l_{z4} -
       2 {l_s} {l_x} l_{z3} l_{z4} -
       2 {l_t} {l_x} l_{z3} l_{z4} +
       2 {l_T} {l_x} l_{z3} l_{z4} -
       4 l_x^2 l_{z3} l_{z4} +
       3 {l_t} l_{z3}^2 l_{z4} -    \\
\nn   &&
       2 {l_T} l_{z3}^2 l_{z4} -
       l_s^2 l_{z4}^2 -
       2 {l_s} {l_t} l_{z4}^2 -
       \frac{9}{2} l_t^2 l_{z4}^2 +
       2 {l_s} {l_T} l_{z4}^2 -
       3 {l_t} {l_T} l_{z4}^2 -
       2 {l_s} {l_u} l_{z4}^2 +
       2 {l_t} {l_u} l_{z4}^2 +     \\
\nn   &&
       {l_T} {l_u} l_{z4}^2 -
       {l_s} {l_x} l_{z4}^2 -
       2 {l_t} {l_x} l_{z4}^2 +
       {l_T} {l_x} l_{z4}^2 -
       l_x^2 l_{z4}^2 +
       {l_s} l_{z3} l_{z4}^2 +
       2 {l_t} l_{z3} l_{z4}^2 -
       2 {l_T} l_{z3} l_{z4}^2 +     \\
\nn  &&
       {l_x} l_{z3} l_{z4}^2 +
       \frac{1}{3} {l_s} l_{z4}^3 +
       \frac{5}{3} {l_t} l_{z4}^3 -
       \frac{2}{3} {l_T} l_{z4}^3 -
       \frac{1}{3} {l_x} l_{z4}^3 -
       2 l_s^3 l_{\beta} - l_s^2 {l_t} l_{\beta} -
       {l_s} l_t^2 l_{\beta} -
       l_s^2 {l_T} l_{\beta} -       \\
\nn  &&
       \frac{3}{2} l_s^2 {l_x} l_{\beta} -
       l_t^2 {l_x} l_{\beta} -
       2 {l_s} {l_T} {l_x} l_{\beta} +
       {l_s} l_x^2 l_{\beta} + {l_t} l_x^2 l_{\beta} -
       {l_T} l_x^2 l_{\beta} -
       \frac{1}{2} l_x^3 l_{\beta} +
       2 l_s^2 l_{z3} l_{\beta} +            \\
\nn  &&
       2 {l_s} {l_t} l_{z3} l_{\beta} +
       2 {l_s} {l_x} l_{z3} l_{\beta} +
       2 {l_t} {l_x} l_{z3} l_{\beta} +
       2 l_s^2 l_{z4} l_{\beta} +
       2 {l_s} {l_x} l_{z4} l_{\beta} -
       2 {l_s} l_{z3} l_{z4} l_{\beta} -                 \\
\nn  &&
       2 {l_x} l_{z3} l_{z4} l_{\beta} -
       2 l_s^2 l_{\beta}^2 - {l_s} {l_t} l_{\beta}^2 -
       2 {l_s} {l_x} l_{\beta}^2 - {l_t} {l_x} l_{\beta}^2 +
       {l_s} l_{z3} l_{\beta}^2 +
       {l_x} l_{z3} l_{\beta}^2 +
       {l_s} l_{z4} l_{\beta}^2 +           \\
\nn   &&
       {l_x} l_{z4} l_{\beta}^2 -
       \frac{2}{3} {l_s} l_{\beta}^3 +
       \frac{2}{3} {l_T} l_{\beta}^3 -
       \frac{5}{3} {l_x} l_{\beta}^3 -
       \frac{2}{3} l_{z3} l_{\beta}^3 +
       \left( l_s^2 - \frac{19}{2} l_t^2 -
          \frac{3}{2} l_T^2 + 4 {l_T} {l_u} - 4 {l_T} {l_x} +   \right. \\
\nn  &&
          5 l_x^2 - 6 {l_T} l_{z3} -
          2 {l_x} l_{z3} - 6 l_{z3}^2 -
          2 {l_T} l_{z4} +
          2 {l_x} l_{z4} -
          12 l_{z3} l_{z4} -
          7 l_{z4}^2 +
          {l_t} ( 9 {l_T} + 21 {l_x} +       \\
\nn   &&
          16 l_{z3} + 6 l_{z4} )  + 4 {l_T} l_{\beta} -
          4 {l_x} l_{\beta} - 4 l_{z3} l_{\beta} +
          {l_s} ( -11 {l_t} -
             4 {l_T} - 4 {l_x} + 2 l_{z3} + 10 l_{z4} +       \\
\nn &&    \left.   \frac{}{} {\rm \hspace{-.2cm}}
          2 l_{\beta} )  \right) \zeta(2) +
      ( 7 {l_s} - 8 {l_t} + 4 {l_T} - 3 {l_x} +
          2 l_{z3} + 4 l_{z4} ) \zeta(3) - \frac{41}{2} \zeta(4)
  + 4 {\rm Li}_2 \left( \frac{-u}{z_4} \right) \times     \\
%
%
\nn  &&
          {l_T} {l_x} -
       2 {\rm Li}_2(x) \left( l_s^2 - l_x l_t
            + l_x l_{z3}  -
          {l_s} {l_t} + 2 l_s {l_T} - l_s l_{z3} \right) +
          2 {\rm Li}_2^2(-x) +           \\
\nn  &&
       {\rm Li}_2 \left( \frac{z_3}{z_4} \right)
        \left( -l_s^2 + 2 {l_s} ( {l_t} - {l_x} )  +
           2 l_x {l_t} + l_x^2  \right)  +
       {\rm Li}_2 \left( \frac{m^2}{z_5} \right)
        \left( -2 l_s^2 - 4 {l_t} {l_T} -
          2 l_x {l_T} +          \right. \\
\nn  &&  \left.
        l_x^2 + 2 {l_s} \left( 2 l_t - l_T - l_x \right) \right)  +
{\rm Li}_2 \left( \frac{T}{m^2} \right)
        \left( -2 {\rm Li}_2 \left( -\frac{t}{s} \right) -
          2 {\rm Li}_2(-x) - \frac{11}{2} l_t^2 -
          2 {l_T} {l_u}        \right.       \\
\nn  &&               \left.
        - 2 l_x^2 - l_{z3}^2 -
          2 l_{z3} l_{z4} - l_{z4}^2
          - l_t {l_T} + 6 l_t ( l_{z3} + l_{z4} )
          - 14 \zeta(2) \right)  +
       {\rm Li}_2 \left( \frac{m^2 s}{D} \right)
        \left( 3 l_s^2 -                \right.   \\
\nn   &&   \left.
          l_t^2 -  2 {l_s} ( 3 {l_t} - {l_T} + {l_u} )  +
          2 {l_t} ( 3 {l_T} + {l_u} )  +
          2 ( {l_T} - {l_x} ) {l_x} - 8 \zeta(2) \right)  +       \\
\nn   &&
       {\rm Li}_2 \left( -\frac{t}{s} \right)
        \left( 3 l_s^2 - 2 {l_s}
           ( 3 {l_t} - 2 {l_T} + {l_u} )  -
          2 \left( l_t^2 + 2 l_T^2 -
             l_t ( 4 {l_T} + {l_u} )  + l_x^2 \right)  -
          8 \zeta(2) \right)           \\
\nn   &&
       + {\rm Li}_2 \left( -\frac{D}{m^2 t} \right)
        \left( -2 {\rm Li}_2 \left( \frac{T}{m^2} \right) - 4 {l_s} {l_t} -
          2 l_t^2 + 6 {l_t} {l_T} - l_x^2 - 4 \zeta(2) \right) -    \\
\nn  &&
{\rm Li}_2 \left( \frac{T}{z_3} \right)
        \left( 2 {\rm Li}_2(-x)    + l_s^2 - 4 {l_s} {l_t} +
          2 {l_t} ( {l_t} + 3 {l_T} - 2 {l_x} -
             l_{z3} - 2 l_{z4} )  + 2 \zeta(2) \right)  +       \\
\nn  &&
       {\rm Li}_2 \left( \frac{m^2 x}{-T} \right)
        \left( - l_x^2 +
          2 {l_t} (-l_t -3 {l_T} - 3 {l_x} + l_{z3} +
             2 l_{z4} )  - 2 \zeta(2) \right)  +       \\
\nn  &&
       {\rm Li}_2 \left( \frac{-t(1 - \beta)}{2\,m^2} \right)
        \left( - 2 {\rm Li}_2(-x)     - 2 l_s^2 - 4 {l_t} {l_T}
          - 2 l_x  {l_T} + l_x^2   +
          2 {l_s} ( 2 {l_t} - {l_T} + {l_x} )  + \right.  \\
\nn  &&   \left.
        8 \zeta(2) \right)  +
        {\rm Li}_2(-x)
        \left( - \frac{9}{4} l_s^2
           - 8 l_x {l_T} + \frac{7}{4} l_x^2 - 2 l_x l_{z3} +
        9 l_t l_x - 2 l_t l_{z3} + 5 l_s l_t - 6 l_s l_T +      \right.  \\
\nn   &&           \left.
          \frac{7}{2} l_s l_x +
             2 l_s l_{z3}   + 12 \zeta(2) \right)  +
       {\rm Li}_3 \left( \frac{m^2}{-t} \right)
        ( -6 {l_s} - l_t + l_T + 2 {l_x} - 4 l_{z4} )  +         \\
\nn  &&
       {\rm Li}_3 \left( \frac{T}{m^2} \right)
        ( -4 {l_s} + 8 {l_t} + 3 {l_T} + 2 {l_x} - 6 l_{z3} - 10 l_{z4} )  +
       2 {\rm Li}_3 \left( \frac{z_3}{t} \right)
        ( l_t + l_T - 4 l_x -                       \\
\nn  &&
        l_{z3} - 2 l_{z4} )  +
       2 \left(   {\rm Li}_3 \left( \frac{m^2(1 - \beta)}{2\,z_5} \right) -
            {\rm Li}_3 \left( \frac{-t(1 - \beta)}{2\,z_5} \right) \right)
        ( l_t + 3 l_T + 3 l_x - l_{z3} -    \\
\nn  &&
         2 l_{z4} )  +
       2 {\rm Li}_3 \left( \frac{z_4}{t} \right)
        ( l_t + l_T + 5 l_x - l_{z3} - 2 l_{z4} )  +
       2 {\rm Li}_3 \left( -\frac{u}{t} \right)
        ( 3 l_s - 3 {l_t} - 2 l_T )  +       \\
\nn  &&
       2 \left(  {\rm Li}_3 \left( -\frac{m^2 t}{D} \right)
          - {\rm Li}_3 \left( \frac{t^2}{D} \right)  \right)
        ( 2 l_s + l_t - 3 l_T )  +
       2 \left(  {\rm Li}_3 \left( -\frac{m^2 u}{s z_5} \right)
     - 2 {\rm Li}_3 \left( \frac{z_3}{-u} \right)     \right. \\
\nn  &&    \left.
       + {\rm Li}_3 \left( -\frac{s z_6}{m^2 u} \right)
       - 2 {\rm Li}_3 \left( \frac{-u}{z_4} \right) \right)
         ( l_s - l_t - l_T )  -
       2 {\rm Li}_3 \left( -\frac{D}{t u} \right) l_T -
       2 \left( {\rm Li}_3 \left( \frac{m^2}{z_5} \right)    \right. \\
\nn  &&  \left.
       + {\rm Li}_3 \left( \frac{z_6}{m^2} \right) \right)
        ( {l_s} - {l_t} + {l_T} )
       - 2 {\rm Li}_3 \left( -\frac{t}{s} \right)
        ( l_t - l_T )  - 2
       {\rm Li}_3 \left( \frac{m^2 s}{D} \right)
        ( l_t - 2 l_T )  +      \\
\nn   &&
       2 {\rm Li}_3(x) ( 3 l_s - l_x )  +
       2 {\rm Li}_3 \left( \frac{s x z_5}{m^2\,t} \right)
        ( l_s - l_x )  -
       2 {\rm Li}_3 \left( \frac{z_4(1 - \beta)}{-2\,m^2} \right)
        ( 2 {l_s} - {l_t} - {l_T} +     \\
\nn  &&
         {l_x} )  +
       2 {\rm Li}_3 \left( \frac{z_3}{z_4} \right) {l_x} +
       2 {\rm Li}_3 \left( \frac{z_6}{z_5} \right) {l_s} +
       2 \left( {\rm Li}_3 \left( \frac{t(1 - \beta)}{2\,z_3} \right)
     + {\rm Li}_3 \left( \frac{z_3}{s \beta} \right)
     + {\rm Li}_3 \left( \frac{z_5}{t \beta} \right)     \right)   \\
\nn  &&
        \times ( l_s + l_x )  +
       2 {\rm Li}_3 \left( \frac{s(1 - \beta)}{-2\,z_3} \right)
        ( -2 l_s + l_t + l_T + l_x )  -
       2 {\rm Li}_3 \left( \frac{1 - \beta}{-2\beta } \right)
        ( l_s + 4 l_T -     \\
\nn  &&
       5 l_x - 4 l_{z3} )  +
       2 {\rm Li}_3 \left( \frac{-x^2}{1 - x^2} \right)
        ( l_s + 2 l_T - 2 l_x - 2 l_{z3} )  +
       6 {\rm Li}_3 \left( \frac{1 - \beta}{2} \right)
        ( l_T - l_x -              \\
\nn  &&
        l_{z3} )  -
       2 {\rm Li}_3 \left( \frac{2\,z_6}{m^2(1 + \beta)} \right)
        ( 2 l_s - l_t - 3 l_T + 2 l_x + l_{z3} + 2 l_{z4} )  +
       2 {\rm Li}_3 \left( \frac{-2\,z_6}{t(1 + \beta)} \right)     \\
\nn  &&
        \times
        ( 2 l_s - l_t - 3 l_T + 2 l_x + l_{z3} + 2 l_{z4} )  +
               2 {\rm Li}_3(-x)
        ( 8 l_s + 4 l_t - 3 l_T - 7 l_x -
          2 l_{z3} +             \\
\nn  &&
         4 l_{z4} )  -
       3 {\rm Li}_4 \left( \frac{m^2}{-t} \right) -
       6 {\rm Li}_4 \left( -\frac{t}{s} \right) +
       3 {\rm Li}_4 \left( \frac{T}{m^2} \right) +
       16 {\rm Li}_4 \left( \frac{T}{t} \right) -
       12 {\rm Li}_4 \left( \frac{1 - \beta}{2} \right)   \\
\nn  &&
     - 12 {\rm Li}_4 \left( \frac{1 + \beta}{2} \right) +
       L_{-++}\left(1,0,0,\frac{m^2}{-T}\right) +
       \frac{9}{2} L_{-++}\left(1,0,0,-\frac{t}{T} \right) -        \\
\nn  &&
       L_{-++} \left(1,0,0,\frac{-2}{1 - \beta} \right) -
       L_{-++} \left(1,0,0,\frac{-2}{1 + \beta} \right) -
       5 L_{-++} \left(1,0,\frac{m^2}{-T},-1 \right) -    \\
\nn  &&
       2 L_{-++} \left(1,0,\frac{m^2}{-T},\frac{1}{x} \right) -
       2 L_{-++} \left(1,0,\frac{m^2}{-T},x \right) -
       \frac{3}{2} L_{-++} \left(1,\frac{m^2}{-T},\frac{m^2}{-T},0 \right) -  \\
\nn  &&
       4 L_{-++} \left( 1,\frac{m^2}{-T},\frac{m^2}{-T},\frac{u}{t} \right) +
       2 L_{-++} \left(1,\frac{m^2}{-T},\frac{m^2}{-T},\frac{1}{x} \right) +
       2 L_{-++} \left(1,\frac{m^2}{-T},\frac{m^2}{-T},x \right)     \\
\nn  &&
       -
       3 L_{-++} \left( \frac{t}{T},0,0,\frac{s}{t} \right) +
       2 L_{+} \left( 0,0,\frac{T}{m^2},\frac{1}{x} \right) +
       2 L_{+} \left( 0,0,\frac{T}{m^2},x \right) -         \\
\nn  &&
       2 L_{+} \left( 0,0,\frac{T}{t},\frac{s}{t} \right) +
       2 L_{+} \left( 0,1,-1,\frac{1}{x} \right) +
       2 L_{+}(0,1,-1,x) +
       4 L_{+}(0,1,\frac{t}{s},\frac{-2}{1 - \beta})         \\
\nn  &&
       +
       4 L_{+} \left( 0,1,\frac{t}{s},\frac{-2}{1 + \beta} \right) +
       8 L_{+} \left( 0,1,\frac{1+\beta}{-2},-\frac{t}{T} \right) +
       8 L_{+} \left( 0,1,\frac{1-\beta}{-2},-\frac{t}{T} \right) +  \\
\nn  &&
       2 L_{+} \left( 0,\frac{m^2}{-t},\frac{T}{t},\frac{m^2}{-T} \right) -
       2 L_{+} \left( 0,\frac{m^2\,t}{-D},\frac{t T}{D},\frac{m^2}{-T} \right) +
       2 L_{+} \left( 0,\frac{m^2\,t}{-D},\frac{tT}{D},\frac{1}{x} \right) +  \\
\nn  &&
       2 L_{+} \left( 0,\frac{m^2\,t}{-D},\frac{t\,T}{D},x \right) +
       2 L_{+} \left( 0,-\frac{t}{s},\frac{t}{s},\frac{m^2}{-T} \right) -
       2 L_{+} \left( 0,\frac{t^2}{D},\frac{t\,T}{-D},-1 \right) -       \\
\nn  &&
       4 L_{+} \left( 0,\frac{T}{m^2},\frac{-T}{m^2},-1) +
       4 L_{+} \left( 0,\frac{T}{m^2},\frac{-T}{m^2},\frac{-2}{1-\beta} \right) +
       4 L_{+}(0,\frac{T}{m^2},\frac{-T}{m^2}, \frac{-2}{1 + \beta} \right)     \\
\nn   &&
     -
       6 L_{+} \left( 0,\frac{T}{t},-\frac{T}{t},-1 \right) +
       4 L_{+} \left( 0,\frac{T}{t},-\frac{T}{t},\frac{1}{x} \right) +
       4 L_{+} \left( 0,\frac{T}{t},-\frac{T}{t},x \right) -         \\
\nn  &&
       2 L_{+} \left( 0,-\frac{u}{s},-\frac{t}{s},-1 \right) +
       2 L_{+} \left( 0,-\frac{u}{s},-\frac{t}{s},\frac{1}{x} \right) +
       2 L_{+} \left( 0,-\frac{u}{s},-\frac{t}{s},x \right) -      \\
\nn  &&
       4 L_{+} \left( 0,\frac{T}{z_6},
         \frac{T(1 + \beta)}{-2\,z_6},\frac{-2}{1 - \beta } \right) -
       4 L_{+} \left( 0,\frac{T}{z_6},
          \frac{T(1 + \beta)}{-2\,z_6},\frac{-2}{1 + \beta} \right) +   \\
\nn   &&
       2 L_{+} \left( 0,\frac{1 - \beta}{2},\frac{1 + \beta}{2},
           \frac{1}{x} \right) +
       2 L_{+} \left( 0,\frac{1 - \beta}{2},\frac{1 + \beta}{2},x \right) +  \\
\nn  &&
       2 L_{+} \left( 0,\frac{m^2(1 - \beta)}{2\,z_5},
                \frac{T(1 - \beta)}{-2\,z_5},\frac{1}{x} \right) +
       2 L_{+} \left( 0,\frac{m^2(1 - \beta)}{2\,z_5},
               \frac{T(1 - \beta)}{-2\,z_5},x \right) +    \\
\nn  &&
       4 L_{+} \left( 0,\frac{T(1 - \beta)}{2\,z_6},
               \frac{T(1 + \beta)}{2\,z_6},-1 \right) -
       2 L_{+} \left( 0,\frac{T(1 - \beta)}{2\,z_6},
               \frac{T(1 + \beta)}{2\,z_6},x \right) +     \\
\nn  &&
       2 L_{+} \left( 0,\frac{1 + \beta}{2},\frac{1 - \beta}{2},\frac{1}{x} \right) +
       2 L_{+} \left( 0,\frac{1 + \beta}{2},\frac{1 - \beta}{2},x \right) -
       2 L_{+} \left( \frac{m^2}{-T},0,-x,x \right)       \\
\nn   &&
       -
       4 L_{+} \left( \frac{m^2}{-T},1,-1,\frac{1}{x} \right) -
       2 L_{+} \left( \frac{m^2}{-T},\frac{T}{t},-\frac{T}{t},-1 \right) -
       2 L_{+} \left( \frac{m^2}{-T},\frac{T}{t},-\frac{T}{t},\frac{u}{t} \right) - \\
\nn   &&
       4 L_{+} \left( \frac{m^2}{-T},\frac{1 - \beta}{2},\frac{1 + \beta}{2},
                \frac{u}{t} \right) -
       4 L_{+} \left( \frac{m^2}{-T},\frac{1 - \beta}{2},\frac{1 + \beta}{2},
                      \frac{1}{x} \right) -      \\
\nn  &&
       4 L_{+} \left( \frac{m^2}{-T},\frac{1 + \beta}{2},\frac{1 - \beta}{2},
                                 \frac{u}{t} \right) -
       4 L_{+} \left( \frac{m^2}{-T},\frac{1 + \beta}{2},\frac{1 - \beta}{2},
                                 \frac{1}{x} \right) -      \\
\nn   &&
       4 L_{+} \left( \frac{1}{x},\frac{1 - \beta}{2},
                \frac{1-\beta}{-2},-1 \right) -
       4 L_{+} \left( x,\frac{1 + \beta}{2},
                \frac{1+\beta}{-2},-1 \right) -    \\
\nn   &&
       L_{+++} \left( 0,0,\frac{m^2}{-T},-1 \right) +
       2 L_{+++} \left( 0,0,\frac{m^2}{-T},\frac{1}{x} \right) +
       2 L_{+++} \left( 0,0,\frac{m^2}{-T},x \right) -     \\
\nn   &&
       2 L_{+++} \left( 0,\frac{m^2}{-T},\frac{u}{t},-1 \right) -
       2 L_{+++} \left( 0,\frac{m^2}{-T},\frac{u}{t},\frac{m^2}{-T} \right) +
       2 L_{+++} \left( 0,\frac{m^2}{-T},\frac{u}{t},\frac{1}{x} \right) +   \\
\nn   &&
       2 L_{+++} \left( 0,\frac{m^2}{-T},\frac{u}{t},x \right) +
       2 L_{+++} \left( 0,\frac{m^2}{-T},\frac{1}{x},\frac{1}{x} \right) +
       L_{+++} \left( 0,\frac{1}{x},\frac{1}{x},\frac{m^2}{-T} \right) -   \\
\nn   &&    \left.
       L_{+++} \left( \frac{m^2}{-T},\frac{m^2}{-T},0,\frac{1}{x} \right) -
       L_{+++} \left( \frac{m^2}{-T},\frac{m^2}{-T},0,x \right) -
       L_{+++} \left( \frac{1}{x},\frac{1}{x},0,\frac{m^2}{-T} \right)
                                     \right];
\ea
\ba
\label{D2imag}
{\rm Im}\,D_2^{(-1)}&=&\pi/(s t), {\rm \hspace{.6in}}
{\rm Im}\,D_2^{(0)}= - 2 \pi l_t /(s t),         \\
\nn
{\rm Im}\,D_2^{(1)}&=& \frac{\pi}{s t} \left[
          -\frac{3}{4} l_s^2 + 2 l_t^2 - \frac{3}{4} l_x^2 -
          l_x l_{z3} - l_t l_x - 2 l_t l_{z3} +
          l_x l_{z4} - l_{z4}^2 +        \right.     \\
\nn  &&   \left.
          l_s ( l_t + \frac{l_x}{2} + l_{z3} +  l_{z4} ) -
          2 \zeta(2) -
          2 {\rm Li}_2 \left( \frac{m^2}{z_5} \right) -
          2 {\rm Li}_2 \left( \frac{t(1 - \beta)}{-2\,m^2} \right)  \right],  \\
\nn
{\rm Im}\,D_2^{(2)}&=&\frac{\pi}{s t} \left[
           \frac{7}{12} l_s^3 - \frac{l_x^3}{12}
           - 2 l_x l_t l_{z3} - l_x^2 l_{z3} + l_{z4}^2
           - l_s \left( 2 l_t^2 - \frac{5}{4} l_x^2 -
             2 l_t ( l_x + l_{z3} )  -        \right.   \right. \\
\nn  &&   \left.
             2 l_x ( l_{z3} - l_{z4} )  - l_{z4}^2 \right)
           - l_s^2 \left( \frac{3}{4} l_x +
             l_{z3} + l_{z4} \right)  +
             4 ( l_t + l_x ) \zeta(2) +
          2 {\rm Li}_2(x)\, l_x +       \\
\nn   &&
       2 {\rm Li}_2 \left( \frac{t(1 - \beta)}{-2\,m^2} \right) ( l_s - l_x ) +
       2 {\rm Li}_2 \left( \frac{z_3}{z_4} \right) l_x +
       2 {\rm Li}_2 \left( \frac{m^2}{z_5} \right) ( l_s + l_x )  -
       2 {\rm Li}_3(x) -     \\
\nn  &&    \left.
       4 {\rm Li}_3 \left( \frac{z_3}{t} \right) +
       2 {\rm Li}_3 \left( \frac{z_3}{z_4} \right) -
       4 {\rm Li}_3 \left( \frac{z_4}{t} \right) -
       2 {\rm Li}_3 \left( \frac{z_6}{z_5} \right) \right].
\ea
\\
\[
\mbox{ \epsfysize = 3.0cm  \epsffile {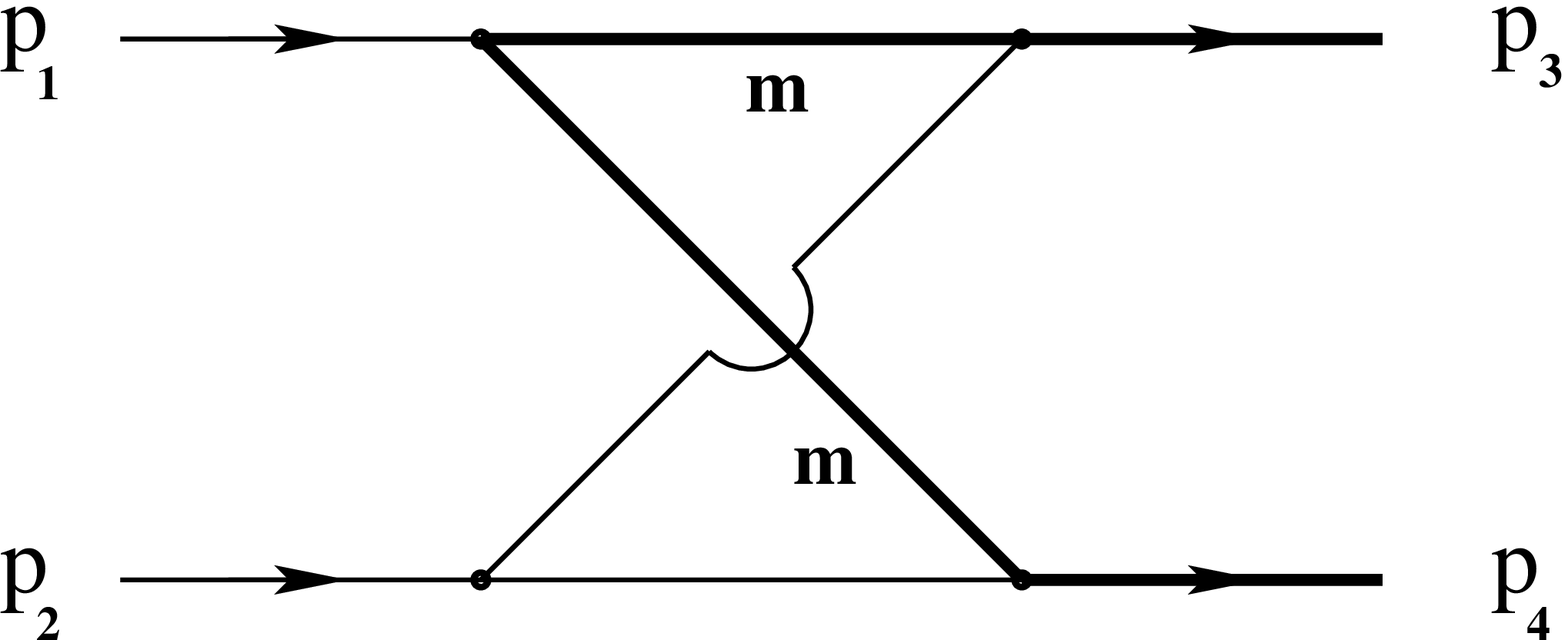}}
\]
{\begin{center}
\tenrm\vglue -0.1cm
FIG.~4. Massive box $D_3$ with two massive propagators.
\end{center}
}
\vglue .1cm

The diagram corresponding to the third four-point function $D_3$ with two massive
propagators is shown in Fig.~4 where we write
\begin{equation}
D_3 \equiv D(-p_2,p_4,-p_1,0,0,m,m)\, .
\end{equation}
The kernel (\ref{kernel}) for $D_3$
can be written as
\be
K = ac\tilde{t} + bd\tilde{u} + (c+d)^2,
\ee
where we have introduced the positive-valued dimensionless variable
\be
\tilde{u} \equiv -\frac{u}{m^2},
{\rm \hspace{.4in}}
\tilde{u}\ge 1, {\rm \hspace{.4in}} \tilde{s}\ge\tilde{u}.
\ee
For the Feynman parameterization we choose $\{a,b,c,d\}$ as $\{x_1(1-x_2), \,
1-x_1, \ x_1x_2(1-x_3), \, x_1x_2x_3\}$, which gives the following integrand:
\[
x_1^2 x_2 \left[ x_1^2 x_2^2 + \tilde{t} x_1^2 x_2 (1-x_2) (1-x_3) + \tilde{u}
x_1 (1-x_1) x_2 x_3 \right]^{-2-\ep}.
\]
The above expression never becomes negative. Therefore, the entire result
for the box $D_3$ does not have an imaginary part. One can set
$\delta=0$ in the kernel from the very beginning.

That the box $D_3$ posesses no imaginary part can be seen in a less technical way
by appealing to the Landau-Cutkosky cutting rules. The diagram corresponding to
the box $D_3$ shown in Fig.~4 does
not admit of any cuts such that the cut lines of the diagram are on their mass
shell simultaneously.

As before we obtain two terms after the first integration over $x_3$.
They are
\ba
\label{ID3x1x2}
I^{D_3}_{x_1x_2}=\frac{x_1^{-1-2\ep} x_2^{-1-\ep} \left[ x_2 + \tilde t(1-x_2)
)\right]^{-1-\ep}}
{(1+\ep)[\tilde u (1-x_1)-\tilde t x_1 (1-x_2)]},         \\
\label{IID3x1x2}
II^{D_3}_{x_1x_2}=-\frac{x_1^{-\ep}x_2^{-1-\ep}\left[ x_1x_2 + \tilde u (1-x_1)
\right]^{-1-\ep}} {(1+\ep)[\tilde u (1-x_1) - \tilde t x_1 (1-x_2)]}.
\ea
Note that the denominators of $I^{D_3}_{x_1x_2}$ and $II^{D_3}_{x_1x_2}$
change sign on the integration path, while the numerators stay positive
(i.e. the relevant integrals have branch cuts). This can easily be seen by
considering the numerators and denominators of the above integrands
at two particular values of the
variable $x_1$, for instance at $x_1=0$ and $x_1=1$. This means that although
the whole $D_3$ box integral does not have an imaginary part, the two terms in
(\ref{ID3x1x2}) and (\ref{IID3x1x2}) separately give rise to unphysical, spurious
imaginary contributions. Of course, these are artefacts of having split the
result into two terms.
On the one hand, this somehow complicates things. On the other hand,
the cancellation of imaginary contributions in the sum of the two terms
(\ref{ID3x1x2}) and (\ref{IID3x1x2}) will serve as a good check for our final
result. To control the
imaginary contributions we do the following replacement in the denominators in
(\ref{ID3x1x2}) and (\ref{IID3x1x2}):
\[
\tilde u (1-x_1) - \tilde t x_1 (1-x_2) \rightarrow
\tilde u (1-x_1) - \tau x_1 (1-x_2),
{\rm \hspace{.4in}} \tau \equiv \tilde t - i\delta.
\]
We start with the $x_2$-integration of the term $I^{D_3}_{x_1x_2}$
Eq.~(\ref{ID3x1x2}):
\be
\label{id3x2}
I^{D_3}_{x_2}=-\frac{\left[ x_2 \, \left( x_2 + \tilde t \, (1-x_2)
\right)  \right]^{-1 -\ep}\,
_{2} F_{1}(1, -2\ep, 1 - 2\ep, \frac{\tilde u + \tau - x_2 \tau}{\tilde u})}
              {2\tilde u\,(1 + \ep) \,\ep }.
\ee
The above expression is singular at the lower integration limit $x_2 = 0$ due
to the term $x_2^{-1 -\ep}$.
To
find a subtraction term, we follow exactly the procedures defined after
Eq.~(\ref{id2x2}). Our subtraction term reads
\be
\label{Ix2D3c}
I_{x_2}^{D_3,s}=\frac{x_2^{-1 - \ep} \,  \tilde t^{-1 - \ep}
 \,[ -1 - 2\ep\, \ln \left(-\frac{\tau}{\tilde u}\right)
+ 4\ep^2 {\rm Li}_{2}\left(\frac{\tilde u + \tau}{\tilde u}\right)  + 8\ep^3\,
{\rm Li}_{3}\left(\frac{\tilde u + \tau}{\tilde u}\right)
+ 16\ep^4 {\rm Li}_{4}\left(\frac{\tilde u + \tau}{\tilde u}\right) ]}
{2\tilde u\,( 1 + \ep ) \,\ep }.
\ee
The above subtraction term can easily be integrated over $x_2$ to obtain
\be
I^{D_3,s}=\frac{\tilde t^{-1-\ep}\,\left[ 1 +
2\ep\,\ln \left(-\frac{\tau}{\tilde u}\right)
- 4\ep^2 {\rm Li}_{2}\left(\frac{\tilde u + \tau}{\tilde u}\right) - 8\ep^3\,
{\rm Li}_{3}\left(\frac{\tilde u + \tau}{\tilde u}\right)
- 16\ep^4 {\rm Li}_{4}\left(\frac{\tilde u + \tau}{\tilde u}\right) \right]}
  {2\, (1 + \ep) \,\ep^2\,\tilde u}
\ee
which can readily be expanded in $\ep$.
As the difference $I^{D_3}_{x_2}-I_{x_2}^{D_3,s}$ does not contain any
poles, we can expand the difference in a series in $\ep$ and perform the
analytical integration over the last variable $x_2$.

To integrate the term (\ref{IID3x1x2}), we split $II_{x_1x_2}^{D_3}$ into two
contributions, i.e.
\ba
II^{D_3}_{x_1x_2}=-\frac{x_2^{-1-\ep}\left[ x_1x_2 + \tilde u (1-x_1)
\right]^{-1-\ep}} {(1+\ep)[\tilde u (1-x_1) - \tau x_1 (1-x_2)]}  \nn \\
\label{split}
-\frac{(x_1^{-\ep}-1) x_2^{-1-\ep}\left[ x_1x_2 + \tilde u (1-x_1)
\right]^{-1-\ep}} {(1+\ep)[\tilde u (1-x_1) - \tau x_1 (1-x_2)]}.
\ea
Then we integrate the first term in (\ref{split}) over $x_1$ to obtain two
hypergeometric functions which are expanded up to $\ep^4$. As was done
previously, we then introduce a subtraction term similar to (\ref{Ix2D3c}) which
is integrated analytically. The finite difference of the original integral and
the subtraction term is then ready to be integrated over the last
integration variable.

For the second term in (\ref{split}) we introduce a subtraction term before the
last two integrations:
\be
\label{lastcount}
II^{D_3,s}_{x_1x_2}=-\frac{(x_1^{-\ep}-1) x_2^{-1-\ep}\left[ \tilde u
(1-x_1)
\right]^{-1-\ep}} {(1+\ep)[\tilde u (1-x_1) - \tau x_1]},
\ee
which is obtained from the second term in (\ref{split}) by the substitution
$x_2=0$ in all terms except for $x_2^{-1-\ep}$.
We first trivially integrate out $x_2$ in (\ref{lastcount}) and expand the
resulting
expression in a series of $\ep$. Because of the factor $(x_1^{-\ep}-1)$
this expansion starts at the order $\ep^0$. Thus, the subtraction term is
finite and ready for the last integration.

As the difference of the second term in (\ref{split}) and its subtraction term
(\ref{lastcount}) does not contain any poles, we expand it up to $\ep^2$ and
integrate over $x_1$. Finally, collecting all the relevant pieces, we perform the
last integration.
Our result for the four-point box diagram $D_3$ reads:
\be
\label{d3}
{\rm Re}\,D_3^{(-2)} = 1/(t u),          {\rm \hspace{.6in}}
   {\rm Re}\,D_3^{(-1)}=-[l_t + l_u]/(t u),   {\rm \hspace{2.0in}}
\ee
\vspace{-.4in}
\ba
\nn
{\rm Re}\,D_3^{(0)}&=& 2 \left[ l_t l_u - 2\,\zeta(2) \right]/(t u),      \\
\nn
{\rm Re}\,D_3^{(1)}&=& \frac{1}{t\,u} \left[
   \frac{l_D^3}{3} - l_D^2 l_t - \frac{l_t^3}{3} +
       2 l_t^2 l_u - 2 {l_t} {l_T} {l_u} - 4 {l_t} l_u^2
+ l_u^3 + {l_s} \left( l_u^2 - l_t^2  \right)  +
       l_D \left( l_t^2 + l_u^2 \right) -     \right.   \\
\nn   &&
2\,l_u^2 l_U +    l_s^2 ( l_t - l_u )  +
       2 ( l_D - 3 {l_s} + {l_t} + 3 {l_u} ) \zeta(2) -
       4 \zeta(3) +
       2 {\rm Li}_2 \left( -\frac{t}{s} \right) ( l_t - l_u ) +       \\
\nn   &&
       2 {\rm Li}_2\left(\frac{m^4}{D}\right) {l_u} -
       2 {\rm Li}_2\left(\frac{T}{m^2}\right) {l_u} -
       2 {\rm Li}_2\left(\frac{U}{m^2}\right) {l_u} +
       2 {\rm Li}_2\left(-\frac{D}{m^2 t}\right) ( {l_u} - {l_t} ) +   \\
\nn   &&
       2 {\rm Li}_3\left(\frac{m^4}{D}\right) -
       2 {\rm Li}_3\left( \frac{m^2}{-t} \right) -
       2 {\rm Li}_3\left( -\frac{m^2 t}{D} \right) -
       2 {\rm Li}_3\left( \frac{m^2}{-u} \right) -
       2 {\rm Li}_3\left( -\frac{m^2 u}{D} \right) +          \\
\nn   &&  \left.
       2 {\rm Li}_3\left( -\frac{u}{t} \right) \right],        \\
\nn
{\rm Re}\,D_3^{(2)}&=& \frac{1}{t\,u} \left[
       \frac{l_D^4}{3} - l_D^2 l_s^2 - \frac{5}{6}\, l_s^4 -
       \frac{4}{3}\,l_D^3 l_t + 4 l_D l_s^2 l_t +
       \frac{1}{6}\,l_s^3 {l_t} + 4 l_D^2 l_t^2 -
       5 l_D l_s l_t^2 - \frac{3}{4}\,l_s^2 l_t^2 -      \right.    \\
\nn   &&
       \frac{11}{3}\,l_D l_t^3 + \frac{3}{2}\,{l_s} l_t^3 +
       \frac{17}{6}\,l_t^4 + \frac{1}{3}\,l_D^3 {l_T} -
       2 l_D l_s^2 {l_T} + l_s^3 {l_T} +
       l_D^2 {l_t} {l_T} + 4 l_D {l_s} {l_t} {l_T} -                 \\
\nn    &&
       2 l_s^2 {l_t} {l_T} - 4 l_D l_t^2 {l_T} +
       3 {l_s} l_t^2 {l_T} + l_s^2 l_T^2 -
       2 {l_s} {l_t} l_T^2 - \frac{4}{3}\,l_D^3 {l_u} +
       2 l_D^2 {l_s} {l_u} - l_D l_s^2 {l_u} +                   \\
\nn    &&
       \frac{5}{2}\,l_s^3 {l_u} - l_D^2 {l_t} {l_u} -
       2 l_D {l_s} {l_t} {l_u} - 2 l_s^2 {l_t} {l_u} +
       3 l_D l_t^2 {l_u} + {l_s} l_t^2 {l_u} -
       \frac{13}{6}\,l_t^3 {l_u} - l_D^2 {l_T} {l_u} +         \\
\nn    &&
       2 l_D {l_s} {l_T} {l_u} - 2 {l_s} {l_t} {l_T} {l_u} +
       l_t^2 {l_T} {l_u} - {l_s} l_T^2 {l_u} -
       \frac{1}{2}\,l_D^2 l_u^2 + l_D {l_s} l_u^2 -
       \frac{9}{4}\,l_s^2 l_u^2 + 2 l_D {l_t} l_u^2 +       \\
\nn    &&
       2 {l_s} {l_t} l_u^2 - \frac{7}{4}\,l_t^2 l_u^2 -
       l_D {l_T} l_u^2 - {l_s} {l_T} l_u^2 +
       5 {l_t} {l_T} l_u^2 + \frac{1}{2}\,l_T^2 l_u^2 -
       \frac{4}{3}\,l_D l_u^3 + \frac{1}{6}\,{l_s} l_u^3 +   \\
\nn    &&
       \frac{3}{2}\,{l_t} l_u^3 - \frac{19}{12}\,l_u^4 +
       2 l_D^2 {l_u} {l_U} - 2 l_D {l_s} {l_u} {l_U} +
       2 l_s^2 {l_u} {l_U} - 2 l_D {l_t} {l_u} {l_U} +
       2 l_t^2 {l_u} {l_U} +                           \\
\nn    &&
       2 l_D {l_T} {l_u} {l_U} - 4 {l_t} {l_T} {l_u} {l_U} -
       2 {l_s} l_u^2 {l_U} +
       2 {l_t} l_u^2 {l_U} + {l_T} l_u^2 {l_U} +
       \frac{13}{3}\,l_u^3 {l_U} + l_D {l_u} l_U^2 +       \\
\nn    &&
       {l_s} {l_u} l_U^2 - 2 {l_t} {l_u} l_U^2 -
       {l_T} {l_u} l_U^2 - 2 l_u^2 l_U^2 -
       \frac{2}{3}\,{l_u} l_U^3 +
       \left( 7 l_D^2 - 2 l_D
              ( 3 {l_s} + 4 {l_t} - {l_T} + 2 {l_u} ) -  \right.      \\
\nn    &&     \left.
          \frac{17}{2}\,l_s^2 - \frac{3}{2}\,l_t^2 +
             3 {l_s} ( 7 {l_t} + 6 {l_u} )  -
             2 {l_t} ( 2 {l_T} + 7 {l_u} )  +
             l_T^2 - 2 {l_T} {l_u} - 7 l_u^2 -
                2 {l_u} {l_U}   \right) \zeta(2) +        \\
\nn    &&
       \left( -3 {l_s} + 21\, {l_t} - 2 {l_T} - 8 {l_u} + 4 {l_U} \right) \zeta(3)
       - \frac{57}{4}\,\zeta(4) +        {\rm Li}_2 \left( \frac{U}{m^2} \right)
        \left( l_D^2 + 2 l_s^2 - 2 l_D {l_t} +     \right.        \\
\nn    &&    \left.
       3 l_t^2 - 2 {l_t} {l_T} + l_T^2 + 2 {l_t} {l_u} +
       2 {l_T} {l_u} + 3 l_u^2 - 2 {l_s} (l_t + l_u)  -
       2 {l_u} {l_U} - 2 \zeta(2) \right) -             \\
\nn    &&
       {\rm Li}_2 \left( \frac{m^4}{D} \right) l_u^2 +
       {\rm Li}_2 \left( -\frac{t}{s} \right)
        \left( 2 {\rm Li}_2 \left( \frac{T}{m^2} \right) +
               2 {\rm Li}_2 \left( \frac{U}{m^2} \right) - 5 l_s^2 -
          \frac{3}{2}\,l_t^2 + 2 {l_t} l_T -    \right.           \\
\nn    &&    \left.
       6 {l_t} l_u  + \frac{5}{2}\,l_u^2 +
       5 {l_s} (l_t + l_u)  + 4 \zeta(2) \right)  +
       {\rm Li}_2 \left( -\frac{D}{m^2 t} \right)
        \left( 2 {\rm Li}_2 \left( \frac{T}{m^2} \right) +
               2 {\rm Li}_2 \left( \frac{U}{m^2} \right) -     \right.    \\
\nn    &&     \left.
          2 l_s^2 - 2 l_D {l_t} +
          3 l_t^2 + 2 {l_t} {l_T} - 2 {l_t} {l_u} - 4 l_u^2 +
          2 {l_s} (l_t + l_u)  + 4 {l_u} {l_U} +
          4 \zeta(2) \right)  +          \\
\nn    &&
          {\rm Li}_2 \left( \frac{m^2 s}{D} \right)
        \left( 2 {\rm Li}_2 \left( \frac{T}{m^2} \right) + l_t^2 -
          l_u^2 + 2 l_u {l_U} + 6\zeta(2) \right)  +
       {\rm Li}_2 \left( \frac{T}{m^2} \right)
        \left( {\rm Li}_2 \left( \frac{T}{m^2} \right) +   \right.       \\
\nn    &&        \left.
          4 {\rm Li}_2 \left( \frac{U}{m^2} \right) + 2 l_D^2 +
          l_s^2 + 3 l_t^2 -
          2 l_D (l_s + 2 l_t)  + 2 {l_s} {l_T} +
          l_u^2 + 2 {l_u} {l_U} + 6 \zeta(2) \right)  +      \\
\nn    &&
       2 {\rm Li}_3\left( \frac{m^4}{D} \right)
        (l_D - l_t + l_u)  +
       2\left( {\rm Li}_3 \left( -\frac{m^2 t}{T u} \right) -
               {\rm Li}_3 \left( \frac{m^2 s}{D}\right) \right)
                           (l_s - l_t - l_u)  -                    \\
\nn    &&
       2 {\rm Li}_3\left( -\frac{m^2 t}{D} \right)
        (l_D - l_s + l_T + l_u)  -
       2 {\rm Li}_3 \left( \frac{U}{m^2} \right) (l_T - l_u)  -
       {\rm Li}_3 \left( -\frac{t}{s} \right)
        (6 {l_s} + l_t -                                \\
\nn    &&
       7 l_u) +     2 {\rm Li}_3 \left( \frac{m^2}{-u} \right)
        (l_D + l_s + 2 l_t - 7 l_T + 2 l_u - 2 l_U)  -
       2 {\rm Li}_3 \left( -\frac{m^2 u}{D} \right)
        (l_D - l_s +                                \\
\nn    &&
       l_T) -         2\left( {\rm Li}_3 \left( \frac{t^2}{D} \right) +
                              {\rm Li}_3 \left (\frac{D}{u^2} \right) \right)
        (l_s - l_t - l_T)  -  {\rm Li}_3 \left( -\frac{u}{t} \right)
        (3 l_s + l_t + 2 l_T + 2 l_u)               \\
\nn    &&
     - 2 {\rm Li}_3 \left( \frac{D}{s U} \right) l_u +
       4 {\rm Li}_3 \left( -\frac{D}{t U} \right) l_u +
       2 {\rm Li}_3 \left( \frac{m^4}{T U} \right) l_u -
       2 {\rm Li}_3 \left( -\frac{t^2}{s T} \right) (l_s - l_t)  +   \\
\nn    &&
       2 {\rm Li}_3 \left( \frac{T}{m^2} \right)
        (2 l_D + l_s - 3 l_t - 2 l_T)  +
       2\left( {\rm Li}_3 \left( \frac{m^2}{-t} \right) +
               {\rm Li}_3 \left( \frac{-D}{t u} \right) \right)
        (l_s - l_t - 2 l_T)                    \\
\nn    &&
     + 2 {\rm Li}_4 \left( \frac{m^2}{-t} \right) +
       {\rm Li}_4 \left( -\frac{t}{s} \right) -
       2 {\rm Li}_4 \left( \frac{T}{m^2} \right) +
       10\, {\rm Li}_4 \left( \frac{T}{t} \right) -
       8 {\rm Li}_4 \left( \frac{m^2}{-u} \right) +      \\
\nn    &&
       7 {\rm Li}_4 \left( -\frac{u}{s} \right) -
       4 {\rm Li}_4 \left( \frac{U}{m^2} \right) -
       4 {\rm Li}_4 \left( \frac{U}{u} \right) +
L_{-++} \left( 1,\frac{m^2}{-T},\frac{m^2}{-T},\frac{-U}{m^2} \right) -     \\
\nn    &&
       2 L_{-++} \left( 1,\frac{m^2}{-T},\frac{u}{t},0 \right) +
       2 L_{-++} \left( 1,\frac{m^2}{-T},\frac{u}{t},\frac{m^2}{-T} \right) -
       \frac{1}{2}\, L_{-++} \left( 1,\frac{t}{u},\frac{t}{u},0 \right) -  \\
\nn    &&
       2 L_{-++} \left( 1,\frac{m^2}{-T},\frac{u}{t},\frac{-U}{m^2} \right)  +
       \frac{1}{2}\,L_{-++} \left( 1,\frac{u}{t},\frac{u}{t},0 \right) -
       \frac{5}{2}\,L_{-++}\left( -\frac{s}{t},0,0,-1 \right) +     \\
\nn    &&
       3 L_{-++} \left( \frac{t}{T},0,0,-1 \right) +
       5 L_{-++} \left( \frac{t}{T},0,0,\frac{u}{m^2} \right) +
       \frac{1}{2}\,L_{-++} \left( -\frac{s}{u},0,0,-1 \right) -     \\
\nn    &&
       2 L_{+} \left( 0,0,-\frac{t}{s},-1 \right) +
       2 L_{+} \left( 0,0,\frac{T}{t},-1 \right) +
       4 L_{+} \left( 0,0,\frac{T}{t},\frac{u}{m^2} \right) +         \\
\nn    &&
       2 L_{+} \left( 0,0,\frac{m^2}{-u},\frac{u}{m^2} \right) -
       2 L_{+} \left( 0,1,\frac{t}{s},-\frac{t}{T} \right) +
       2 L_{+} \left( 0,1,\frac{t}{s},\frac{u}{m^2} \right) +        \\
\nn    &&
       2 L_{+} \left( 0,\frac{t^2}{D},\frac{t T}{-D},-1 \right) -
       2 L_{+} \left( 0,\frac{t^2}{D},\frac{t T}{-D},-\frac{t}{T} \right) +
       2 L_{+} \left( 0,\frac{t^2}{D},\frac{t T}{-D},\frac{u}{m^2} \right) + \\
\nn    &&
       2 L_{+} \left( 0,\frac{T}{m^2},\frac{-T}{m^2},\frac{s}{t} \right) -
       2 L_{+} \left( 0,\frac{T}{m^2},\frac{-T}{m^2},-\frac{t}{T} \right) +
       2 L_{+} \left( \frac{m^2}{-T},0,\frac{T}{m^2},\frac{u}{t} \right)  -   \\
\nn    &&    \left.
         2 L_{+} \left( \frac{m^2}{-T},\frac{U}{u},\frac{m^2}{-u},
                      \frac{-U}{m^2} \right)
                                                  \right];          \\
\label{D3imag}
{\rm Im}\,D_3^{(j)}&=& 0.
\ea
The non-planar topological structure of the four-point function $D_3$
implies that $D_3$ has to be ($t\! \leftrightarrow \! u$)--symmetric
(see Fig.~4).
This can best be seen by exchanging the momenta $p_3 \leftrightarrow p_4$
in Fig.~4 followed by a twist of the r.h.s. of Fig.~4 \footnote{The
($t \! \leftrightarrow \! u$)--symmetry is not so easy to see when exchanging
$p_1 \leftrightarrow p_2$ in Fig.~4. In this case the
($t \! \leftrightarrow \! u$)--symmetry becomes apparent only after Feynman
parametrization.}. The
($t \! \leftrightarrow \! u$)--symmetry
provides for a check on our results for $D_3$. The
symmetry is obviously satisfied for ${\rm Re} D_3^{(-2)}$, ${\rm Re} D_3^{(-1)}$
and ${\rm Re} D_3^{(0)}$ but is not manifest for
${\rm Re}D_3^{(1)}$ and ${\rm Re}D_3^{(2)}$ in
(\ref{d3}). However, it is quite
straightforward to verify numerically that the ($t \! \leftrightarrow \! u$)--symmetry
indeed holds for all coefficient functions in (\ref{d3}).

Apart from the internal checks mentioned earlier the most important
check on our four-point function results has been a comparison with
numerical results provided to us by M.M.~Weber $[$\ref{Private}$]$ for several
phase space points. Within numerical errors we have found
complete agreement with the results of M.M.~Weber for each of the three
four-point functions. It is important to emphasize that the approach of
M.M.~Weber to numerically evaluate the
four-point functions is completely different from ours $[$\ref{Weber}$]$.

\vglue 1cm
\renewcommand{\theequation}{4.\arabic{equation}}
\begin{center}\begin{large}\begin{bf}
V. SUMMARY AND CONCLUSIONS
\end{bf}\end{large}\end{center}
\vglue .3cm

We have presented analytical results up to ${\cal O}(\ep^2)$
for all the massive scalar
one-loop
integrals that arise in the calculation of one-loop matrix elements in heavy
flavor hadroproduction.
Many of our results are new (see Table~\ref{t:tab1}).
The one-loop scalar integrals are needed for that part of the NNLO
hadroproduction of heavy flavours which is obtained from
the product of one-loop contributions called loop-by-loop contribution.

What remains to be done in order to obtain
the full one-loop amplitude structure is to take into
account positive powers of $\ep$ (up to ${\cal O}(\ep^2)$) resulting from the
Passarino--Veltman decomposition and the spin algebra.
The full one-loop
amplitudes to order $\ep^0$ were given in $[$\ref{KM}$]$.
The missing results for the $\ep$- and $\ep^2$-coefficients of the one-loop
amplitudes will be presented in a forthcoming publication $[$\ref{second}$]$.
In a last step the amplitudes themselves have to be squared, which, in the
case of gluon-initiated production, will generate further positive powers
of $\ep$ in dimensional regularization.
The calculation of the loop-by-loop contributions in Fig.~1 is a
necessary starting point in the evaluation of the NNLO contributions
to heavy quark pair production in hadronic interactions. It is very
likely that the calculation of the other three classes of diagrams
in Fig.~1 will proof to be very difficult. This holds true in
particular for the massive two-loop box contributions.

In the Laurent series expansion
of the scalar one-loop integrals the successive coefficient functions
increase in length and complexity with each order of $\ep$. The reason is that the
$\ep$--expansion of the integrand before the last parametric
integration itself generates coefficient functions with increasing complexity
with each order of $\ep$. The most complex expressions arise from the
box contributions where one encounters multiple polylogarithms
up to weight and depth four at ${\cal O}(\ep^2)$.


In a
numerical NNLO evaluation of heavy hadron production the various contributing
pieces will have to be evaluated at many values of the kinematical variables. This
requires efficiency in the numerical codes for each of the contributing pieces. We
believe that we have provided for such numerical efficiency in the loop-by-loop
portion of the NNLO calculation by presenting results in analytical form which are
fast to evaluate
numerically.
All our results are available in electronic form $[$\ref{EPAPS}$]$.
We are planning to present our results in terms of multiple polylogarithms
in the near future.
In recent years number of new methods were developed for
semi-numerical evaluation of general Feynman diagrams (see e.g.
$[$\ref{Weber},\ref{Passarino},\ref{Binoth}$]$).
First numerical tests have shown that our representation in terms of
the L-functions perform better than the present implementation of
the flexible all-purpose approach described in
$[$\ref{Weber},\ref{Passarino}$]$.

The analytical results presented in this paper cover the
whole kinematical domain with a single expression.
They evaluate numerically very fast and efficiently.
Further advantages of having the results in analytical form are that they allow
one to investigate various limiting cases as well as their
analyticity properties.
Also, when analytical results are available
the mathematical structure of the results becomes manifest which would not
be visible in a purely numerical approach.

The full calculation of the NNLO corrections to heavy hadron production at
hadron colliders will be a very difficult task to complete. It involves the
calculation of very many Feynman diagrams of many different topologies. The
problem is further complicated by the fact that heavy hadron
production is a multi-scale problem with three mass scales provided by the
kinematic variables $s$ and $t$ in the loop expressions, and the mass of the
heavy quark.
It is clear that an undertaking of this dimension will have to
involve many theorists and cannot be done by a single group alone.
In this sense the present calculation is a first step
(or second step $[$\ref{Bernreuther}$]$)
in the direction of obtaining NNLO results on heavy hadron production at hadron
colliders. The present calculation allows one to obtain a first glimpse of the
mathematical and computational complexity that is waiting for us in the full
NNLO calculation. This complexity does in fact already reveal itself in terms
of a very rich polylogarithmic and multiple polylogarithmic structure of the
Laurent series expansion of the scalar one-loop integrals as shown in this paper.

\vglue 1cm
\begin{center}\begin{large}\begin{bf}
ACKNOWLEDGMENTS
\end{bf}\end{large}\end{center}
\vglue .3cm

We are grateful to M.M.~Weber for providing us with his numerical results on
massive three-- and four--point functions which served as another check on the
correctness of our analytical results.
We thank A.~Davydychev, G.~Heinrich and M.~Kalmykov for communications, and
G.~Heinrich and J.~Gegelia for discussions.
Z.~M. would like to thank the Particle Theory Group of the Institut
f{\"u}r Physik, Universit{\"a}t Mainz for hospitality. The work of Z.~M.
was supported by a DFG (Germany) grant under contract 436 GEO 17/3/03.
M.~R. was supported by the DFG through the Graduiertenkolleg
``Eichtheorien'' at the University of Mainz.

\vglue 1cm
\begin{center}\begin{large}\begin{bf}
APPENDIX~A
\end{bf}\end{large}\end{center}
\vglue .3cm

\setcounter{equation}{0}
\renewcommand{\theequation}{A\arabic{equation}}

In this Appendix we evaluate a special two-point integral which is needed for the
calculation of the
one-loop fermion self-energy diagram insertion into the massive external fermion
line.
This integral is also needed for the definitions of the fermion mass and wave
function renormalization constants in the on-shell
renormalization scheme. In particular, we need to evaluate the integral
\be
I_1 \equiv B(p,0,m) = \mu^{2\ep}\int \frac{d^nq}{(2\pi)^n}\frac{1}
{q^2 [(q+p)^2-m^2]}
\ee
up to ${\cal O}(p^2-m^2)$. We therefore Taylor expand $I_1$ around $p^2=m^2$:
\ba
\label{Taylor}
\nn
I_1&=&I_1\biggl |_{p^2=m^2} + \,\frac{dI_1}{dp^2}\biggl |_{p^2=m^2} (p^2-m^2)
+ \ldots   \\
&\equiv& E_0 + E_1 (p^2-m^2) + \ldots
\ea
Note that the expansion coefficients $E_i$ in (\ref{Taylor}) are
functions of $\ep$.
The first coefficient $E_0$ is nothing but $B_3$ obtained in Section~II.
The second coefficient $E_1$ is proportional to the sum of a scalar and a vector
integral obtained by differentiating $I_1$ w.r.t. $p^\mu$. One obtains
\be
E_1=\frac{1}{2p_{\mu}} \frac{dI_1}{dp^{\mu}} = - I_2 - \frac{p^\mu I_{2\mu}}{m^2},
\ee
where
\be
I_{\{2,2\mu\}}= \mu^{2\ep}\int \frac{d^nq}{(2\pi)^n} \frac{\{1,q_{\mu}\}}
{q^2 [(q+p)^2-m^2]^2} =
                      i C_{\ep}(m^2) \frac{1}{m^2} \left\{ \frac{1}{2\ep},
\frac{p_{\mu}}{1-2\ep} \right\}.
\ee
One finally has
\be
\label{I1}
I_1=i C_{\ep}(m^2) \frac{1}{\ep(1-2\ep)} \left(1 - \frac{p^2-m^2}{2m^2}\right) +
{\cal O}[(p^2-m^2)^2].
\ee
The result for the integral $I_1$ in the form (\ref{I1}) was used
in $[$\ref{KMC}$]$ to evaluate external heavy quark self-energy diagrams
and obtain heavy quark wave function renormalization constants in the NLO
calculation.

\vglue 1cm
\begin{center}\begin{large}\begin{bf}
APPENDIX~B
\end{bf}\end{large}\end{center}
\vglue .3cm

\setcounter{equation}{0}
\renewcommand{\theequation}{B\arabic{equation}}

In this Appendix we shall demonstrate how the $L$-functions introduced
in Eqs.~(\ref{Lfunction}) and (\ref{Lpfunction}) are related to multiple
polylogarithms as defined in $[$\ref{Multilogs}$]$.
Multiple polylogarithms are defined as a limit of Z-sums, e.g.
\ba
\label{zsum}
Li_{m_{k},...,m_{1}}(x_{k},...,x_{1})=\lim_{n_1 \rightarrow \infty } \sum_{
n_{1} > n_{2}...> n_{k} > 0} \frac{x_{1}^{n_{1}}x_{2}^{n_{2}}...x_{k}^{n_{k}}
}{n_{1}^{m_{1}}n_{2}^{m_{2}}...n_{k}^{m_{k}} }.
\ea
The number $w=m_{1}+...+m_{k}$ is called the weight and $k$ is called
the depth of the multiple polylogarithm. The
power series (\ref{zsum}) is convergent for $|x_{i}|<1$, and can be analytically
continued via the iterated integral representation:
\ba
\label{intrepr}
Li_{m_{k},...,m_{1}}(x_{k},...,x_{1})=\int \limits_{0}^{x_{1}x_{2}...x_{k}} \left(
\frac{dt}{t} \circ \right)^{m_{1}-1}
\frac{dt}{x_{2}x_{3}...x_{k}-t} \circ
\nonumber \\ \left( \frac{dt}{t} \circ \right)^{m_{2}-1}
\frac{dt}{x_{3}...x_{k}-t} \circ ...  \circ  \left( \frac{dt}{t} \circ
\right)^{m_{k}-1}   \frac{dt}{1-t},
\ea
where the following notation is used for the iterated integrals:
\ba
\int \limits_{0}^{\lambda} \frac{dt}{a_{n}-t}\circ ...\circ
\frac{dt}{a_{1}-t} = \int \limits_{0}^{\lambda} \frac{dt_{n}}{a_{n}-t_{n}}
\int \limits_{0}^{t_{n}} \frac{dt_{n-1}}{a_{n-1}-t_{n-1}} \times...\times \int
\limits_{0}^{t_{2}}\frac{dt_1}{a_{1}-t_{1}}.
\ea
Note that the classical polylogarithms
\be
\label{classic}
{\rm Li}_{n}(z)\equiv \int \limits_{0}^{z}
\frac{ {\rm Li}_{n-1}(\xi ) }{\xi} d\xi, \,\, n\geq 2 ;
{\rm \hspace{.4in}}
{\rm Li}_{1}(z)\equiv-\ln(1-z)
\ee
are a subset of multiple polylogarithms. Examples of this statement can be
found in the subsequent discussion.

We start by considering the single-index $L$-function Eq.~(\ref{Lpfunction}):
\be
L_{\sigma_1}(\alpha_1,\alpha_2,\alpha_3,\alpha_4)=
\int_0^1 dy\, \frac{\ln (\alpha_1+\sigma_1 y)} {\alpha_4+y} {\rm
Li}_2(\alpha_2+\alpha_3 y).
\ee
After changing the integration variable $y=(t_1-\alpha_2)/\alpha_3$ we get
\be
L_{\sigma_1}=\int \limits_{\alpha_2}^{\alpha_2+\alpha_3} \frac{dt_1}{\alpha_3}
\frac{\ln (\alpha_1+\sigma_1 \frac{t_1-\alpha_2}{\alpha_3})}
{\alpha_4+\frac{t_1-\alpha_2}{\alpha_3}} {\rm Li}_2(t_1) =
\int \limits_{\alpha_2}^{\alpha_2+\alpha_3}
dt_1 \frac{\ln \frac{\sigma_1}{\alpha_3} + \ln (\frac{\alpha_1
\alpha_3}{\sigma_1} - \alpha_2 + t_1)}{\alpha_3 \alpha_4 - \alpha_2 + t_1} {\rm
Li}_2(t_1).
\ee
The integration interval can be split into two pieces, $[\alpha_2, 0]$ and
$[0, \alpha_2 + \alpha_3]$. One can then write $L_{\sigma_1}$ as a sum of four
terms:
\be
\label{sum4}
L_{\sigma_1} = \ln \frac{\sigma_1}{\alpha_3} \left( \int
\limits_{0}^{\alpha_2+\alpha_3}  - \int \limits_{0}^{\alpha_2} \,\, \right)
\frac{dt_1} {\gamma+t_1} {\rm Li}_{2}(t_1)
+ \left( \int \limits_{0}^{\alpha_2+\alpha_3}
                              - \int \limits_{0}^{\alpha_2} \,\,\right)
dt_1 \frac{\ln(\alpha+t_1)} {\gamma+t_1} {\rm Li}_{2}(t_1),
\ee
where we have introduced the notation
\be
\label{algama}
\alpha = \frac{\alpha_1 \alpha_3}{\sigma_1} - \alpha_2, {\rm \hspace{0.4in}}
\gamma = \alpha_3 \alpha_4 - \alpha_2.
\ee
Regarding Eq.~(\ref{sum4}) it is clear that there are only two different
types of integrals to be dealt with:
\be
\label{types}
\int \limits_{0}^{t_m} \frac{dt_1} {\gamma+t_1} {\rm Li}_{2}(t_1)
{\rm \hspace{0.4in}  and}    {\rm \hspace{0.4in}}
\int \limits_{0}^{t_m} dt_1 \frac{\ln(\alpha+t_1)} {\gamma+t_1} {\rm Li}_{2}(t_1).
\ee
with upper limits $t_m = \alpha_2 + \alpha_3$ or $t_m = \alpha_2$.
The first integral can be evaluated analytically in terms
of standard logarithms and classical polylogarithms up to ${\rm Li}_3$. However,
the same integral can also be expressed in terms of multiple polylogarithms via
the integral representation (\ref{intrepr}), e.g.
\be
\label{multi1}
\int \limits_{0}^{t_m}\frac{dt_{1}}{\gamma+t_{1}}{\rm Li}_{2}(t_{1})=
\int \limits_{0}^{t_m}\frac{dt_{1}}{\gamma+t_{1}}
\int \limits_{0}^{t_1} \frac{dt_{2}}{t_2}
\int \limits_{0}^{t_2} \frac{dt_3}{1-t_3} =
-Li_{2,1}\left(-\gamma,\frac{-t_m}{\gamma}\right).
\ee
We now deal with
the second integral in (\ref{types}). Consider the following multiple
polylogarithm of weight four:
\ba
Li_{2,1,1} \left( -\gamma, \frac{\alpha}{\gamma},\frac{t_m}{-\alpha} \right)=
\int \limits_{0}^{t_m}  \frac{dt_{2}}{-\alpha-t_{2}}  \int \limits_{0}^{t_{2}}
\frac{dt_{1}}{-\gamma-t_{1}}{\rm Li}_{2}(t_{1})   =
\int \limits_{0}^{t_m}\frac{dt_{1}}{\gamma+t_{1}}{\rm Li}_{2}(t_{1}) \int
\limits_{t_{1}}^{t_m}  \frac{dt_{2}}{\alpha+t_{2}}    \nn \\
%
%
=\int \limits_{0}^{t_m}\frac{dt_{1}}{\gamma+t_{1}}{\rm Li}_{2}(t_{1})
\ln(\alpha+t_m) - \int
\limits_{0}^{t_m}\frac{dt_{1}}{\gamma+t_{1}}{\rm Li}_{2}(t_{1}) \ln(\alpha+t_{1}).
\ea
In the first step we have used the usual trick to change the order of
integration.
As already noted before (see Eq.~(\ref{multi1})) the first term on the second line
can be expressed through a multiple polylogarithm of weight three. Thus one has
\be
\label{multi2}
 \int \limits_{0}^{t_m} dt_1 \frac{\ln(\alpha+t_1)} {\gamma+t_1} {\rm Li}_{2}(t_1)
=
 - Li_{2,1,1}\left(-\gamma, \frac{\alpha}{\gamma},\frac{t_m}{-\alpha}\right)  -
Li_{2,1}\left(-\gamma,\frac{-t_m}{\gamma}\right) \ln (\alpha+t_m).
\ee
Finally, substituting Eqs.~(\ref{multi1}) and (\ref{multi2}) into
Eq.~(\ref{sum4}) we arrive at the desired relation
\ba
L_{\sigma_1}(\alpha_1,\alpha_2,\alpha_3,\alpha_4)=
Li_{2,1,1}\left(-\gamma, \frac{\alpha}{\gamma},\frac{\alpha_2}{-\alpha}\right) -
Li_{2,1,1}\left(-\gamma, \frac{\alpha}{\gamma},\frac{\alpha_2+\alpha_3}
{-\alpha}\right)                                \\
\nn
+ Li_{2,1}\left(-\gamma,\frac{-\alpha_2}{\gamma}\right) \left(
\ln \frac{\sigma_1}{\alpha_3} + \ln (\alpha + \alpha_2) \right)   \\
\nn
- Li_{2,1}\left(-\gamma,\frac{-\alpha_2-\alpha_3}{\gamma}\right) \left(
\ln \frac{\sigma_1}{\alpha_3} + \ln (\alpha + \alpha_2 + \alpha_3) \right),
\ea
where $\alpha$ and $\gamma$ are defined in Eq.~(\ref{algama}).

Next we turn to the triple-index $L$-function Eq.~(\ref{Lfunction})
\be
\label{l3gen1}
L_{\sigma_1\sigma_2\sigma_3}(\alpha_1,\alpha_2,\alpha_3,\alpha_4)=
\int_0^1 dy \frac{\ln (\alpha_1+\sigma_1 y) \ln (\alpha_2+\sigma_2 y)
\ln (\alpha_3 + \sigma_3 y)} {\alpha_4 + y}
\ee
We again change the
integration variable $y=(1-\alpha_3-t_1)/\sigma_3$ and obtain
\ba
\label{l3gen2}
L_{\sigma_1\sigma_2\sigma_3}=- \int \limits_{1-\alpha_3}^ {1-\alpha_3-\sigma_3}
\frac{dt_1}{\sigma_3} \, \ln (\gamma_1 - \frac{\sigma_1}{\sigma_3} t_1) \ln
(\gamma_2 - \frac{\sigma_2} {\sigma_3} t_1) \,
\frac{\ln (1 - t_1)} {\gamma_3 - t_1/\sigma_3},
\ea
where
\be
\label{gamma123}
\gamma_1 = \alpha_1 + \frac{\sigma_1}{\sigma_3} (1-\alpha_3),  {\rm \hspace{.4in}}
\gamma_2 = \alpha_2 + \frac{\sigma_2}{\sigma_3} (1-\alpha_3),  {\rm \hspace{.4in}}
\gamma_3 = \alpha_4 +  \frac{1 - \alpha_3} {\sigma_3}.
\ee
At this point we
take $\sigma_1=\sigma_2=-\sigma_3=1$, which
does not affect the generality of our further discussion.
Splitting the integration interval in (\ref{l3gen2}) into two pieces as before, we
arrive at
\begin{equation}
\label{l3gen3}
L_{++-} = \left( \int \limits_{0}^ {1-\alpha_3-\sigma_3}
- \int \limits_{0}^ {1-\alpha_3} \,\, \right)
dt_1 \, \ln (\gamma'_1 + t_1) \ln (\gamma'_2 + t_1) \,
\frac{\ln (1 - t_1)} {\gamma'_3 + t_1} ,
\end{equation}
with
\be
\gamma'_1 = \alpha_1 - 1 + \alpha_3,  {\rm \hspace{.4in}}
\gamma'_2 = \alpha_2 - 1 + \alpha_3,  {\rm \hspace{.4in}}
\gamma'_3 = \alpha_4 - 1 + \alpha_3.
\ee
Now consider the following multiple polylogarithm of weight four:
\ba
\label{l3gen4}
- Li_{1,1,1,1}\left(-\gamma_3, \frac{\gamma_2}{\gamma_3},
\frac{\gamma_1}{\gamma_2}, \frac{-t_m}{\gamma_1} \right) =
 \int \limits_{0}^{t_m} \frac{dt_{4}}{\gamma_1 +t_{4}} \int \limits_{0}^{t_{4}}
\frac{dt_{3}}{\gamma_2 +t_{3}}
 \int \limits_{0}^{t_{3}}  \frac{dt_{2}}{\gamma_3 +t_{2}} \int \limits_{0}^{t_{2}}
\frac{dt_{1}}{1-t_{1}}   =             \\
\nn
- \int \limits_{0}^{t_m} \frac{dt_{4}}{\gamma_1 +t_{4}} \int \limits_{0}^{t_{4}}
\frac{dt_{3}}{\gamma_2 +t_{3}}
  \int \limits_{0}^{t_{3}} d t_{2} \frac{\ln(1-t_{2})}{\gamma_3 +t_{2}}  =
- \int \limits_{0}^{t_m} \frac{dt_{4}}{\gamma_1 +t_{4}} \int \limits_{0}^{t_{4}}
d t_{2} \frac{\ln(1-t_{2})}{\gamma_3 +t_{2}}
 \int \limits_{t_{2}}^{t_{4}} \frac{dt_{3}}{\gamma_2 +t_{3}}=       \\
\nn
- \int \limits_{0}^{t_m} dt_{4} \frac{\ln(\gamma_2 +t_{4})}{\gamma_1 +t_{4}}  \int
\limits_{0}^{t_{4}} d t_{2} \frac{ \ln(1-t_{2})}{\gamma_3 + t_{2}} +
 \int \limits_{0}^{t_m} \frac{dt_{4}}{\gamma_1 + t_{4}} \int \limits_{0}^{t_{4}}
dt_{2} \frac{\ln(1-t_{2}) \ln(\gamma_2 + t_{2})}{\gamma_3 +t_{2}}=\label{big}  \\
\nn
-I'(t_m) + \int \limits_{0}^{t_m} dt_{2}  \frac{ \ln(1-t_{2}) \ln(\gamma_2 +
t_{2})} {\gamma_3 +t_{2}}  \int \limits_{t_{2}}^{t_m}  \frac{dt_{4}}{\gamma_1
 + t_{4}}=                           \\
\nn
-I'(t_m) + I''(t_m)
 - \int \limits_{0}^{t_m} dt_{2} \frac{ \ln(\gamma_1 + t_{2})
\ln(\gamma_2 + t_{2})\ln(1-t_{2})} {\gamma_3 +t_{2}},
\ea
where we have introduced the notation
\ba
\label{iprimes}
I'(t_m) =  \int \limits_{0}^{t_m} dt_{4} \frac{\ln(\gamma_2 +t_{4})}{\gamma_1
+t_{4}}  \int \limits_{0}^{t_{4}} d t_{2} \frac{ \ln(1-t_{2})}{\gamma_3 + t_{2}},
\\    \nn
I''(t_m) = \ln(\gamma_1 + t_m) \int \limits_{0}^{t_m} dt_{2} \frac{ \ln(1-t_{2})
\ln(\gamma_2 + t_{2})} {\gamma_3 + t_{2}}.
\ea
The third term on the last line of (\ref{l3gen4}) is the integral of
the required type needed to express $L_{++-}$ in Eq.~(\ref{l3gen3}). We can
write:
\be
\label{l3gen5}
\int \limits_{0}^{t_m} dt_{1} \ln(\gamma_1 + t_{1}) \ln(\gamma_2 + t_{1})
                                    \frac{\ln(1-t_{1})} {\gamma_3 + t_{1}} =
Li_{1,1,1,1}\left(-\gamma_3, \frac{\gamma_2}{\gamma_3}, \frac{\gamma_1}
{\gamma_2}, \frac{-t_m}{\gamma_1} \right) - I'(t_m) + I''(t_m).
\ee
To express the last two terms of (\ref{l3gen5}) in terms of multiple
polylogarithms we proceed as follows. Consider first
\ba
-Li_{1,1,1} \left( -\gamma_3,\frac{\gamma_2}{\gamma_3},\frac{-t_m}{\gamma_2}
\right)=
 \int \limits_{0}^{t_m}  \frac{dt_{2}}{\gamma_2 + t_{2}} \int \limits_{0}^{t_{2}}
dt_{1} \frac{\ln(1-t_{1})} {\gamma_3 + t_{1}}=
 \int \limits_{0}^{t_m}  dt_{1} \frac{\ln(1-t_{1})} {\gamma_3 + t_{1}}  \int
\limits_{t_{1}}^{t_m} \frac{dt_{2}}{\gamma_2 + t_{2}}=             \\
\nn
\ln(\gamma_2 + t_m)  \int \limits_{0}^{t_m} dt_{1} \frac{ \ln(1-t_{1})} {\gamma_3
+ t_{1}}
- \int \limits_{0}^{t_m} dt_{1} \frac{ \ln(1-t_{1}) \ln(\gamma_2 + t_{1})}
{\gamma_3 + t_{1}}  =                                  \\
\nn
\ln(\gamma_2 + t_m) Li_{1,1} \left( -\gamma_3, \frac{-t_m}{\gamma_3} \right)
- \int \limits_{0}^{t_m} dt_{1} \frac{ \ln(1-t_{1})\ln(\gamma_2 + t_{1})}
{\gamma_3 + t_{1}},
\ea
from which we immediately conclude that
\be
\label{sub1}
I''(t_m)/\ln(\gamma_1 + t_m) =
Li_{1,1,1} \left( -\gamma_3,\frac{\gamma_2}{\gamma_3},\frac{-t_m}{\gamma_2}
\right) +
\ln\left(\gamma_2 + t_m\right) Li_{1,1} \left( -\gamma_3, \frac{-t_m}{\gamma_3}
\right).
\ee
Note that the above multiple polylogarithms of weight three and weight two can
also be expressed in terms of
logarithms and classical polylogarithms by direct evaluation of the corresponding
integrals.

For the first integral in (\ref{iprimes}) we write
\be
I'(t_m) =  \int \limits_{0}^{t_m} dt_{4} \frac{\ln(\gamma_2 +t_{4})}{\gamma_1
+t_{4}}  \int \limits_{0}^{t_{4}} d t_{2} \frac{ \ln(1-t_{2})}{\gamma_3 + t_{2}} =
\int \limits_{0}^{t_m} dt_{4} \frac{\ln(\gamma_2 + t_{4})} {\gamma_1 + t_{4}}
Li_{1,1}\left(-\gamma_3,\frac{-t_{4}}{\gamma_3}\right) .
\ee
On the other hand one has
\ba
Li_{1,1,1,1} \left( -\gamma_3, \frac{\gamma_1}{\gamma_3}, \frac{\gamma_2}
{\gamma_1}, \frac{-t_m}{\gamma_2} \right) =
\int \limits_{0}^{t_m} \frac{ dt_{2}}{\gamma_2 + t_{2}}
\int \limits_{0}^{t_{2}} \frac{dt_{1}} {\gamma_1 + t_{1}}
Li_{1,1} \left( -\gamma_3, \frac{-t_{1}}{\gamma_3} \right)=        \\
\nn
 \int \limits_{0}^{t_m} \frac{dt_{1}}{\gamma_1 + t_{1}}
Li_{1,1} \left( -\gamma_3, \frac{-t_{1}}{\gamma_3} \right)
 \int \limits_{t_{1}}^{t_m}  \frac{ dt_{2}}{\gamma_2
+ t_{2}}=                                  \\
\nn
\ln(\gamma_2 + t_m) \int \limits_{0}^{t_m} \frac{dt_{1}} {\gamma_1 + t_{1}}
Li_{1,1} \left( -\gamma_3, \frac{-t_{1}}{\gamma_3} \right)
- \int \limits_{0}^{t_m} dt_{1} \frac{ \ln(\gamma_2 + t_{1})}{\gamma_1 + t_{1}}
Li_{1,1}\left( -\gamma_3, \frac{-t_{1}}{\gamma_3} \right) =               \\
\nn
- \ln(\gamma_2 + t_m) Li_{1,1,1}\left( -\gamma_3, \frac{\gamma_1}{\gamma_3},
\frac{-t_m}{\gamma_1} \right) - I'(t_m).
\ea
We then conclude that
\be
\label{sub2}
I'(t_m) = - Li_{1,1,1,1} \left( -\gamma_3, \frac{\gamma_1}{\gamma_3},
\frac{\gamma_2} {\gamma_1}, \frac{-t_m}{\gamma_2} \right) - \ln(\gamma_2 + t_m)
Li_{1,1,1} \left( -\gamma_3, \frac{\gamma_1}{\gamma_3}, \frac{-t_m}{\gamma_1}
\right).
\ee
Finally, substituting Eqs.~(\ref{sub1}) and (\ref{sub2}) into Eq.~(\ref{l3gen5})
we write down the result for the integral
\ba
\int \limits_{0}^{t_m} dt_1 \ln(\gamma_1 + t_1) \ln(\gamma_2 + t_1)
\frac{\ln(1 - t_1)} {\gamma_3 + t_1} =
\ln(\gamma_1 + t_m) \ln(\gamma_2 + t_m)
Li_{1,1}\left( -\gamma_3, \frac{-t_m} {\gamma_3} \right) +
\\           \nn
  \ln(\gamma_2 + t_m) Li_{1,1,1}\left( -\gamma_3, \frac{\gamma_1} {\gamma_3},
\frac{-t_m}{\gamma_1} \right)
+ \ln(\gamma_1 + t_m)
Li_{1,1,1}\left( -\gamma_3, \frac{\gamma_2} {\gamma_3}, \frac{-t_m}
{\gamma_2} \right) +                                \\
\nn
 Li_{1,1,1,1}\left( -\gamma_3, \frac{\gamma_2} {\gamma_3}, \frac{\gamma_1}
{\gamma_2}, \frac{-t_m}{\gamma_1} \right)
+ Li_{1,1,1,1} \left( -\gamma_3, \frac{\gamma_1} {\gamma_3}, \frac{\gamma_2}
{\gamma_1}, \frac{-t_m}{\gamma_2} \right)
\ea
which demonstrates how the two integrals in Eq.~(\ref{l3gen3}) representing
$L_{++-}$ are related to multiple polylogarithms.
The discussion of the other
cases for the $L_{\sigma_1\sigma_2\sigma_3}$ functions proceeds along
similar lines.

\vglue 1cm
\begin{center}\begin{large}\begin{bf}
APPENDIX~C
\end{bf}\end{large}\end{center}
\vglue .3cm

\setcounter{equation}{0}
\renewcommand{\theequation}{C\arabic{equation}}

In Appendix~C we discuss properties and identities involving the single- and
triple-index $L$-functions in
Eqs.~(\ref{Lfunction}) and (\ref{Lpfunction}). There are two different categories
of identities which we discuss in turn.
We start by considering the simplest identities originating from
symmetries related to permutations in the indices and arguments. We then present
further identities based on integration-by-parts techniques.
\vglue .5cm
\begin{center}\begin{bf}
1. Symmetry properties
\end{bf}\end{center}

We start with the single-index function
$L_{\sigma_1}(\alpha_1, \alpha_2, \alpha_3, \alpha_4)$.
One notices that a change of the integration variable $y\rightarrow 1-y$ results
in the identity
\be
L_{\sigma_1} (\alpha_1, \alpha_2, \alpha_3, \alpha_4) = -
L_{-\sigma_1} (\alpha_1 + \sigma_1, \alpha_2 + \alpha_3, -\alpha_3, -\alpha_4
- 1)
\ee
which implies that $L_{-}$ can always be related to $L_{+}$, and
vice versa. We have thus written our results for the three-point and four-point
functions in the main text only in terms of the $L_{+}$ functions.

Next we turn to the triple-index $L$-function.
Note that $L_{\sigma_1 \sigma_2 \sigma_3} (\alpha_1, \alpha_2, \alpha_3, \alpha_4)$
is symmetric under permutations of any
two pairs of indices and arguments $\{\sigma_i, \alpha_i\}$ and $\{\sigma_j,
\alpha_j\}$ for $(i\ne j)$. The same change of variables as above $y\rightarrow
1-y$ results in
\be
\label{symmetry2}
L_{\sigma_1 \sigma_2 \sigma_3} (\alpha_1, \alpha_2, \alpha_3, \alpha_4) = -
L_{-\sigma_1 -\sigma_2 -\sigma_3} (\alpha_1 + \sigma_1, \alpha_2 + \sigma_2,
\alpha_3 + \sigma_3, -\alpha_4 - 1).
\ee
Therefore, from the eight functions $L_{---}$, $L_{--+}$, $L_{-+-}$,
$L_{+--}$, $L_{-++}$, $L_{+-+}$, $L_{++-}$, $L_{+++}$ only two are independent. We
have chosen to write our results
in terms of $L_{-++}$ and $L_{+++}$.
\vglue .5cm
\begin{center}\begin{bf}
2. Integration-by-parts identities
\end{bf}\end{center}

The triple- and single-index $L$-functions $L_{+++}, L_{-++}$ and
$L_{+}$ defined in Eqs.~(\ref{Lfunction}) and (\ref{Lpfunction}) have been
devised such
that they have neither  branch cuts nor  poles on the integration path
$y \in [0,1]$.
This also implies that the $L_{+++}, L_{-++}$ and $L_{+}$ functions are real.
Remember that the
branch cuts for the $\ln$ and ${\rm Li}_{2}$ functions are
$(-\infty, 0]$ and $(1, +\infty)$, respectively. The domains of the
functions
$L_{+++}, L_{-++}$ and $L_{+}$ are
\ba
\label{domain}
 \begin{array}{r@{\quad:\quad}l}
 L_{+++}(\alpha_{1},\alpha_2, \alpha_{3}, \alpha_{4})  &  \alpha_{1}>0
 ,  \alpha_2 > 0, \alpha_{3} > 0, \alpha_{4}<-1 \quad  {\rm or}\quad
 \alpha_{4}>0;  \\
 L_{-++}(\alpha_{1},\alpha_2, \alpha_{3}, \alpha_{4})  &  \alpha_{1} >
 1,  \alpha_2 > 0, \alpha_{3}> 0, \alpha_{4}<-1\quad {\rm or}\quad
 \alpha_{4}>0;  \\
 L_{+}(\alpha_{1},\alpha_2, \alpha_{3}, \alpha_{4})  &  \alpha_{1} >0
 ,  \alpha_2 \leq 1,\alpha_{2}+ \alpha_{3}\leq 1, \alpha_{3}\neq 0,
 \alpha_{4}<-1\quad {\rm or}\quad  \alpha_{4}>0.
  \end{array}
\ea
Looking at the definition of the triple-index $L$--function in
(\ref{Lfunction}) one concludes that the boundary points $\alpha_1=0$
and/or $\alpha_2=0$ and/or $\alpha_3=0$ can be included in the domain of
definition
for $L_{+++}$. The same holds true for  $\alpha_1=1$ and/or $\alpha_2=0$
and/or $\alpha_3=0$ for $L_{-++}$. Also, from the definition of the single-index
function $L_+$ in (\ref{Lpfunction}) on concludes that
the boundary point $\alpha_1=0$ can be added to its domain of definition.

The points $\alpha_{4}=\{-1, 0\}$ can also be included in the domain if the
values taken by other parameters $\alpha_{i}$ guarantee the convergence of
the integral. In what follows we assume everywhere in this appendix that
the conditions (\ref{domain}) are satisfied.
Nevertheless, it is always possible to analytically continue the parameters to the
complex plane.

In order to obtain integration-by-parts identities one makes use of the standard
integration-by-parts formula
\be
\label{parts}
\int \limits_{0}^{1}  {U V' d y} = U V{\Big |}_{0}^{1} - \int
\limits_{0}^{1}  V U' d y \, .
\ee

We start with the triple-index functions $L_{-++}$ and $L_{+++}$ defined in
Eq.~(\ref{Lfunction}). Setting $U$ equal to the numerator
$\left[ \ln (\alpha_1+\sigma_1 y) \ln(\alpha_2+\sigma_2 y) \ln (\alpha_3+\sigma_3
y)\right]$
and $V'$ equal to the remainder $(\alpha_4+y)^{-1}$ we then arrive at
\ba
\label{forLppp}
L_{+++}(\alpha_{1},\alpha_2, \alpha_{3}, \alpha_{4})  = \left\{
\begin{array}{rl}  \alpha_{4}>0: &
   \ln(\alpha_{1}+y)\ln(\alpha_{2}+y)\ln(\alpha_{3}+y)\ln(\alpha_{4}+y)
{\Big |}_{0}^{1} \\ &
-L_{+++}(\alpha_{4},\alpha_2, \alpha_{3},\alpha_{1})
-L_{+++}(\alpha_{1},\alpha_4, \alpha_{3}, \alpha_{2}) \\ &
-L_{+++}(\alpha_{1},\alpha_2, \alpha_{4}, \alpha_{3}); \\
 \alpha_{4}<-1: &
\ln(\alpha_{1}+y)\ln(\alpha_{2}+y)\ln(\alpha_{3}+y)\ln(-\alpha_{4}-y)
{\Big |}_{0}^{1} \\
& -L_{-++}(-\alpha_{4},\alpha_2, \alpha_{3},
\alpha_{1})-L_{-++}(-\alpha_{4},\alpha_1,\alpha_{3}, \alpha_{2})\\ &
-L_{-++}(-\alpha_{4},\alpha_1, \alpha_{2}, \alpha_{3});
  \end{array}   \right.
\ea
and
\ba
\label{forLmpp}
L_{-++}(\alpha_{1},\alpha_2, \alpha_{3}, \alpha_{4})  = \left\{
\begin{array}{rl} \alpha_{4}>0 :  &
\ln(\alpha_{1}-y)\ln(\alpha_{2}+y)\ln(\alpha_{3}+y)\ln(\alpha_{4}+y)
{\Big |}_{0}^{1}\\
 & -L_{+++}(\alpha_{4},\alpha_2, \alpha_{3},- \alpha_{1})
-L_{-++}(\alpha_{1},\alpha_4, \alpha_{3}, \alpha_{2})\\  &
-L_{-++}(\alpha_{1},\alpha_2, \alpha_{4}, \alpha_{3});
\\   \alpha_{4}<-1: &
\ln(\alpha_{1}-y)\ln(\alpha_{2}+y)\ln(\alpha_{3}+y)\ln(-\alpha_{4}-y)
{\Big |}_{0}^{1} \\ &
-L_{-++}(-\alpha_{4},\alpha_2, \alpha_{3},- \alpha_{1})\\ &
+L_{-++}(\alpha_{3}+1,\alpha_1-1,- \alpha_{4}-1, -\alpha_{2}-1)\\ &
+L_{-++}(\alpha_{2}+1,\alpha_1-1,- \alpha_{4}-1, -\alpha_{3}-1).
  \end{array}   \right. \quad .
\ea
For the second part of Eq.(\ref{forLmpp}) we have made use of relation
(\ref{symmetry2}).

There are some special cases when some of the $\alpha_i$ take values on the
boundary of the domain of definition where one can still make use of the
identities (\ref{forLppp}) and (\ref{forLmpp}) even if the conditions
(\ref{domain}) are not met. For example, for the case
$\{\alpha_1=0, \alpha_4=-1\}$ the identitity
(\ref{forLmpp}) is still valid. There are similar special cases for further
identities to be derived below.

The integration-by-parts identities for the single-index $L_{+}$
function are more involved.
To prepare ourselves we first write down the derivative of the dilog
function in the integrand of (\ref{Lpfunction}). One has
\be
\frac{d  {\rm Li_{2}}(\alpha_{2}+\alpha_{3}y)}{dy}=
-\frac{\ln(1-\alpha_{2}-\alpha_{3}y)}
{\frac{\alpha_{2}}{\alpha_{3}}+y   } .
\ee
In the case of the single-index function it will prove important to consider two
different choices for $U$. We start by setting
the whole numerator $\left[\ln (\alpha_1+\sigma_1 y) {\rm Li}_2(\alpha_2+
\alpha_3 y)\right]$ in the integrand of Eq.~(\ref{Lpfunction}) to $U$. For $V'$ we
then have $(\alpha_4+y)^{-1}$. One obtains
\ba
L_{+}(\alpha_{1},\alpha_2, \alpha_{3}, \alpha_{4})  =  \nn
\ea
\vglue -.4in
\ba
\label{type1}
  \left\{  \begin{array}{rl}
\alpha_{4}>0, \alpha_{3}>0 : &      \ln(\alpha_{1}+y){\rm
Li}_{2}(\alpha_{2} +\alpha_{3}y) \ln(\alpha_{4}+y) {\Big |}_{0}^{1} \\ &
- L_{+}(\alpha_{4},\alpha_2, \alpha_{3}, \alpha_{1})\\
 & +L_{-++}(\frac{1-\alpha_2}{\alpha3},\alpha_1, \alpha_4,
\frac{\alpha_2}{\alpha_3 })+
 \ln(\alpha_{3})  \int \limits_{0}^{1} \frac{\ln(\alpha_{1}+y)
\ln(\alpha_{4}+y)   }{\frac{\alpha_{2}}{\alpha_{3}}+y  } dy;   \\
\alpha_{4}>0, \alpha_{3}<0 : &   \ln(\alpha_{1}+y){\rm Li}_{2}(\alpha_{2}
+\alpha_{3}y) \ln(\alpha_{4}+y) {\Big |}_{0}^{1} \\ &
- L_{+}(\alpha_{4},\alpha_2, \alpha_{3}, \alpha_{1})\\
 &  +L_{+++}(\frac{\alpha_2-1}{\alpha3},\alpha_1, \alpha_4,
\frac{\alpha_2}{\alpha_3 } )+
 \ln(-\alpha_{3})  \int \limits_{0}^{1} \frac{\ln(\alpha_{1}+y)
\ln(\alpha_{4}+y)   }{\frac{\alpha_{2}}{\alpha_{3}}+y  }  dy;   \\
\alpha_{4}<-1, \alpha_{3}>0 : &  \ln(\alpha_{1}+y){\rm Li}_{2}(\alpha_{2}
+\alpha_{3}y) \ln(-\alpha_{4}-y) {\Big |}_{0}^{1}\\
 &  +L_{+}(-\alpha_{4}-1,\alpha_2+ \alpha_{3},- \alpha_{3},-
\alpha_{1}-1) \\ &
-L_{-++}(\alpha_{1}+1,\frac{1-\alpha_{2}-\alpha_{3}}{\alpha_{3}} ,
-\alpha_{4}-1,-\frac{\alpha_{2}+\alpha_{3}}{\alpha_{3}} )\\ & +
 \ln(\alpha_{3}) \int \limits_{0}^{1} \frac{\ln(\alpha_{1}+y)
\ln(-\alpha_{4}-y)    }{\frac{\alpha_{2}}{\alpha_{3}}+y  } dy;   \\
\alpha_{4}<-1, \alpha_{3}<0 : &   \ln(\alpha_{1}+y){\rm
Li}_{2}(\alpha_{2} +\alpha_{3}y) \ln(-\alpha_{4}-y) {\Big |}_{0}^{1}\\
 &  +L_{+}(-\alpha_{4}-1,\alpha_2+ \alpha_{3},- \alpha_{3},-
\alpha_{1}-1) \\
 & +L_{-++}(-\alpha_{4}, \alpha_1, \frac{\alpha_2-1}{\alpha3},
\frac{\alpha_2}{\alpha_3 })
+ \ln(-\alpha_{3}) \int \limits_{0}^{1} \frac{\ln(\alpha_{1}+y)
\ln(-\alpha_{4}-y)    }{\frac{\alpha_{2}}{\alpha_{3}}+y  } dy.
\end{array}    \right.
\ea
An additional condition for (\ref{type1}) has to be explicated because
it does not follow automatically from (\ref{domain}), namely the
parameters $\alpha_{2}$ and $\alpha_{3}$ are restricted by
 \be
\label{ratiocond}
 \frac{\alpha_{2}}{\alpha_{3}} <-1 \quad {\rm or} \quad
\frac{\alpha_{2}}{\alpha_{3}} > 0  .
\ee
The integrals in  (\ref{type1}) are simple enough to be evaluated in
terms of classical polylogarithms up to
${\rm Li} _{3}$.
We do not provide explicit results for these integrations since they are rather
lengthy and, in addition, depend on relations between the parameters.

A second choice for $U$ in (\ref{parts}) provides
further identities for $L_{+}$. In this case one sets
${\rm Li}_{2}(\alpha_{2}+\alpha_{3}y)$ to $U$ and
$\ln (\alpha_1 + y)/(\alpha_4 + y)$ to $V'$.
To calculate $V$ one has to differentiate between three cases for the set
of parameters $\alpha_1$ and $\alpha_4$. One has
\be
\label{V}
V=\int \frac{\ln(\alpha_{1}+y)}{\alpha_{4}+y} dy=   \left\{
\begin{array}{rl} \alpha_{1}<  \alpha_{4}: &
\ln(\alpha_{1}+y) \ln(\frac{\alpha_4+y}{\alpha_4-\alpha_1})+{\rm
Li}_{2}(\frac{\alpha_1+y}{\alpha_1-\alpha_{4} });    \\
\alpha_{1}> \alpha_{4}> 0: &
\ln(\alpha_1-\alpha_{4})\ln(\alpha_{4}+y) - {\rm Li}_2
(\frac{\alpha_4+y}{\alpha_4-\alpha_{1} } );    \\
 \alpha_{4}<-1: & \ln(\alpha_1-\alpha_{4})\ln(-\alpha_{4}-y) - {\rm Li}_2
(\frac{\alpha_4+y}{\alpha_4-\alpha_{1} } ).
\end{array}     \right. \,
\ee
Using Eq.~(\ref{V}) one then obtains
\ba
L_{+}(\alpha_{1},\alpha_2, \alpha_{3}, \alpha_{4})  =      \nn
\ea
\vglue -0.2in
\ba
\label{type2}
   \left\{ \begin{array}{rl}
\alpha_{1}<\alpha_{4}, \alpha_{3}>0: &
{\rm Li}_{2}(\alpha_{2}+\alpha_{3}y)   \left( \ln(\alpha_{1}+y)
\ln(\frac{\alpha_4+y}{\alpha_4-\alpha_1}) +{\rm
Li}_{2}(\frac{\alpha_1+y}{\alpha_1-\alpha_{4} })\right){\Big
|}_{0}^{1}\\ &
+L_{-++}(\frac{1-\alpha_2}{\alpha3},\alpha_1, \alpha_4,
\frac{\alpha_2}{\alpha_3 })\\ &
-L_{+}(\frac{1-\alpha_{2}-\alpha_3}{\alpha_{3}},
\frac{\alpha_1+1}{\alpha_{1}-\alpha_{4}},
\frac{1}{\alpha_{4}-\alpha_{1}}, -\frac{\alpha_{2}+\alpha_3}{\alpha_3}) \,+ \\
&
\int \limits_{0}^{1} \frac{\ln(\alpha_3) \left [{\rm
Li}_2(\frac{\alpha_{1}+y}{\alpha_{1}-\alpha_4}) +
\ln(\alpha_1+y)\ln(\frac{\alpha_{4}+y}{\alpha_{4}-\alpha_1})\right ]
-
\ln(\alpha_4-\alpha_1)\ln(\alpha_1+y)\ln(\frac{1-\alpha_{2}}{\alpha_{3}}-y )
              }{ \frac{\alpha_2}{\alpha_3}+y}dy ;          \\
\alpha_{1}<\alpha_{4}, \alpha_{3}<0: &
{\rm Li}_{2}(\alpha_{2}+\alpha_{3}y)   \left( \ln(\alpha_{1}+y)
\ln(\frac{\alpha_4+y}{\alpha_4-\alpha_1}) +{\rm
Li}_{2}(\frac{\alpha_1+y}{\alpha_1-\alpha_{4} })\right){\Big
|}_{0}^{1}
\\ & +L_{+++}(\frac{\alpha_2-1}{\alpha3},\alpha_1, \alpha_4,
\frac{\alpha_2}{\alpha_3 })
\\ & +L_{+}(\frac{\alpha_{2}-1}{\alpha_{3}},
\frac{\alpha_1}{\alpha_{1}-\alpha_{4}},
\frac{1}{\alpha_{1}-\alpha_{4}}, \frac{\alpha_{2}}{\alpha_{3}}  ) \,+
\\ &
\int \limits_{0}^{1} \frac{\ln(-\alpha_3) \left [{\rm
Li}_2(\frac{\alpha_{1}+y}{\alpha_{1}-\alpha_4}) +
\ln(\alpha_1+y)\ln(\frac{\alpha_{4}+y}{\alpha_{4}-\alpha_1})\right ]
- \ln(\alpha_4-\alpha_1)\ln(\alpha_1+y)\ln(\frac{\alpha_{2}-1}
{\alpha_{3}}+y )              }{ \frac{\alpha_2}{\alpha_3}+y}dy ;     \\
\begin{array}{c}
\alpha_{1}>\alpha_{4}>0, \alpha_{3}>0
\end{array}
: &   {\rm Li}_{2}(\alpha_{2}+\alpha_{3}y) \left(
\ln(\alpha_{1}-\alpha_{4}) \ln(\alpha_4+y) -{\rm
Li}_{2}(\frac{\alpha_4+y}{\alpha_4-\alpha_{1} })\right){\Big
|}_{0}^{1}
\\ &
+L_{+}(\frac{1-\alpha_{2}-\alpha_{3}}{\alpha_{3}},
\frac{\alpha_4+1}{\alpha_{4}-\alpha_1},
\frac{1}{\alpha_1-\alpha_4},-\frac{\alpha_2+\alpha_3}{\alpha_3}  ) \,+
\\ &  \int \limits_{0}^{1} \frac{-\ln(\alpha_{3}){\rm
Li}_2(\frac{\alpha_{4}+y}{\alpha_{4}-\alpha_1})
+ \ln(\alpha_1-\alpha_4)\ln(\alpha_4+y)\ln(1-\alpha_{2}-\alpha_3 y )
}{ \frac{\alpha_2}{\alpha_3}+y}dy ;         \\
\begin{array}{c}
 \alpha_{4}<-1 ,  \alpha_{3}>0
\end{array}
: &   {\rm Li}_{2}(\alpha_{2}+\alpha_{3}y) \left(
\ln(\alpha_{1}-\alpha_{4}) \ln(-\alpha_4-y) -{\rm
Li}_{2}(\frac{\alpha_4+y}{\alpha_4-\alpha_{1} })\right){\Big
|}_{0}^{1}
\\ & +L_{+}(\frac{1-\alpha_{2}-\alpha_{3}}{\alpha_{3}},
\frac{\alpha_4+1}{\alpha_{4}-\alpha_1},
\frac{1}{\alpha_1-\alpha_4},-\frac{\alpha_2+\alpha_3}{\alpha_3}   )
\\ & + \int \limits_{0}^{1} \frac{-\ln(\alpha_{3}){\rm
Li}_2(\frac{\alpha_{4}+y}{\alpha_{4}-\alpha_1})
+  \ln(\alpha_1-\alpha_4)\ln(-\alpha_4-y)\ln(1-\alpha_{2}-\alpha_3 y )
}{ \frac{\alpha_2}{\alpha_3}+y}dy ;        \\
\begin{array}{c}
\alpha_{1}>\alpha_{4}>0,  \alpha_{3}<0
\end{array}
: &  {\rm Li}_{2}(\alpha_{2}+\alpha_{3}y) \left(
\ln(\alpha_{1}-\alpha_{4}) \ln(\alpha_4+y) -{\rm
Li}_{2}(\frac{\alpha_4+y}{\alpha_4-\alpha_{1} })\right) {\Big
|}_{0}^{1}
\\ & -L_{+}(\frac{\alpha_{2}-1}{\alpha_{3}},
\frac{\alpha_{4}}{\alpha_{4}-\alpha_{1}},
\frac{1}{\alpha_{4}-\alpha_{1}}, \frac{\alpha_{2}}{\alpha_3}  ) \,+
\\ &  \int \limits_{0}^{1} \frac{-\ln(-\alpha_{3}){\rm
Li}_2(\frac{\alpha_{4}+y}{\alpha_{4}-\alpha_1})
+ \ln(\alpha_1-\alpha_4)\ln(\alpha_4+y)\ln(1-\alpha_{2}-\alpha_3 y )
}{ \frac{\alpha_2}{\alpha_3}+y}dy ;          \\
\begin{array}{c}
 \alpha_{4}<-1 ,  \alpha_{3}<0
\end{array}
:  & {\rm Li}_{2}(\alpha_{2}+\alpha_{3}y) \left(
\ln(\alpha_{1}-\alpha_{4}) \ln(-\alpha_4-y) -{\rm
Li}_{2}(\frac{\alpha_4+y}{\alpha_4-\alpha_{1} })\right){\Big
|}_{0}^{1}
\\ &  -L_{+}(\frac{\alpha_{2}-1}{\alpha_{3}},
\frac{\alpha_{4}}{\alpha_{4}-\alpha_{1}},
\frac{1}{\alpha_{4}-\alpha_{1}}, \frac{\alpha_{2}}{\alpha_3}  ) \,+
\\ &   \int \limits_{0}^{1} \frac{-\ln(-\alpha_{3}){\rm
Li}_2(\frac{\alpha_{4}+y}{\alpha_{4}-\alpha_1})
+  \ln(\alpha_1-\alpha_4)\ln(-\alpha_4-y)\ln(1-\alpha_{2}-\alpha_3 y )
}{ \frac{\alpha_2}{\alpha_3}+y}dy .
\end{array}     \right.
 \ea
In deriving (\ref{type2}) it is important to take into account condition
(\ref{ratiocond}).
As was the case in Eq.~(\ref{type1}) the integrals in (\ref{type2}) can be
evaluated in
terms of classical polylogarithms up to ${\rm Li} _{3}$.

There is also one special case of the last identity (\ref{type2}) when the first
and fourth arguments of the single-index $L_{+}$ function are equal, e.g.
$\alpha_1 = \alpha_4$. In this case $L_{+}$ can be expressed only in terms of
the functions $L_{-++}$ or $L_{+++}$ as follows:
\ba
L_{+}(\alpha_{1},\alpha_2, \alpha_{3}, \alpha_{1})  = \nonumber
\ea
\vglue -0.4in
\ba
\left\{ \begin{array}{rl}
\alpha_{3}>0: & \frac{1}{2}\ln^{2}(\alpha_{1}+y){\rm
Li}_{2}(\alpha_{2}+\alpha_{3}y){\Big
|}_{0}^{1} +\frac{1}{2}L_{-++}(\frac{1-\alpha_{2}}{\alpha_{3}}, \alpha_{1},
\alpha_{1},\frac{\alpha_{2}}{\alpha_{3}}  )
\\ & +\frac{\ln(\alpha_{3})}{2}\int \limits_{0}^{1}
\frac{ \ln^2(\alpha_1+y)}   {    \frac{\alpha_2 }{\alpha_3 }+y   } dy ,  \\
\alpha_{3}<0: &  \frac{1}{2}\ln^{2}(\alpha_{1}+y){\rm
Li}_{2}(\alpha_{2}+\alpha_{3}y){\Big
|}_{0}^{1}  +\frac{1}{2} L_{+++}(\frac{\alpha_{2}-1}{\alpha_{3}}, \alpha_{1},
\alpha_{1},\frac{\alpha_{2}}{\alpha_{3}} )+
\\ & +\frac{\ln(-\alpha_{3})}{2}\int \limits_{0}^{1}
\frac{ \ln^2(\alpha_1+y)}   {    \frac{\alpha_2 }{\alpha_3 }+y       } dy \, .
\end{array}     \right.
\ea
The third and last identity for the $L_{+}$ function is obtained from the
definition (\ref{Lpfunction}) without
making direct use of the integration-by-parts identity Eq.~(\ref{parts}).
Nevertheless it can still be called
an integration-by-parts identity because it makes use of the well-known
identity
\be
\label{dilog}
{\rm Li}_{2}(z)=\zeta(2)-\ln(z)\ln(1-z)-{\rm Li}_{2}(1-z), \quad  z \in C
\ee
which in turn is derived from the definition of the ${\rm Li}_{2}$--function
(\ref{classic})
with the help of the integration-by-parts identity (\ref{parts}).
After transforming ${\rm Li}_{2}(\alpha_{2}+\alpha_{3}y)$
according to (\ref{dilog})
one gets
\ba
L_{+}(\alpha_{1},\alpha_2, \alpha_{3}, \alpha_{4})  = \nn
\ea
\vglue -.4in
\ba
\label{type3}
\left\{ \begin{array}{rl}
\begin{array}{c} \\
\alpha_{3}>0
\\ 0 \leq \alpha_{2} \leq 1, \alpha_{2}+\alpha_{3} \leq 1
\end{array}:  & -L_{+}(\alpha_{1}, 1-\alpha_{2}, -\alpha_2, \alpha_{4}
)-L_{-++}(\frac{ 1-\alpha_{2}  }{\alpha_{3} },\alpha_{1} ,
\frac{\alpha_{2}}{\alpha_{3}}, \alpha_{4} ) +
\\  &
\int \limits_{0}^{1} \frac{\ln(\alpha_{1}+y)   }{\alpha_{4}+y   }\left(
\zeta(2)-\ln(\alpha_{3})[\ln(\alpha_{2}+\alpha_{3}y)
+\ln(\frac{1-\alpha_{2}}{c}-y )   ]   \right) dy;    \\
\begin{array}{c} \\
\alpha_{3}<0\\
0 \leq \alpha_{2} \leq 1, 0 \leq \alpha_{2}+\alpha_{3}
\end{array} : & -L_{+}(\alpha_{1}, 1-\alpha_{2}, -\alpha_2, \alpha_{4} )
-L_{-++}(-\frac{\alpha_{2}} {\alpha_{3}} ,\alpha_{1} ,  \frac{
\alpha_{2}-1  }{\alpha_{3} }  , \alpha_{4} ) +
\\ & \int \limits_{0}^{1} \frac{\ln(\alpha_{1}+y)   }{\alpha_{4}+y
}\left( \zeta(2)-\ln(-\alpha_{3})[\ln(\alpha_{2}+\alpha_{3}y)
+\ln(\frac{\alpha_{2}-1}{c}+y )   ]   \right) dy.
\end{array}     \right.
\ea
A few final comments are appropiate. In spite of the rather complicated
appearance of the identities (\ref{forLppp}), (\ref{forLmpp}), (\ref{type1}),
(\ref{type2}), (\ref{type3})
these turn out to be very useful to reduce the length of the results presented
in the main text. The first step in the chain of reductions is to write
everything in terms of the functions $L_{+}$, $L_{-++}$ and $L_{+++}$.
In a second step one uses the identities written down in this Appendix to
find the set of arguments of $L$-functions  for which  the number of  the
functions $L_{+}$, $L_{-++}$ and $L_{+++}$ is minimal.
We have devised several programs for the MATHEMATICA computer algebra system
which automatically find minimal sets of the single- and triple-index
$L$-functions.
With the help of these programs we have been able to greatly reduce the number
of $L$-functions appearing in our results and have thereby greatly reduced their
length. The same was done for the logarithms and classical polylogarithms using
standard one- and two-variable identities for classical polylogarithms as given
e.g. in $[$\ref{Lewin}$]$.

\newpage
\begin{center}\begin{large}\begin{bf}
REFERENCES
\end{bf}\end{large}\end{center}
\vglue .3cm

   \begin{list}{$[$\arabic{enumi}$]$}
    {\usecounter{enumi} \setlength{\parsep}{0pt}
     \setlength{\itemsep}{3pt} \settowidth{\labelwidth}{(99)}
     \sloppy}
\item \label{Dawson:1988}
P.~Nason, S.~Dawson and R.~K.~Ellis,
Nucl.\ Phys.\ {\bf B303}, 607 (1988); {\bf B327}, 49 (1989); {\bf B335}, 260(E)
(1990).
\item \label{Been}
W.~Beenakker, H.~Kuijf, W.~L.~van Neerven and J.~Smith, Phys. Rev. D
{\bf 40}, 54 (1989);
W.~Beenakker, W.~L.~van Neerven, R.~Meng, G.~A.~Schuler, and J.~Smith,
Nucl. Phys. {\bf B351}, 507 (1991).
\item \label{LO:1978}
M.~Gl\"uck, J.F.~Owens and E.~Reya, Phys. Rev. D {\bf 17}, 2324 (1978); \\
B.~L.~Combridge, Nucl.\ Phys.\ {\bf B151} (1979) 429; \\
J.~Babcock, D.~Sivers and S.~Wolfram, Phys. Rev. D {\bf 18}, 162 (1978); \\
K.~Hagiwara and T.~Yoshino, Phys. Lett. {\bf 80B}, 282 (1979); \\
L.~M.~Jones and H.~Wyld, Phys. Rev. D {\bf 17}, 782 (1978); \\
H.~Georgi {\it et al.}, Ann. Phys. (N.Y.) {\bf 114}, 273 (1978).
\item \label{CDF}
F.~Abe {\it et al.}, [CDF Collaboration], Phys. Rev. Lett. {\bf 71}, 500
(1993);                \\
D.~Acosta {\it et al.}  [CDF Collaboration], Phys.\ Rev.\ D {\bf 65},
052005 (2002)
$[$ArXiv: hep-ph/0111359$]$.
\item \label{D0}
S.~Abachi {\it et al.}  [D0 Collaboration], Phys. Rev. Lett. {\bf 74},
3548 (1995).
\item \label{Italians}
M.~Cacciari, S.~Frixione, M.~L.~Mangano, P.~Nason, G.~Ridolfi, JHEP {\bf 0407},
033 (2004)  $[$ArXiv: hep-ph/0312132$]$.
%
%
%
\item \label{DREG}
G.~'t~Hooft and M.~Veltman, Nucl.\ Phys.\ {\bf B44}, 189 (1972);  \\
C.~G.~Bollini and J.~J.~Giambiagi, Phys.\ Lett.\ {\bf 40B}, 566 (1972); \\
J.~F.~Ashmore, Nuovo\ Cimento\ Lett.\ {\bf 4}, 289 (1972).
\item \label{ibyparts}
K.~G.~Chetyrkin and F.~V.~Tkachov,
       Nucl.\ Phys.\ {\bf B192}, 159 (1981);  \\
F.~V.~Tkachov,
  Phys.~Lett.\ {\bf 100B}, 65 (1981).
\item \label{Andrei}
E.~E.~Boos, A.~I.~Davydychev, Theor. Math. Phys. {\bf 89}, 1052 (1991); \\
A.~I.~Davydychev, J. Math. Phys. {\bf 33}, 358 (1992).
%
%
\item \label{AndreiExp}
A.~I.~Davydychev and M.~Yu.~Kalmykov, Nucl. Phys. {\bf B605}, 266 (2001)
$[$ArXiv: hep-th/0012189$]$.
\item \label{Binomial}
A.~I.~Davydychev and M.~Yu.~Kalmykov, Nucl. Phys. {\bf B699}, 3 (2004)
$[$ArXiv: hep-th/0303162$]$.
\item \label{Tarasov}
J.~Fleischer, F.~Jegerlehner and O.V.~Tarasov, Nucl. Phys. {\bf B672}, 303
(2003)
$[$ArXiv: hep-ph/0307113$]$.
\item \label{KM}
J.~G.~K\"{o}rner and Z.~Merebashvili, Phys. Rev. D {\bf 66}, 054023 (2002)
$[$ArXiv: hep-ph/0207054$]$.
%
%
%
\item \label{Multilogs}
A.~B.~Goncharov, Math. Res. Lett. {\bf 5}, 497 (1998) $[$available at
http://www.math.uiuc.edu/K-theory/0297$]$;          \\
%
S.~Moch, P.~Uwer and S.~Weinzierl, J. Math. Phys. {\bf 43}, 3363 (2002)
$[$ArXiv: hep-ph/0110083$]$.
\item \label{Weinzierl}
J.~Vollinga and S.~Weinzierl, hep-ph/0410259.
\item \label{Nierste}
U.~Nierste, D.~M\"{u}ller, M.~B\"{o}hm, Z.~Phys. {\bf C57}, 605 (1993).
\item \label{Private}
M.M.~Weber, private communication.
\item \label{Weber}
M.M.~Weber, Acta~Phys.~Polon. {\bf B35}, 2655 (2004)
$[$ArXiv: hep-ph/0410166$]$.
\item \label{second}
J.~G.~K\"{o}rner, Z.~Merebashvili and M.~Rogal. To be published.
\item \label{EPAPS}
See EPAPS Document No.~E-PRVDAQ-71-025507 for our
analytical results for the integrals in
MATHEMATICA format. A direct link to this document may be found in the
online article's HTML reference section. The document may also be reached via
the EPAPS homepage (http://www.aip.org/pubservs/epaps.html) or from ftp.aip.org
in the directory /epaps/. See the EPAPS homepage for more information.
\item \label{Passarino}
A.~Ferroglia, M.~Passera, G.~Passarino and S.~Uccirati, Nucl. Phys. {\bf
B680}, 199 (2004) $[$ArXiv: hep-ph/0311186$]$;
{\bf B650}, 162 (2003) $[$ArXiv: hep-ph/0209219$]$.
\item \label{Binoth}
T.~Binoth and G.~Heinrich, Nucl. Phys. {\bf B680}, 375 (2004)
$[$ArXiv: hep-ph/0305234$]$.
\item \label{Bernreuther}
W.~Bernreuther {\it et al.}, Nucl. Phys. {\bf B706}, 245 (2005)
$[$ArXiv: hep-ph/0406046$]$.
\item \label{KMC}
B.~Kamal, Z.~Merebashvili, and A.P.~Contogouris, Phys. Rev. D {\bf 51}, 4808
(1995); {\bf 55}, 3229(E) (1997).
\item \label{Lewin}
L.~Lewin, {\it Polylogarithms and Associated Functions} (North Holland, Amsterdam,
1981).
\end{list}

\end{document}